\documentstyle[11pt]{article}
\newcommand\pow[1]{{\cal P}(#1)}

\newcommand\limp{\Rightarrow}
\newcommand\rimp{\Rightarrow}
\newcommand{\liff}{\Leftrightarrow}
\newcommand{\citeyear}{\cite}

\newcommand\la{(}
\newcommand\ra{)}
\newcommand{\commentout}[1]{}

\newcommand\lang{{\cal L}}
\newcommand\s{\mbox{\rm 's}\ }
\newcommand\says{~{\it says}~}
\newcommand\cert{~{\it cert}~}
\newcommand\saysn{{\it says}}
\newcommand\certn{{\it cert}}
\newcommand\spk{\longmapsto} % speaks for
\newcommand\speaksfor{\spk}
\newcommand\bdto{\spk} % bound to
\newcommand\self{{\tt Self}}

\newcommand\be{\begin{enumerate}}
\newcommand\ee{\end{enumerate}}

\newcommand\bi{\begin{itemize}}
\newcommand\ei{\end{itemize}}
\newcommand\trans[2]{\tau_{#1}(#2)}

\newtheorem{THEOREM}{Theorem}[section]
\newenvironment{theorem}{\begin{THEOREM} \hspace{-.85em} {\bf :} }%
                        {\end{THEOREM}}
\newtheorem{LEMMA}[THEOREM]{Lemma}
\newenvironment{lemma}{\begin{LEMMA} \hspace{-.85em} {\bf :} }%
                      {\end{LEMMA}}
\newtheorem{COROLLARY}[THEOREM]{Corollary}
\newenvironment{corollary}{\begin{COROLLARY} \hspace{-.85em} {\bf :} }%
                         {\end{COROLLARY}}
\newtheorem{PROPOSITION}[THEOREM]{Proposition}
\newenvironment{proposition}{\begin{PROPOSITION} \hspace{-.85em} {\bf :} }%
                            {\end{PROPOSITION}}
\newtheorem{DEFINITION}[THEOREM]{Definition}
\newenvironment{definition}{\begin{DEFINITION} \hspace{-.85em} {\bf :} \rm}%
                            {\end{DEFINITION}}
\newtheorem{CLAIM}[THEOREM]{Claim}
                            {\end{CLAIM}}
\newtheorem{EXAMPLE}[THEOREM]{Example}
\newenvironment{example}{\begin{EXAMPLE} \hspace{-.85em} {\bf :} \rm}%
                            {\end{EXAMPLE}}
\newtheorem{REMARK}[THEOREM]{Remark}
\newenvironment{remark}{\begin{REMARK} \hspace{-.85em} {\bf :} \rm}%
                            {\end{REMARK}}
\newcommand{\thm}{\begin{theorem}}
\newcommand{\lem}{\begin{lemma}}
\newcommand{\pro}{\begin{proposition}}
\newcommand{\dfn}{\begin{definition}}
\newcommand{\rem}{\begin{remark}}
\newcommand{\xam}{\begin{example}}
\newcommand{\cor}{\begin{corollary}}
\newcommand{\prf}{\noindent{\bf Proof:} }
\newcommand{\ethm}{\end{theorem}}
\newcommand{\elem}{\end{lemma}}
\newcommand{\epro}{\end{proposition}}
\newcommand{\edfn}{\bbox\end{definition}}
\newcommand{\erem}{\bbox\end{remark}}
\newcommand{\exam}{\bbox\end{example}}
\newcommand{\ecor}{\end{corollary}}
\newcommand{\eprf}{\bbox\vspace{0.1in}}
\newcommand{\tn}{{\tt n}}
\newcommand{\tp}{{\tt p}}
\newcommand{\tg}{{\tt g}}
\newcommand{\tk}{{\tt k}}

\newcommand{\bbox}{\vrule height7pt width4pt depth1pt}

\newcommand{\intension}[2]{[\![ #1 ]\!]_{#2}}
\newcommand{\emph}[1]{{\em #1}}

\newcommand{\tq}{{\tt q}}
\newcommand\union{\cup}

\newcommand\sat{\models}

\newcommand\W{{\cal W}}

\newcommand\lna{{\it LNA}}
\newcommand\lub{\sqcup}
\newcommand\Nat{{\bf N}}
\newcommand{\setcpr}[2]{\{ #1 \;:\; #2 \}} %set comprehension

\newcommand{\false}{\mbox{\it false}}

\newcommand\llnc{LLNC} % Our logic, the Logic of Local Name Containment
\newcommand\llncs{LLNC$^{s}$}
\newcommand{\reft}{\mbox{{\rm REF2}}} % our version of ref2!
\newcommand{\tr}{{\tt r}}
\newcommand{\omodels}{\models_{{\rm o}}}
\newcommand{\cmodels}{\models_{{\rm c}}}
\newcommand{\NR}{\mbox{{\it NR}}}

\newenvironment{oldthm}[1]{\par\noindent{\bf Theorem #1:} \em \noindent}{\par}
\newenvironment{oldlem}[1]{\par\noindent{\bf Lemma #1:} \em \noindent}{\par}
\newenvironment{oldcor}[1]{\par\noindent{\bf Corollary #1:} \em \noindent}{\par}
\newenvironment{oldpro}[1]{\par\noindent{\bf Proposition #1:} \em \noindent}{\par}
\newcommand{\othm}[1]{\begin{oldthm}{\ref{#1}}}
\newcommand{\eothm}{\end{oldthm} \medskip}
\newcommand{\olem}[1]{\begin{oldlem}{\ref{#1}}}
\newcommand{\eolem}{\end{oldlem} \medskip}
\newcommand{\ocor}[1]{\begin{oldcor}{\ref{#1}}}
\newcommand{\eocor}{\end{oldcor} \medskip}
\newcommand{\opro}[1]{\begin{oldpro}{\ref{#1}}}
\newcommand{\eopro}{\end{oldpro} \medskip}
\newcommand{\AXinf}{\mbox{AX}_{\it inf}}
\newcommand{\AXfin}{\mbox{AX}_{\it fin}}
\newcommand{\AXinfself}{\mbox{AX}_{\it inf}^{\it self}}
\newcommand{\AXfinself}{\mbox{AX}_{\it fin}^{\it self}}

\begin{document}

%joe10
%joe6: IEEE style
\begin{titlepage}
%joe4: I didn't like ``complete logic'' (logic aren't complete, only
%axiomatizations).  We can still change the title later.
%\title{On Abadi's Logic for SDSI's Linked  Local Name Spaces}
%joe14: moved the \thanks (which was here) to a separate
%acknowledgements section at the end
%joe8: added in case it makes a difference for copyright reasons
\title{A Logic for SDSI's Linked  Local Name Spaces}
%joe10: no longer preliminary
%Preliminary Version%
\author{Joseph~Y. Halpern\\
Cornell University\\
Dept. of Computer Science\\
Ithaca, NY 14853\\
\verb=halpern@cs.cornell.edu=\\
%joe7: cut for IEEE version, so names fit on one line
%joe10: resinstated
\verb=http://www.cs.cornell.edu/home/halpern=\\
\and
Ron van der Meyden\\
%School of Computing Sciences\\
%University of Technology, Sydney\\
%PO Box 123, Broadway N.S.W. 2007\\
%Australia\\
%\verb+ron@socs.uts.edu.au+\\
%\verb+http://www-staff.socs.uts.edu.au/+$\sim$\verb+ron+\\
%ron12: put new address 
School of Computer Science and Engineering, \\
University of New South Wales, \\ 
Sydney 2052,\\ 
Australia. \\
\verb=meyden@cse.unsw.edu.au= \\ 
\verb=http://www.cse.unsw.edu.au/=$\sim$\verb=meyden=
}
\date{\today}
%joe6: IEEE style
\setcounter{page}{0}
\maketitle
\thispagestyle{empty}

\begin{abstract}
%ron3: STILL TO COME ...
Abadi
%joe5
%recently
has
introduced a logic to explicate the meaning of
local names in SDSI, the Simple Distributed Security Infrastructure
proposed by Rivest and Lampson. Abadi's logic does not correspond
precisely to SDSI, however; it draws conclusions about local names
that
%joe5
%are not warranted by SDSI.
do not follow from SDSI's name resolution algorithm.
Moreover, its semantics is somewhat
unintuitive. This paper presents the Logic of Local Name Containment,
%joe5:
%an alternative logic that
which
does not suffer from these deficiencies. It
has a clear semantics and provides a tight characterization of SDSI
name resolution.  The semantics is shown to be closely related to that
of logic programs, leading to an approach to the efficient
implementation of queries concerning local names. A complete
axiomatization of the logic is also provided.
\end{abstract}
\end{titlepage}

\section{Introduction}

Rivest and Lampson \cite{RL96}
%joe5
%recently
introduced SDSI---a Simple
Distributed Security Infrastructure---to facilitate the construction of
secure systems.%
%ron10: added footnote
\footnote{SDSI now forms the basis for the Simple Public Key
Infrastructure (SPKI) standardization work \cite{SPKI}. SPKI
simplifies some SDSI features (e.g., it eliminates groups) but adds
many others. We focus in this paper on the core naming features
of SDSI---there are some minor differences in the way that
SPKI has chosen to handle these features, but we believe that our work
is equally relevant to the the fragment of SPKI dealing with naming.}
In SDSI, principals (agents) are identified with public
keys.  In addition to principals, SDSI allows other names, such as {\tt
poker-buddies}.  Rather than having a global
name space, these names are interpreted {\em locally\/}, by each
principal.  That is, each principal associates with each name a set of
principals.  Of course, the interpretation of a name such as {\tt
poker-buddies} may be different for each agent.  However, a principal
can ``export'' his bindings to other principals.
%SDSI allows
%expressions such as {\tt Joe's poker-buddies} (where {\tt Joe} is a name),
%which are interpreted as the poker buddies of the principals associated
%with the name {\tt Joe}.
Thus, Ron may receive  a message from the principal he names {\tt Joe}
describing a set of principals Joe associates with {\tt poker-buddies}.
Ron may then refer to
%ron10: this set of principals .... could mean the ones in the message
the principals Joe associates with {\tt poker-buddies}
by the expression
{\tt Joe's poker-buddies}.

%ron1: perhaps better to use poker-budies rather than my-poker-budies above

Rivest and Lampson \cite{RL96} give an operational account of local
names;
%that is,
they provide a name-resolution algorithm that, given a
principal $\tk$
and a name $\tn$, computes the set of principals associated with $\tn$
according to $\tk$.
%Of course, to compute the set of principals
%associated with a name such as {\tt Joe's poker-buddies} in {\tt Ron}'s
%name space may involve computing the set of principals associated with
%${\tt poker-buddies}$ in {\tt Joe}'s name space.
Abadi \cite{Abadi98} has provided a logic that, among
other things, gives a more semantic account of local names.
%Its purpose, according to Abadi ``is to explain local names in a general,
According to Abadi, its purpose  ``is to explain local names in a general,
self-contained way, without requiring reference to particular
implementations.''
% Moreover,
Abadi shows that the SDSI
name-resolution algorithm can be captured in terms of a collection of
sound proof rules in his logic.

%joe5: moved the next sentence down and did some rewriting, to be
%sensitive to Martin's concerns
%We very much subscribe to Abadi's goal of using a logic to give a
%general account of naming.
%However, we find Abadi's semantics somewhat
%unclear, partly because he seems to be trying to capture too many
%features of naming with his constructs.
Abadi's focus is on axioms.  He constructs a semantics, not with the
goal of capturing the intended meaning of his constructs, but rather,
with the goal of showing that certain formulas are not
derivable from his axioms.  (In particular, he shows that {\em false\/}
is not derivable, showing that his axioms are consistent.)
%ron5: Perhaps due to
While adequate for
Abadi's restricted goals, his semantics
validates some formulas that we certainly would not expect to be valid.
 One consequence of this is
that, while he is able to pinpoint some potential concerns with the
logic, the resolution of these concerns is less
%ron3: clear. - overused
satisfactory.
For example,
he observes that adding two seemingly reasonable axioms to his logic
allows us to reach quite an unreasonable conclusion.   However, it is not
%ron3: clear
obvious
from
%ron5: his semantics
% .. well, that actually makes a decision!
the semantic intuitions provided by Abadi
which (if either) of the axioms is
unreasonable, or why it is unreasonable.  Moreover, while he proves that
this particular unreasonable conclusion is not derivable in his
framework, as we show, a closely related (and equally unreasonable)
conclusion is in fact
%joe5: actually, I don't think it's derivable, although it is valid
%derivable.
valid.  This means we have no assurance that it or other similar
formulas cannot be derived from Abadi's axioms.

%joe5: moved down from above
We very much subscribe to Abadi's goal of using a logic to give a
general account of naming.
In this paper, we provide a logic whose syntax is very similar to
Abadi's, but whose semantics is quite different and, we believe,
%joe5
%clearer.
captures better the meaning we intend the constructs to have.
Nevertheless, all but one of Abadi's name space axioms are sound in our
system.

%joe5: added, to stress point.
We remark that, in a sense, our task is much easier than Abadi's, since
we give the constructs in the logic a somewhat narrower reading than he
does.  Abadi tends to intertwine and occasionally identify issues of
naming with issues of rights and delegation.  (Such an identification is
also implicitly made to some extent in designs such as
PolicyMaker \cite{BFL96}.)
We believe that it is important to treat these issues separately.
Such a separation allows us to both give a cleaner semantics for each of
the relevant notions and to clarify a number of subtleties.
This paper focuses on naming, which we carefully separate from the other
issues; a companion paper \cite{HMS} considers authority and delegation.

We believe that our approach has a number of significant
advantages:
\begin{itemize}
\item We can still simulate the SDSI's name resolution algorithm;
Abadi's extra axiom is unnecessary.  In fact, our logic captures SDSI's
name resolution more accurately than Abadi's.  Abadi's logic can draw
conclusions that SDSI's name resolution cannot; our logic, in a precise
sense, draws exactly the same conclusions as SDSI's name resolution
algorithm.
\item According to our semantic intuition, one of Abadi's
proposed additional axioms is in fact quite unreasonable; it does not
hold under our semantics, and it is quite clear why.
\item
We are able to provide a sound and complete axiomatization of
our logic.
%joe5
Thus, unlike Abadi, we have a proof system that corresponds precisely to
our semantics.
%(which Abadi did not do for his).
%joe3: this will be meaningless in the intro
%joe5
This will allow us to prove stronger results than Abadi's about formulas
that cannot be derived in our framework.
%and to prove a much stronger
%result about the kinds of conclusions that can be derived in our
%framework.
%ron10: added
Our completeness proof also yields a
%joe11
%decision procedure for validity of
(provably optimal) NP-complete decision procedure for satisfiability of
formulas in the logic.
\item Our logic is closely related to Logic Programming.  This allows us
to translate queries about names to Logic Programming queries, and thus
use all the well-developed Logic Programming technology to deal with
such queries.
\item Our approach opens the road to a number of generalizations,
which allow us to deal with issues like permission, authority, and
delegation
%joe5: we've already mention that HMS is a companion paper.
%we discuss these in a companion paper \cite{HMS}.
\cite{HMS}.
\end{itemize}

The rest of this paper is organized as follows.  In Section~\ref{abadi},
we review Abadi's logic and, in the process, describe SDSI's naming
scheme.  We also point out what we see as the problems with Abadi's
approach.  In Section~\ref{sec:llnc}, we give the syntax and semantics of our
logic,
%ron3: ... [[STILL TO COME]]
and present a complete axiomatization. In Section~\ref{sec:nras}
we show that our logic provides a tight characterization of
SDSI name resolution. Section~\ref{sec:lp} deals with the connection
between our account of SDSI name resolution and logic programming, and
Section~\ref{sec:self} concerns $\self$, an additional
construct considered by Abadi. Section~\ref{sec:concl} concludes.

\section{SDSI's  Name Spaces and Abadi's Logic}\label{abadi}

In this section, we briefly review SDSI's naming scheme and Abadi's
logic, and discuss our criticism of Abadi's logic.
%The reader is referred to the original sources for more details.
%ron5: Note that SDSI is still evolving.
% .. Martin suggested that SPKI had settled
Like Abadi, we are basing
our discussion on  SDSI 1.1 \cite{RL96}.

%ron1: SDSI has bee incorprated into SPKI, so probably further
%   developments will happen there rather than to SDSI -
%   check with Rivest

\subsection{SDSI's Name Spaces}

%SDSI has {\em global names\/} and {\em local names}.
%ron2: this is Abadi's terminology, so clarified
SDSI has {\em local names} and
a set of reserved names, which we
%joe3: technically, it should be "shall".  Cutting it avoids the problem
%will
refer to as
{\em global names}.
Both are
associated with sets of principals, but the set of principals associated
with a local name depends on the principal
%ron10: added
owning the local name space,
while the set of principals
associated with a global name does not.
%Principals are in fact a subset of global names.
We denote the set of global names by $G$ with
generic element $\tg$, the set of local names by $N$ with generic element
$\tn$, and the set of keys (principals) by $K$ with generic element
$\tk$.  We assume that all these sets are pairwise disjoint
%joe10
and that $K$ is nonempty.
%joe2: as we discussed
%assume that $G$ and $N$ are disjoint, and $K$ is a subset of $G$.
%joe3: added
{\em Global identifiers} are either keys or global names.%
%joe4: added
\footnote{Note that Abadi uses $G$ for global identifier; thus, his $G$
corresponds to our $G \union K$.}

The elements of $K \union G \union N$ are said to be {\em simple names}.
We form {\em principal expressions\/} from simple names inductively.
%ron10: reworked this para for correctness
%: Simple
%names are principal expressions, and if $\tp_1, \ldots, \tp_m$ are
%principal expressions,
%then so is $({\tt ref:} \tp_1, \ldots, \tp_m)$, which is more
%typically written as $\tp_1$'s \ldots $\tp_{m-1}$'s $\tp_m$.%
%%joe5: added footnote and following two sentences
%\footnote{Like Abadi, we allow $m$ to be 0, taking ({\tt ref:}) to be
%the current principal.  In Section~\ref{sec:self}, we follow Abadi by
%considering an expression $\self$ that represents ({\tt ref:}).}
%Abadi's semantics (and ours) makes ${\tt ref}$ associative, in that
%($\tp\s\tq)\s \tr$, $\tp\s \tq\s \tr$, and
%$\tp\s (\tq\s\tr)$ are all the same.  In light of this, we can
%ignore parenthesization when writing such expressions.
%%joe1: should we call them principal expressions too?
%%ron1: that's what I've been doing below. I think it makes sense:
%%    principal expressions = P
%%    local names = N
%%    global names = G
%%    principal expressions = things of the form (p's q)
%%joe2: let's discuss; in any case, added sentence below
%joe5: added this to tie it in to previous part of paragraph
Simple names are principal expressions, and if $\tp$ and $\tq$ are
principal expressions, then so is $(\tp\s \tq)$.  Abadi's semantics
(and ours) makes the latter operation associative, in that
$((\tp\s\tq)\s \tr)$ and $(\tp\s (\tq\s\tr))$ have the same meaning.
In light
of this, we can ignore parenthesization when writing such expressions.
The expression $\tp_1$'s \ldots $\tp_{m-1}$'s $\tp_m$ is written in
SDSI as $({\tt ref:} \tp_1, \ldots, \tp_m)$.\footnote{%
%ron10: Like Abadi, we allow $m$ to be 0,
SDSI allows $m$ to be 0,
taking ({\tt ref:}) to be
the current principal.  In Section~\ref{sec:self}, we follow Abadi by
considering an expression $\self$ that represents ({\tt ref:}).}
We remark for future reference that
SDSI has a
special global name denoted ``${\tt DNS!!}$'', which represents the root of
the DNS (Internet mail) hierarchy; this allows us to express an email
address such as {\tt bob@fudge.com} as ${\tt DNS!!}$'s {\tt com}'s {\tt
fudge}'s
%ron8: {\tt Bob}.
{\tt bob}.

%joe5: changed to singular.
%SDSI allows principals to issue {\em certificates} of the form
%$\tn \spk \tp$, signed with their keys.%
SDSI allows a principal to issue {\em certificates} of the form
$\tn \spk \tp$, signed with its key.
%joe5: moved footnote, since we haven't talked about binding yet.
%\footnote{This is a simplification of the SDSI notation; SDSI also
%allows other forms of binding that we do not consider here.}
If $\tk$ issues such a certificate, it has the effect of binding local
name $\tn$ in $\tk$'s name space to the
%ron5: cut to further address Martins complaint: set of
principals denoted by the principal expression $\tp$.%
%joe5: moved footnote
\footnote{SDSI also
allows other forms of binding that we do not consider here.  Our
notation is also a simplification of that used by SDSI.}
Notice that only principals
issue certificates, and that these certificates bind a local name (not a
global name) to some set of principals.  In general, a local name may be
bound to a unique principal, no principal, or many principals.  SDSI
allows a principal $\tk$ to issue certificates
%ron3: $\tn \gets \tp_1$ and $\tn \gets \tp_2$.
$\tn \spk \tp_1$ and $\tn \spk \tp_2$.
This has the effect of binding $\tn$ to (at least)
%ron5: cut: the union of
the principals denoted by $\tp_1$ and $\tp_2$.

SDSI provides a name-resolution algorithm for computing the set of 
%joe14: oops!
%names bound to a principal.  
principals bound to a name.
The core of the algorithm consists of a
nondeterministic procedure
%joe2: slight rewriting here; added B, for consistency with Martin
%joe3: redid
REF2.
For ease of exposition, we take REF2 to have four arguments: a principal
$\tk$, a function $c$ that associates with each principal $\tk'$ a set
of bindings (intuitively, ones that correspond to certificates signed by
$\tk'$), a function $\beta$ which associates with each global name
$\tg$ a set of principals (intuitively, the ones bound to $\tg$), and a
principal expression $\tp$.
REF2($\tk$,$\beta$,$c$,$\tp$)
%a set of bindings $B$ in $\tk$'s name space,
%a collection of signed certificates $C$,
%a set $\beta$ of bindings of global names to keys,
%and a principal expression
returns the principal(s) bound to $\tp$ in $\tk$'s name
space, given the bindings $\beta$ and the certificates $c$.
%Since
REF2 is nondeterministic;
%, we assume that
the set of possible outputs of REF2 is
%precisely
taken to be the set of
principals bound to $\tp$ in $\tk$'s name space.  REF2 is described in
Figure~\ref{fig:reft}.%
\footnote{Our version of REF2 is similar, although not identical, to
Abadi's.   Like Abadi's,
it is simpler than that in \cite{RL96}, in that we do not deal
with a number of issues, such as quoting or encrypted objects, dealt
with by SDSI.   Our presentation of REF2 differs from Abadi's mainly in
its treatment of global names.  Abadi assumes that REF2 takes only two
arguments,
{\tt o} and {\tt p}, where {\tt o} is either a
%joe3: I think this is right
global identifier (i.e., an element of $G \union K$)
or {\tt current principal}, denoted {\tt cp}. Although he does not write
$c$ explicitly as an argument, he does assume that there is a set he
denotes assumptions({\tt o}) that includes bindings corresponding to
signed certificates.  In addition,
it includes bindings for {\tt cp}.  We do not have a distinguished
current principal; rather, if the current principal is $\tk$, then
for uniformity we assume that all of the current principal's bindings
are also described by the bindings in $c(\tk)$.
More significantly, if {\tt g} is a global name, then Abadi's REF2({\tt
o},{\tt g}) would return {\tt g}, while ours would return some principal
{\tt k} to which {\tt g} is bound in $\beta$.  Our approach seems
more consistent with the SDSI presentation of REF2, but this difference
is minor, and all of Abadi's results hold for our presentation of REF2.}
%joe6: made it figure*, so it cuts across columns
\begin{figure*}
\begin{center}
\parbox{5in}{
\begin{tabbing}
pp\= \kill
%joe3: slight rewriting for consistency
%\reft($w$,\tp,\tk) \\
\reft(\tk,$\beta$,$c$,\tp) \\
\> if $\tp \in K$ then return(\tp) \\
\> else \=if $\tp \in G$\\
\> \> then \=if $\beta(\tp) = \emptyset$ then fail\\
%ron3: \>\>\>else return(q) for some $q \in \beta(tp)$\\
% to emphasize the output is a key
\>\>\>else return($\tk'$) for some $\tk' \in \beta(\tp)$\\
\> else if $\tp$ is a local name $\tn$ in $N$ \\
\> \> then if $c(\tk)= \emptyset$ then fail \\
\> \> \> else for some $\tn \spk \tq\in c(\tk)$
return(\reft(\tk,$\beta$,$c$,\tq))\\
\> else if $\tp$ is of the form $\tq\s \tr$ \\
\> \> then return(\reft(\reft(\tk,$\beta$,$c$,\tq),$\beta$,$c$,\tr))
\end{tabbing}
}
\end{center}
\caption{Procedure \reft \label{fig:reft}}
\end{figure*}

\subsection{Abadi's Logic: Syntax, Semantics, and Axiomatization}

The formulas in Abadi's logic are formed by starting with a set of
primitive propositions and formulas of the form $\tp \speaksfor \tp'$,
where $\tp$ and $\tp'$ are principal expressions.
%ron2: added
%joe3: Abadi does distinguish keys and global names.  *Global
%identifiers* includes both.
%(We remark that Abadi
%%does not make a distinction between keys and global names, and
%views the set of principals as being a subset of the set of global
%names. We return to this point below.)
More
complicated formulas are formed by closing off under conjunction,
negation,
and formulas of the form $\tp\says \phi$, where $\phi$ is a formula.

%joe5: rewrote paragraph, in light of Martin's comments.
%Abadi discusses two possible interpretations for $\speaksfor$:  One is
%that $\tp \speaksfor \tp'$ means that $\tp$ is bound to $\tp'$ in the
Abadi views $\tp \speaksfor \tp'$ as meaning that
$\tp$ is ``bound to'' $\tp'$.  He considers two possible
interpretations
of ``bound to''.  The first is equality; however, he rejects this as
being
inappropriate.  (In particular, it does not satisfy some of his axioms.)
%However,
%since SDSI allows only local (simple) names to be bound to principal
%expressions, it is not clear, under this interpretation, what $\tp
%\speaksfor \tp'$ should mean if $\tp$ is a principal expression.%
%\footnote{Abadi tries to give some intuition for allowing arbitrary
%principal expressions on the left-hand side of $\speaksfor$ in
%\cite[Section~2.5]{Abadi98}, but we find this intuition difficult to
%follow.}
The second is that $\tp \speaksfor \tp'$ means
%joe5: we had this backwards, given Martin's semantics; changed the rest
%consistently.
%$\tp$ ``speaks-for'' $\tp'$, in the sense discussed in
$\tp'$ ``speaks-for'' $\tp$, in the sense discussed in
\cite{ABLP93,LABW92}.
Roughly speaking, this says that any message certified by $\tp'$ should
be viewed as also having been certified by $\tp$.  While the
``speaking-for'' interpretation is the one favored by Abadi, he does not
commit to it.  Note that under Abadi's ``speaking-for'' interpretation,
it makes sense to write $\tp \speaksfor \tp'$ for arbitrary principal
expressions $\tp$ and $\tp'$.  However,
SDSI allows only local (simple) names to be bound to principal
expressions.  We shall make a similar restriction in our logic (and,
indeed, under our semantic interpretation of binding, it would not make
sense to allow an arbitrary principal expression to be bound to another
one.)

%joe5
The ``speaks-for'' interpretation intertwines issues of delegation with
those of naming.  As we suggested in the introduction, we believe these
issues should be separated.  We shall give $\speaksfor$ a different
interpretation that we believe is simpler and more in the spirit of
binding.   We believe that the
``speaks-for'' relation of \cite{ABLP93,LABW92}
should have quite different semantics than that
of binding names to principals.  (We hope to return to this issue in
future work.)

Abadi
%ron1: added
interprets
$\tp \says \phi$ as ``the principal denoted by $p$
%joe5
%makes statement $\phi$''.  In the case where $\tp$ is a key (i.e,
makes a statement that implies $\phi$''.  In the case where $\tp$ is a
key (i.e.,
principal) $\tk$, this could mean that $\tk$ signs a statement  saying
$\phi$.
%joe5: I agree with Martin that this is relatively clear under his
%interpretation.
%It is not clear to us what this formula should mean (at
%least in the context of SDSI) if $\tp$ is not a principal.
%joe5
Under our more restrictive interpretation, this is exactly how we
interpret our analogue to $\saysn$.

In any case, note that Abadi
%%joe3
%%translates SDSI's binding $\tn \gets \tp$ as
%%$\tn \speaksfor \tp$ and translates $\tk$ signing a certificate saying
%captures SDSI's binding $\tn \gets \tp$ by the formula
%$\tn \speaksfor \tp$ and captures $\tk$ signing a certificate saying
%$\tn \gets \tp$ by the formula $\tk\says \tn \speaksfor \tp$.
%ron3: to conform to the above change
translates SDSI's local name $\tn$ being bound to $\tp$ as
$\tn \speaksfor \tp$ and captures $\tk$ signing a certificate saying
$\tn$ is bound to $\tp$ by the formula $\tk\says \tn \speaksfor \tp$.
%joe2:
For future reference, it is worth noting that, in order to capture
the binding of names to principals, no use is made of primitive
propositions.

%joe10
%Abadi's interprets formulas in his logic with respect to a tuple
Abadi interprets formulas in his logic with respect to a tuple
$(\W,\alpha,\rho,\mu)$.  The function $\alpha$ maps global identifiers
($G \union K$) to subsets of $\W$.  The function $\rho$ maps $N \times
\W$ to subsets of $\W$.  Finally, $\mu$ associates with each world
(principal) $\tk$ and primitive proposition $p$ a truth value
$\mu(p,\tk)$.

Abadi does not provide any intuition for his
semantics, but suggests that $\W$ should be thought of as a set of
possible worlds, as in modal logic. However, he also suggests
%joe7: I don't think he actually says this in the paper
[private communication, 1999]
that his
semantics was motivated by the work of Grove and Halpern
\cite{GroveH2}, in which the corresponding set contains {\em pairs}
consisting of a world and an agent.  Some of Abadi's definitions make
more intuitive sense
%joe7
%for the former, some for the latter.
if we think of $\W$ as a set of agents, while others make more sense if
we think of $\W$ as a set of worlds.
We elaborate on this point below.
%joe9: cut, as you requested, but added comment below
%We remark that, in general, we believe it will be necessary to follow
%the approach of Grove and Halpern and interpret formulas with respect to
%both worlds and agents (where worlds contain information about such
%things as the truth value of primitive propositions, messages that were
%sent, and so on); in fact, that is what we do in the semantics of our
%logic.
%ron7: I'm not completely happy with this sentence, and prefer to cut it:
% a) necessary for what purposes? the same as in this paper?
% b) the last part is slightly misleading - from teh perspective of
% modal logic we effectively have a fixed world  in our semantics and vary only the agent.

Given $\tk \in \W$ and $\tp \in P$, Abadi defines
$\intension{\tp}{\tk}$
%---intuitively, the set of principals bound to
%principal expression $\tp$ according to $\tk$---
inductively, as follows:
\begin{itemize}
\item $\intension{\tg}{\tk} = \alpha(\tg)$, for
%joe4
%$\tg \in G$
$\tg \in G \union K$
\item $\intension{\tn}{\tk} = \rho(\tn,\tk)$ for $\tn \in N$
\item $\intension{\mbox{$\tp_1$'s $\tp_2$}}{\tk} = \union
\{\intension{\tp_2}{\tk'}: \tk' \in \intension{\tp_1}{\tk}\}$
\end{itemize}
%joe7
%ron7: Notice here that our notation suggests that we are thinking of the
% ... I want to stay neutral on that, and not even sugggest it!
Here we have used a notation corresponding to the interpretation of
the ``worlds'' in $\W$ as agents.
%ron7: more neutral form
%; this is because we think of
Using this interpretation we may think of
$\intension{\tp}{\tk}$ as the set of principals bound to principal
expression $\tp$ according to $\tk$.
%joe7: moved this back from below
%Under this interpretation
%$\intension{\tp}{\tk}$ is the set of principals that principal $\tk$
%refers to using the expression $\tp$.
The clause for
$\intension{\mbox{$\tp_1$'s $\tp_2$}}{\tk}$ then says that if
$\tk'$ is one of the principals referred to by $\tk$ as $\tp_1$, then
$\tk$ uses $\tp_1\s\tp_2$ to refer to any principal referred to by
$\tk'$ as $\tp_2$.
%joe7: cut
%As we shall see, the latter interpretation
%is the one we adopt in our logic, although to make it coherent we need
%to take a different appraoch to the semantics of the other
%constructs. We discuss this in Section~\ref{sec:llnc:sem}.

Abadi also defines what it means for a formula $\phi$ to be true
at
%according to principal
world $\tk\in \W$, written $\tk \sat \phi$, inductively, by
\begin{itemize}
\item $\tk \sat p$ iff $\mu(p,\tk) = {\bf true}$, if $p$ is a primitive
proposition
\item $\tk \sat \phi \land \psi$ iff $\tk \sat \phi$ and $\tk \sat \psi$
\item $\tk \sat \neg \phi$ iff $\tk \not \sat \phi$
\item $\tk \sat \tp \speaksfor \tp'$ iff $\intension{\tp}{\tk} \subseteq
\intension{\tp'}{\tk}$
\item $\tk \sat \tp \says \phi$ iff $\tk' \sat \phi$ for all
$\tk' \in \intension{\tp}{\tk}$.
\end{itemize}

These clauses defining $\sat$ are quite intuitive if one interprets
$\W$ to be a set of worlds and considers $\intension{\tp}{\tk}$ to be
the set of worlds consistent with what principal $\tp$ has said at world
$\tk$.  In particular, under this interpretation, the clause for
$\saysn$ can be read as stating that $\tp \says \phi$ if $\phi$ holds
in all worlds consistent with what $\tp$ has said.  The clause for
$\speaksfor$ also has quite a plausible reading under the
``speaks-for'' interpretation of this construct: it states that $\tp'$
speaks for $\tp$ if all worlds consistent with what $\tp$ has said are
consistent with what $\tp'$ has said, i.e., $\tp$ is constrained to
speak consistently with what $\tp'$ has said.
%ron7: On the other hand,
% -- avoid repetition
However, it seems rather difficult to extend this intuitive
reading to encompass the
%joe7
inductive
definition of $\intension{\tp}{\tk}$. In
particular, it is far from clear to us what intuitive understanding to
assign to the clause for $\intension{\mbox{$\tp_1$'s $\tp_2$}}{\tk}$
on this reading.

%joe7:
On the other hand, note that if we interpret the worlds as agents, then
we can think of $\tk \sat \phi$ as saying that $\phi$ is true
%ron7: added next line, else agents decide every formula
when local names are interpreted
according
to agent $\tk$.
%ron7: added explanation of what is wrong with this!
But this reading of the clauses, when combined with the
intuitive reading of $\intension{\tp}{\tk}$ as the set of principals
that $\tk$ refers to using $\tp$,
also has its difficulties.
Intuitively, when $\tn$ is bound to $\tp$ in principal $\tk$'s
local name space, the principals that $\tk$ refers to using
$\tp$ should be a subset of the principals that $\tk$  refers to
using $\tn$.
Abadi interprets $\tn$ being bound to $\tp$ as
$\tn \speaksfor \tp$;
%ron8: cut
%but under this reading,
this holds with respect to principal $\tk$
when $\intension{\tp}{\tk}$ is a {\em superset\/} of
$\intension{\tn}{\tk}$.
This is precisely the opposite of what we would expect.
Thus, neither the interpretation of $\W$ as a set of worlds nor the
interpretation of $\W$ as a set of agents gives a fully satisfactory
justification for Abadi's semantics.
As we shall see, in our semantics, the interpretation of a principal
expression  $\tp$ according to an agent will be a set of agents, but we
use the reverse of Abadi's  containment to represent binding.

Abadi provides an axiom system for his logic, which has three components:
\begin{enumerate}
\item The standard axioms and rules of propositional logic.
\item The standard axiom and rule for modal logic for the $\saysn$ operator:
\[ (\tp \says (\phi \limp \psi)) \limp ((\tp \says \phi) \limp
(\tp \says \psi))\]
\[ \begin{array}{c}
   \phi \\
\hline
\tp \says \phi
\end{array}
\]
\item New axioms dealing with linked local name spaces, shown in
Figure~\ref{fig:aax}.
\end{enumerate}
He shows that this axiomatization is sound, but conjectures it is not
complete.

\begin{figure*}
\[ \begin{array}{ll}
\mbox{Reflexivity:} & \tp \spk \tp \\
\mbox{Transitivity:} & (\tp \spk \tq) \limp ((\tq \spk \tr) \limp
(\tp
\spk \tr))\\
\mbox{Left Monotonicity:} & (\tp \spk \tq) \limp ((\tp\s \tr) \spk (\tq\s
\tr))\\
\mbox{Globality:} & (\tp\s \tg) \spk \tg
                 \mbox{ if $\tg$ is a global identifier} \\
\mbox{Associativity:} & ((\tp\s \tq)\s \tr) \spk (\tp\s (\tq\s \tr)) \\
                     & (\tp\s (\tq\s \tr)) \spk ((\tp\s \tq)\s \tr) \\
\mbox{Linking:} & (\tp \says (\tn \spk \tr) \limp ((\tp\s \tn) \spk
(\tp\s
\tr))\\
               & \mbox{ if $\tn$ is a local name} \\
\mbox{Speaking-for:} & (\tp \spk \tq) \limp ((\tq \says \phi) \limp \tp
\says \phi)
\end{array}
\]
\caption{Abadi's axioms for linked local name spaces}\label{fig:aax}
\end{figure*}

%joe3
%\subsection{Relating Abadi's Logic to SDSI}
\subsection{Name Resolution in Abadi's Logic}

Abadi proves a number of interesting results relating his logic to SDSI.
First, he shows that in a precise sense his logic can simulate REF2.
%Consider the name-resolution rules described in Figure~\ref{NR-rules}.
%[[COPY FROM PAPER:]]
%joe3:
He provides a collection of name-resolution rules $\NR$
%joe10: decided to cut.  What do you think?
%ron10: no problem with this.
%(which we present in the full paper \cite{HM99full})
 and proves the following results:%
\footnote{The results stated here are a variant of those stated in
Abadi's paper, since our version of REF2 differs slightly from his.
%ron8: don't need next line in this version as we don't say what they are
%Our collection $\NR$ of rules also differs slightly from his.
Nevertheless, the proofs of the results are essentially identical.}

%joe3: I agree
%[[ given the current form of section 4, which is largely independent
%of the name resolution rules, we should perhaps reconsider the
%significance of and the extent to which we need to go into the name
%resolution rules!  Perhaps we can simplify this to a discussion of how
%REF2 is sound with respect to Abadis' logic, but mention that
%he provides a sound set of rules that are complete. ]]
%Abadi then proves the following two results:

\pro \label{pro:abadi1}
Given a collection of $c$ of bindings corresponding to signed
certificates and a set $\beta$ of bindings of global names to keys,
let $E$ be the conjunction of the formulas $\tk \says \tn\spk \tq$ for
each certificate $\tn\spk \tq \in c(\tk)$ and the
formulas
%ron3: $\tg \gets \tk$
$\tg \spk \tk$
for each $\tk \in \beta(\tg)$.
%$\tp$ and all global identifiers $\tf$ and $\tg$, we have
%\begin{itemize}
%\item $E\limp (\tp\spk \tf)$ is provable with the name resolution
%rules if and only if REF2(\cp,\tp) yields $\tf$.
%\item $E\limp ((\tg\s \tp)\spk \tf)$ is provable with the name
%resolution rules if and only if REF2(\cp,\tp) yields $\tf$.
Then $E\limp ((\tk\s \tp)\spk \tk_1)$ is provable with the name
resolution rules $\NR$ if and only if $\reft(\tk,\beta,c,\tp)$ yields
$\tk_1$.
\epro

\pro \label{pro:abadi2}
The name resolution rules are sound with respect to the logic.
%joe3
%Thus, given $E$ as in Proposition~\ref{pro:abadi1}, and any principal
That is, given $E$ as in Proposition~\ref{pro:abadi1} and any principal
expression~$\tp$, if $E\limp (\tp\spk \tk)$ is provable
using $\NR$ then $E\limp (\tp\spk \tk)$ is also provable in the logic.
\epro

%joe3: cut; we say it in English later, and that should suffice
%Together, these results have the following corollary:
%\cor \label{cor:abadi3}
%The name-resolution rules are sound with respect to the logic.
%That is, given $E$ as in Proposition~|ref{pro:abadi1}
%and any principal expression $\tp$, if REF2(\cp,\tp) yields
%a global identifier $\tf$, then
%$E \limp (\tp\spk \tf)$ is provable in the logic.
%\ecor

These results show that
any bindings of names to principals that can be
deduced using REF2 can also be deduced using Abadi's logic.  However,
Abadi shows that his logic is actually more powerful than REF2, by
giving two examples of conclusions that can be deduced from his logic
but not
%from
using REF2:
%joe3: labeled the examples
\xam\label{counterexample1}
%joe3: rewrote to be consistent with presentation above
%\item using the Globality axiom and  we can immediately get ({\tt
%\tt Lampson's
Using the Globality, Associativity, and Transitivity axioms, if $\tk$
and $\tk'$ are keys, we immediately get
$\tk\s ({\tt Lampson}\s
\tk') \speaksfor \tk'$.  This result does not follow from the
REF2 algorithm.  That is, $\reft(\tk,\beta,c,{\tt Lampson}\s \tk'$) does
not necessarily yield $\tk'$ for arbitrary $c$ and $\beta$ (in
particular, it will not do so if
{\tt Lampson} is not bound to anything in $c$).
\exam
\xam\label{counterexample2}
%joe3: again, rewrote for consistency
%bindings $B = \{{\tt Lampson}\speaksfor {\tt KL1},~
%{\tt Lampson}\speaksfor {\tt KL2}\}$ and the certificates $C = \{{\tt
%KL1} \says ({\tt Ron}\speaksfor {\tt Rivest}), ~ {\tt KL2}\says ({\tt
%Rivest} \speaksfor \tk')\}$ (where $\tk'$ is an arbitrary key), we can can
%joe2: rewrote slightly
Suppose $c$ consists of the four certificates that correspond to the
following formulas:
$\tk \says ({\tt Lampson}\speaksfor \tk_1)$,
$\tk \says ({\tt Lampson}\speaksfor \tk_2)$,
$\tk_1 \says ({\tt Ron}\speaksfor {\tt Rivest})$,
and $\tk_2 \says ({\tt Rivest} \speaksfor \tk_3)$
(where $\tk$, $\tk_1$, $\tk_2$, and $\tk_3$ are keys).
Using the Speaking-for axiom, it is not hard to show that we
can conclude that $\tk\s (\mbox{{\tt Lampson's Ron}}) \speaksfor \tk_3$.
It is easy to show that REF2 cannot reach this conclusion; that is,
$\reft(\tk,\beta,c,\mbox{{\tt Lampson's Ron}})$ does not yield $\tk_3$
for any
$\beta$.%
%ron10: added footnote
\footnote{SPKI certificates and SDSI certificates have a slightly
different syntactic form. A SPKI certificate issued by $\tk$ to bind
$\tn$ to $\tp$ could be expressed in the logic as $\tk \says (\tk\s
\tn \speaksfor \tp)$.
Abadi has remarked [private communication
1999], that if we rewrite the example using assertions in this form,
the corresponding conclusion of this example would not follow in
%joe11
%Abadi's logic.
his logic.
We have followed the SDSI format for certificates in
this paper, but note that after some minor changes to the definitions,
%ron10: all it takes is changing defn of consistency of l, I think
%joe11: not Example 2.4
%all our results would still apply to SPKI certificates.}
all the results in Sections 3--5 would still apply to SPKI certificates.}
%joe2: cut this
%We prove this
%formally in the next section by showing that (a) the conclusion does not
%follow in our logic and (b) our logic draws precisely the same
%conclusions as REF2.
\exam

%joe5: expanded this paragraph
In reference to Example~\ref{counterexample1}, Abadi \citeyear{Abadi98}
says that ``it is not clear whether [these conclusions] are harmful, and they
might in fact be useful''.  In general, he views it as a feature of his
logic that it allows reasoning about names without knowing their
bindings [private communication, 1999].  While we agree that, in
general, reasoning about names without knowing their bindings is a
powerful feature, we believe it is important to make clear exactly which
conclusions are desirable and which are not.  This is what a good
semantics
can provide.  Under our semantics, neither of these two conclusions are
valid.
%As we shall see, neither of these two conclusions follows in our logic.
In fact, our logic draws precisely the same conclusions as REF2.
Of course, the conclusions of Examples~\ref{counterexample1}
and~\ref{counterexample2} are valid under Abadi's semantics but, as we
observed earlier, Abadi's semantics is not really meant to be used as a
guide to which conclusions are acceptable (and, indeed, as we shall see,
it validates a number of conclusions that do not seem so acceptable).

Abadi also considers the effect of extending his axiom system.  In
particular, he considers adding the following two axioms:
\begin{itemize}
\item the converse of Globality: $\tg \speaksfor (\tp\s \tg)$
\item a generalization of Linking: $(\tp\says (\tp_1 \speaksfor
\tp_2)) \rimp (\tp\s \tp_1 \speaksfor \tp\s \tp_2)$, for an arbitrary
principal
%%ron3: $\tp_1$.
%$\tp$.
%ron10: ....hmm, you were right here, changed it back
$\tp_1$ (instead of a local name).
\end{itemize}
%joe3
%Interestingly,
The generalization of Linking is in fact sound under
Abadi's semantics.  The converse of Globality is not, but
% the only reason that we may not have $\tk \sat \tg \speaksfor (\tp's \tg)$ is that
only because we may have $\intension{\tp}{\tk} = \emptyset$.
%ron2: If $\intension{\tp}{\tk} \ne \emptyset$, then we have Note.
%joe2: rewrote, in line with your intuition
%This means that if we slightly extended the
%syntax of Abadi's logic to allow formulas of the form $\tp \ne
%\emptyset$, with the obvious semantics, then the following variant of
%the converse of Globability would be sound under Abadi's semantics:
%$\tp \ne \emptyset \rimp (\tg \speaksfor (\tp's \tg))$.
%ron1: $p = \emptyset$ is expressible as $p \says false$ !
%ron2:  $\tk \sat \tg \speaksfor (\tp's \tg)$
Note that $\intension{\tp}{\tk} = \emptyset$ iff $\tk \sat \tp \says
\false$; thus, the following variant of
the converse of Globability is sound under Abadi's semantics:
$\neg(\tp \says \false) \rimp (\tg \speaksfor (\tp\s \tg))$.

This is quite relevant to our purposes because Abadi shows that if we
added the two axioms above to his system, then from
$\tk \says ({\tt DNS!!} \speaksfor \tk)$, we can conclude ${\tt DNS!!}
\speaksfor \tk$.
Thus, just from $\tk$ saying that ${\tt DNS!!}$ is bound to $\tk$, it
follows that ${\tt DNS!!}$ is indeed bound to $\tk$.  This is particularly
disconcerting under Abadi's ``speaks-for'' interpretation, where ${\tt
DNS!!} \speaksfor \tk$ becomes ``$\tk$ speaks for {\tt DNS!!}''.  We
certainly do not want an arbitrary principal to speak for the name
server!

Abadi proves a result showing that such conclusions are not derivable
from
%joe3
hypotheses of a certain type in
his logic (which does not have these two axioms).
%More precisely, he proves the following result.

\pro\label{assurance} {\rm \cite{Abadi98}} Let $\tk$ and $\tk'$ be distinct global
names; let $\phi$ be a formula of the form $(\tk'\says(\tn_1
\speaksfor \tp_1)) \land \ldots \land (\tk'\says(\tn_k \speaksfor
\tp_k))$, where
$\tn_1, \ldots, \tn_k$ are local names and $\tp_1, \ldots, \tp_k$ are
principal expressions; let $\psi$ be a formula of the form $(\tk \says
\psi_1) \land \ldots \land (\tk\says\psi_m)$, where $\psi_1, \ldots,
\psi_m$ are arbitrary formulas.  Then $\phi \land \psi \rimp (\tk'
\speaksfor \tk)$ is not valid.%
\footnote{Abadi's result actually says ``$\phi \land \psi \rimp (\tk'
\speaksfor \tk)$ is not derivable''; since his axiomatization is sound,
but not necessarily complete, the claim that it is not valid is
stronger, and that is what Abadi's proof shows.}
\epro

While Proposition~\ref{assurance} provides some assurance that
undesirable formulas are not derivable in the logic, it does not provide
much.  Indeed, if we allow the $\psi$ to include the formula $\neg (\tk'$
$\says \false)$, then the result no longer holds.  In fact, it follows
from our earlier discussion that the formula
$$(\tk \says ({\tt DNS!!} \speaksfor \tk)) \land \neg(\tk \says
 \false) \rimp ({\tt DNS!!} \speaksfor \tk)$$
is valid.
Moreover, it does not seem so unreasonable to allow
conjuncts such as $\neg(\tk \mbox { $\says$ } \false)$ as part of $\psi$.
%ron3: cut and replaced - not sure how this connects to the above
%Principals may indeed know that certain global names have
%nonempty bindings (indeed, they will often know exactly what principals
%various global names are bound to).
%joe5
%Indeed, it would seem that the typical situation is one in
%which principals' statements are consistent.
We certainly want to be able to use the logic to be able to say that if
a principal's statements are not blatantly inconsistent, then certain
conclusions follow.

%ron2: \section{An Alternative Logic}
\section{The Logic of Local Name Containment}\label{sec:llnc}

In this section we propose the Logic of Local Name Containment
(henceforth \llnc) as an alternative to Abadi's logic.
%ron3: cut to save space
% Rivest and Lampson state that
%\begin{quote}
%A local name can be undefined, or it can be bound to some value
%(i.e., some object). The principal may assign a value to a local name
%by issuing a corresponding certificate. If the local name already has a
%valid name/value certificate, the new certificate augments the old
%one, in the sense that an SDSI application is deemed to act correctly
%if it uses the name/value binding given in either certificate. (\cite{RL96},
%Section 5.2)
%\end{quote}
%\llnc\ seeks to capture these intuitions by interpreting local names
%as sets of principals and interpreting SDSI certificates as stating
\llnc\ interprets local names
as sets of principals and interprets SDSI certificates as stating
containment relationships between these sets.  We define the syntax in
Section~\ref{sec:llnc:syn}.  In Section~\ref{sec:llnc:sem} we
describe two distinct semantics for the logic.
%ron3:
Section~\ref{sec:llnc:axioms} presents a complete axiomatization.

\subsection{Syntax} \label{sec:llnc:syn}

%Our alternate logic
\llnc\ has syntactic elements that are closely related to
the syntactic elements of Abadi's logic.  However,
%joe3
%as our semantic intuitions are different
%are not  we will denote these elements differently, and
%moreover make a number of restrictions on their range of
%applicability.
our notation differs slightly from Abadi's to help emphasize some of the
differences in intuition.

%There are three types of basic principal expressions.  We assume three
%mutually disjoint sets $K$,~$G$ and~$N$.  The set~$K$ represents the
%set of \emph{keys}, which may be used to sign certificates. The
%set~$G$ is the set of \emph{global names} and~$N$ is the set of
%\emph{local names}. Semantically, local and global names will
%represent sets of keys.
%More complex principal expressions are formed exactly as in Abadi's
%logic. The set of {\em principal expressions}~$P$ is the smallest set
%containing~$K$,~$G$ and~$N$, and such that if $\tp,\tq \in P$ then $(\tp\s
%\tq) \in P$.
%ron2: Just as with Abadi's logic,
% ... Abadi does not make these distinctions!
%As above,
Again, we start with keys~$K$, global names~$G$, and
local names~$N$, and form principal expressions from them.
%ron3: following seems not to have been said anywhere..
%ron10: For technical reasons, we make the (quite reasonable)
%       assumption that~$K$ is a finite set.
%joe11: cut this here; I think it's the wrong place
%We will consider below both the case that $K$ is restricted to be
%finite and the case that $K$ is infinite.
%joe6: cut
%but that~$N$ is an infinite set.
%ron2: added
%joe3: so does Martin
%(Note that
%unlike Abadi, we distinguish between
%global names and keys, and assume these are disjoint.)
%More complex principal expressions are formed exactly as in Abadi's
%logic.
%The set of {\em principal expressions}~$P$ is the smallest set
%containing~$K$,~$G$ and~$N$, and such that if $\tp,\tq \in P$ then $(\tp\s
%\tq) \in P$.
The formulas of our language are formed as follows:
\bi
\item If $\tp$ and $\tq$ are principal expressions then
$\tp \bdto \tq$ is a formula.

\item If $\tk\in K$ and $\phi$ is a formula then
$\tk \cert \phi$ is a formula.%
%ron10: added footnote
\footnote{For our account of SDSI naming, it would suffice to restrict
this clause to formulas of the form $\tk \cert \tn \bdto \tp$ where
$\tn \in N$ and $\tp \in P$: our semantics will treat more general
certificates as irrelevant to the meaning of principal expressions.
We allow the more general form for purposes of discussion and because
we envisage generalizations of the logic in which other types of
certificates will be required.}
%[[ could restrict this to  $\tk \cert \tn \bdto \tp$
%but see section~\ref{sec:gen}.]]

\item If $\phi_1$ and $\phi_2$ are formulas, then so are
$\neg \phi_1$ and $\phi_1 \land \phi_2$. As usual,
$\phi_1 \lor \phi_2$ is an abbreviation for
$\neg (\neg \phi_1 \land \neg \phi_2)$ and
$\phi_1 \limp \phi_2$ is an abbreviation for
$\neg \phi_1 \lor \phi_2$.
\ei
%
%joe3: Americanized formulae throughout
%We write $\lang$ for the set of all formulae.
We write $\lang$ for the set of all formulas.
%ron2:
(For simplicity, we  omit primitive propositions, although we
could easily add them. They play no role in Abadi's account of SDSI names,
nor will they in ours.)
%[[ omit basic statements - Abadi has them but makes no use of them ]]
%joe1: I agree, but we should comment on this
%ron1: perhaps best to do it already in the exposition of Abadi's logic

%%joe3
%%We read the expression $\tp \bdto \tq$ as ``$\tp$ contains $\tq$.''
%We read the expression $\tp \bdto \tq$ as ``$\tq$ is bound to $\tp$.''
%ron3: reinstated: we can still read it as contains even if
%      the notation differs! It supports our intuitions.
%  if you do change it back, note it should be ``$\tp$ is bound to $\tq$.''
%  I'm constantly getting confused on this myself, but SDSI terminology
%  seems to be that a local name is bound to an expression.
We read the expression $\tp \bdto \tq$ as ``$\tp$ contains $\tq$'';
%joe4: added
we intend for it to capture the fact that
%joe5: I think this is confusing.  How about
%$\tp$ is bound to $\tq$.
all the keys bound to $\tq$ are also bound to $\tp$.
%joe3
%This expression is our equivalent for Abadi's expression $\tp \spk \tq$;
%we use it to capture the binding of $\tq$ to $\tp$.
However, our
intuitions about the meaning of $\tp \bdto \tq$ are quite different from
Abadi's.
%joe3: since we're using the same symbol
%intuitions for $\tp \spk \tq$.
In particular, we do not
wish to interpret $\tp \bdto \tq$ as
%ron3: ``$\tp$ speaks for $\tq$.''  %that's the wrong way around
``$\tq$ speaks for $\tp$.''
We consider
the ``speaks for'' relation as being about rights and delegation,
which requires a more sophisticated semantics than we wish to consider
here.
%joe5
(See \cite{HMS} for a logic for reasoning about rights and delegation.)
The expression $\tp \bdto \tq$ should be understood as simply
asserting a containment relationship between the denotations of
principal expressions $\tp$ and $\tq$;
%joe3
this is exactly what our semantics will enforce.

We read the expression $\tk \cert \phi$ as ``$\tk$ has certified that
$\phi$.''  This corresponds roughly to Abadi's $\tk \says \phi$.
%joe5: expanded again.
There are two significant differences, however. For one thing, we
do not allow arbritrary principal expressions on the left-hand side;
only keys may certify a formula $\phi$.
%joe5: added
For another, our interpretation of $\certn$ is more restrictive
than Abadi's
$\saysn$, in that $\certn$ is treated quite syntactically; it refers to
an actual
%ron5: certification
certificate
issued by a principal, while $\saysn$ considers
logical consequences of such
%ron5: certifications
certificates.
As a consequence,
%joe3: I don't think this is the time to talk about semantic properties
%Another difference,
%resulting from the semantics to be introduced below, will be that that
%whereas ``$\tk \says$'' is closed under logical consequence, ``$\tk
%\cert$'' is not.
whereas $\saysn$ satisfies standard properties of modal operators (e.g.,
closure under logical consequence), $\certn$ does not.

\subsection{Semantics} \label{sec:llnc:sem}

Our semantics is designed to model the SDSI principle that
%joe5: Martin preferred this the other way around.  I think of binding
%as somewhat symmetric, but I slightly prefer Martin's reading.
%principals bind values (keys) to local names by issuing certificates.
principals bind names in their local name space to values by issuing
certificates.
The interpretation of a local name depends on the principal and
%joe3
%is a function of the certificates that have been issued
%by that concerning the local name.
the certificates that have been issued.
As the principal may rely on others for its
interpretation of local names, the certificates issued by other principals
also play a role.  The interpretation of global names and keys will be
independent of both the principal and the certificates that have been
issued.

%joe3
%A {\em world} will be a
A {\em world} is a
%structure
%joe3: using \beta to distinguish from Abadi's alpha
pair $w = (\beta, c)$, where $\beta:G
%joe3: I think it's worth doing this (although we don't really need it)
%\rightarrow \pow{K}$ and $c:K \rightarrow \pow{\lang}$.
%joe10: I think this is what was really intended:
\rightarrow \pow{K}$ and $c:K \rightarrow \pow{\lang}$
%where $\pow{X}$ (resp., $\powf{X}$) denotes the set of subsets (resp.,
%finite subsets) of $X$.
(where $\pow{X}$ denotes the set of subsets of $X$) and $\union_{\tk \in
K} c(\tk)$ is finite.  Intuitively,
the function $\beta$ interprets global (or fixed) names as sets of
keys. The intended interpretation of the function $c$ is that it
associates with every key $\tk$ the set of formulas $c(\tk)$ that have
been certified using this key. That is, if $\phi\in c(\tk)$ then,
intuitively, a certificate asserting $\phi$ has been signed using
$\tk$.\footnote{We make the simplifying assumption that certificates do
not have expiration dates. It is not difficult to extend the logic to
take into account certificate expiration;
%joe10: added pointer to our new paper, more intution.
see \cite{HM00}.  The assumption that $\union_{\tk \in K}c(\tk)$ is finite
is meant to enforce the intuition that only finitely many certificates
are issued.  None of our later results depend on this assumption, but it
seems reasonable given the intended application of the logic.}

%[[ could tighten $\pow{\lang}$ to a smaller set, but see
%Section~\ref{sec:gen}]]

Formulas of the logic will be interpreted in a world
with respect to a key. Intuitively, this key indicates the
principal from whose perspective we interpret principal expressions.
%joe4: it depends on all certificates, so this could be confusing
%In particular, the interpretation of local names will
%depend on the certificates issued by this principal.

%ron2: It is convenient to structure the presentation around
%two distinct satisfaction relations. The first of these
%involves an additional semantic construct. A
To interpret local names, we introduce an additional semantic
construct. A {\em local name assignment} will be a function $l:K\times
N \rightarrow \pow{K}$ associating each key and local name with a set
of keys. Intuitively, $l(\tk,\tn)$ is the set of keys represented by
principal $\tk$'s local name $\tn$. We write $\lna$ for the set of all
local name assignments.

Given a world $w=( \beta,c)$, a local name assignment $l$, and a key
$\tk$, we may assign to each principal expression $\tp$ an interpretation
$\intension{\tp}{w,l,\tk}$, a set of keys.
The definition is much like that of Abadi's $\intension{p}{\tk}$:
% This set is defined by the following recursion:
%
\bi
\item $\intension{\tk'}{w,l,\tk} = \{\tk'\}$, if $\tk'\in K$ is a key,

\item $\intension{\tg}{w,l,\tk} = \beta(\tg)$, if $\tg\in G$ is a global name,

\item $\intension{\tn}{w,l,\tk} = l(\tk,\tn)$, if $\tn\in N$ is a local name,

\item $\intension{\tp\s \tq}{w,l,\tk} = \bigcup \{
\intension{\tq}{w,l,\tk'} ~|~ \tk'\in \intension{\tp}{w,l,\tk} \}$,
for principal expressions $\tp,\tq\in P$.
\ei
%

%ron10: say a bit more about intuitions ...
Our intuitions for $\intension{\tp}{w,l,\tk}$ are essentially the same
as for the ``agent-based'' reading of Abadi's logic, discussed
above. That is, $\intension{\tp}{w,l,\tk}$ is the set of keys
associated with the expression $\tp$ in $\tk$'s local name space, when
local names are interpreted according to $l$.  With respect to
principal $\tk$, the expression $\tp\s \tq$ denotes the set of principals
that principals referred to by $\tk$ as $\tp$ refer to as $\tq$.

%Using this interpretation of principal expressions,
%ron2: revised - define open semantics a little later
%we may now define
%the first of our semantics, which we will refer to as the {\em open}
%semantics. This semantics is
%we define a satisfaction relation.  We write
We now define what it means for a formula $\phi$ to be true at a world
$w = (\beta,c)$ with respect to a local name assignment $l$ and
key $\tk$, written $w,l,\tk \models \phi$,
%joe3:
%(We restrict this relation below to yield our first semantics for
%\llnc.) The definition is by means of the following recursion on the
by induction on the
structure of~$\phi$.%
%joe9
\footnote{Note that our semantics is thus in the spirit of that of Grove
and Halpern \citeyear{GroveH2}, in that the truth of a formula depends
on both an agent and some features of the world (captured by $w$ and
$l$).}
\bi
%ron2: models_o -> \models here
%
\item $w,l,\tk \models  \tp \bdto \tq$
if $\intension{\tp}{w,l,\tk} \supseteq \intension{\tq}{w,l,\tk}$
\item $w,l,\tk \models \tk' \cert \phi$ if $\phi \in c(\tk')$
\item $w,l,\tk \models \neg \phi_1$ if not
$w,l,\tk \models \phi_1$
\item $w,l,\tk \models \phi_1 \land \phi_2$ if
$w,l,\tk \models \phi_1$ and $w,l,\tk \models \phi_2$.
\ei
%joe5: added paragraph
Note that the semantics of $\certn$ reinforces its syntactic nature.  To
determine if $\tk' \cert \phi$ is true at $(w,l,\tk)$, we check whether
a certificate has been issued in world $w$ by $\tk'$ certifying
$\phi$.  Moreover, as we shall see, while we allow any formula to be
certified by $\tk$, the only formulas whose certification has a
nontrivial semantic impact are those of the form $\tn \bdto \tp$, where
$\tn$ is a local name.  We return to this issue below.

%joe3: rewrote
%Note that, in this semantics,
We do not consider all pairs $w,l$ as being appropriate on the left-hand
side of $\sat$.  If $w = (\beta,c)$, we expect
the local name assignment $l$ to respect the certificates that
have been issued in $c$.  That is, if $c(\tk)$ includes the binding $\tn
\bdto \tp$, we would expect that $l(\tk,\tn)$ would include all the keys
bound to $\tp$ in $\tk$'s name space.  The question is whether there can
be
other keys bound to $\tn$ in $\tk$'s name space beyond those forced by
the certificates.  How we answer this question depends on our intuitions
for $c$.  For example, we could view $c$ as the set of certificates
received by one of the principals.
This would be particularly appropriate if we wanted to reason about the
knowledge and belief of the agents, an extension we plan to explore in
future work.  With this viewpoint, we could view $l$ as consisting of
all the bindings, including ones that the principal does not know about.
Thus, $l$ would at least have all the bindings forced by
$c$, but perhaps others as well.
Alternatively, we could view $c$ as consisting of all the
certificates that have been issued.  In this case, we would want $l$ to
be in some sense {\em
minimal}, and have no bindings beyond those forced by the certificates
in $c$.  We now present two different semantics, which reflect each of
these two intuitions.  We then show that, as far as validity is
concerned, the semantics are equivalent; that is, they have the same
proof theory.

% so far, there is no connection between the
%interpretation of local names and the component $c$
%modelling the certificates that have been issued. However, having defined
%the relation $\models$, we may establish such a connection.
A local name assignment $l$ is {\em consistent\/} with a
world $w = (\beta,c)$ if, for all keys $\tk$, local names
$\tn$, and principal expressions $\tp$, if the formula $\tn \bdto \tp$
is in $c(\tk)$, then $w,l,\tk \models \tn \bdto \tp$.
%[[ perhaps this notion should be called ``$l$ \emph{consistent with local
%name binding certificates in} $w$'' if we later talk about
%more general  types of consistency.]]
%
%[[could unpack that definition and avoid use of $\models_o$ ]]
%
%[[ discussion of how this relates to SDSI and principals
%telling the truth ]]
%
%[[ add more intuition -- o-validity corresponds to principals having
%made only true assertions about their bindings, but allows them to
%have told less than the whole truth.  ]]
%ron2: revised this para
Intuitively, assignments that are not consistent with a world provide
an inappropriate basis for the interpretation of local names, since
%joe3
%they allow that principals have issued bindings that are
%not reflected in their interpretations of local names.\footnote{One
%could also
%interprete this constraint by taking the viewpoint that local name
%assignments exist prior to
%certificates, and the the certificates issued should be truthful.
%We discuss this interpretation in Section~\ref{sec::nras}.}
the certificates issued by principals are not necessarily reflected in
their local bindings.
%joe3
%Thus, we now restrict the definition of satisfaction to local name assignments
%consistent with the world. This yields the first of our semantics,
We obtain our first semantics,
called the {\em open} semantics,
by restricting to
%joe4
%local name assignments consistent with the world.
consistent local name assignments.
%joe3: unnecessary
%For worlds $w$, keys $\tk$ and local name assignments $l$,
We write
$w,l,\tk\omodels \phi$ if  $w,l,\tk\models\phi$ and
$l$ is consistent with $w$. The formula $\phi$
%joe3
%will be said to be
is
{\em o-satisfiable\/} if there exists a triple $w,l,\tk$ such that
$w,l,\tk\omodels \phi$ and
%joe3
%We say that a formula $\phi$ is
%valid with respect to the open semantics, or simply
%that
$\phi$ is {\em o-valid}, denoted $\omodels \phi$, if
there does not exist a triple $w,l,\tk$ such that
$w,l,\tk\omodels \neg \phi$.
%ron8: cut stray text
%and have no bindings beyond those forced by the certificates
%in $c$.

%joe5: added paragraph
Although our syntax
%ron8: allow
allows
$\tk$ to certify arbitrary formulas,
it is easy to
see that, according to the semantics just introduced (as well as the one
we are about to introduce), only the certification of formulas of the
form $\tn \bdto \tp$ has any impact on consistency; all other formulas
certified by $\tk$ are ignored.  There is a good reason for this
restriction.  We are implicitly assuming that when $\tk'$
certifies $\tn \bdto \tp$, that very act causes all the keys bound
to $\tp$ to also be bound to $\tn$ in $\tk$'s name space.
Thus, if $\tn \bdto \tp \in c(\tk)$, then we want $\tn \bdto \tp$ to be
true in $(w,l,\tk)$.
But if $\tk$ certifies a formula like $\tk_1$'s $\tn \bdto \tk_3$
where $\tk_1 \ne \tk$, then we cannot conclude that
this formula is true in $(w,l,\tk)$
unless we are prepared to make additional
assumptions about $\tk$'s truthfulness.  We feel that if such
assumptions are to be made, then they should be modeled explicitly in
the logic, not hidden in the semantics.

It does seem reasonable to extend the notion of $l$
being consistent with $w$ to require that if $\tk$ certifies a formula
$\psi$ which is a Boolean combination of formulas of the form $\tn
\bdto \tp$ then $(w,l,\tk) \sat \psi$.
%ron10: cut following sentence, I'm not at all sure its right,
%       but we talk about it..
%       (example: k1 cert n\bdto k2\s m, k1 \cert \not(n \bdto k3)
%                 k2 cert m \bdto k3
%                 - this is inconsistent, whereas there are no
%                 such inconsistencies on the current story.
%        the right way for k1 to do it would be
%                  k1 cert n\bdto (k2\s m) \minus k3
%       -- this relates to revocation, by the way.
%%In fact, nothing would change in our
%%results if we did this for formulas $\psi$ that are conjunctions of
%%formulas of the form $\tn \bdto \tp$ and $\neg(\tn \bdto \tp)$.
However, once we allow more general Boolean combinations (in
particular, once we allow disjunctions), there will be problems making
sense out of the intuition of our next semantics, that 
%joe14
there are 
``no bindings beyond those forced by the certificates in $c$''.
We consider this issue next.

%ron2: added para
According to the open semantics, it is possible for a local name~$\tn$
of principal~$\tk_1$ to be bound to a key~$\tk_2$ even when no
certificate concerning~$\tn$ has been issued.
Arguably, this is not in
accordance with the
%joe4: no longer quoted
%intuitions of Rivest and Lampson quoted above.
intentions of SDSI.
To better capture these intentions, we define a second semantics,
that restricts the name bindings to those forced by the
certificates issued.

%For our second semantics,
To do so, we first establish
that the open semantics satisfies
a kind of ``minimal model'' result.
Define the ordering $\leq$ on the space $\lna$ of
local name assignments by
$l_1 \leq l_2$ if $l_1(\tk,\tn) \subseteq l_2(\tk,\tn)$ for all
%joe3
%keys
$\tk\in
K$ and $\tn\in N$.
It is readily seen that $\lna$ is given the structure of a
% complete partial order \cite{?} by this relation. (Indeed,
% $\lna$ is a complete lattice.)
complete lattice \cite{Birkhoff67}  by this relation.
Say that a local name assignment $l$ is {\em minimal} in a set of
local name assignments $L$ if $l\in L$ and $l\leq l'$ for all $l'\in L$.

\thm \label{thm:min:lna}
Given a world $w$, there exists a unique local name assignment
$l_w$ minimal in the set of all local name assignments consistent
with $w$.
%joe3: combined with later proposition
%Let $w$ be a world,
Moreover, if $\tp$ is a principal expression and $\tk_1$ and
$\tk_2$ are keys, then
%joe5
%$w,l_w,\tk_1 \cmodels \tp \bdto \tk_2$ iff,
%$w,l,\tk_1 \omodels \tp \bdto \tk_2$ for all local name assignments $l$
%consistent with $w$.
$w,l_w,\tk_1 \omodels \tp \bdto \tk_2$ iff,
for all local name assignments $l$
consistent with $w$, we have $w,l,\tk_1 \omodels \tp \bdto \tk_2$.
\ethm
%joe3: moved actual proof to after \end{document}; it should go in an
%appendix anyway.
%We describe the proof of this result shortly.
The proof of this result
%joe10:
%(which, like that of all the other theorems in
%this extended abstract, can be found in the full paper \cite{HM99full})
(which, like that of all the technical results in this paper, is
deferred to the appendix)
 uses standard techniques
%joe5: I think its more the theory of fixed points
%of recursion theory.
from the theory of fixed points.
%joe3: notation now introduced in theorem (which I agree isn't great)
%We write $l_w$ for
%the local name assignment whose existence is promised by
%Theorem~\ref{thm:min:lna}.
%joe3
%Our second semantics for \llnc, which we call the
%{\em closed} semantics, is based on this assignment. This semantics

We now define our second semantics, called the {\em closed semantics}.
It attempts to capture the intuition
that the only bindings in $l$ should be those required by the
certificates in $c$, using the minimal assignment promised by
Theorem~\ref{thm:min:lna}.
%uses a satisfaction relation $\cmodels$ that interprets the language
%with respect to a world $w$
%joe3
% and a key $\tk$. For $\phi \in \lang$, we
%ron2: revised presentation
We write $w,\tk\cmodels \phi$ if $w,l_w,\tk \models \phi$.  We say that
$\phi$ is {\em c-satisfiable\/} if there exists a world $w$ and key $\tk$
such that $w,\tk\cmodels \phi$ and
%We say that $\phi$ is
%valid with respect to the closed semantics, or
that $\phi$ is {\em c-valid\/}, denoted $\cmodels
\phi$, if $w,\tk \cmodels \phi$ for all worlds $w$ and principals
$\tk$.  Note that by Theorem~\ref{thm:min:lna},
%ron2: we have $w,\tk\cmodels \phi$ iff $w,l_w, \tk\omodels \phi$,
the assignment $l_w$ is consistent with $w$, so
c-satisfiability implies
o-satisfiability.
%joe3
%Moreover, if $\omodels \phi$ then $\cmodels \phi$.
Thus, if $\omodels \phi$ then $\cmodels \phi$.
%joe10: moved the result, as we discussed
%As the following result shows,
As we shall soon see (Theorem~\ref{same}),
somewhat surprisingly, the converse holds as well.

%joe5: I no longer believe this (which I think means that we can drop
%the restriction of N being infinite.  Let's discuss.
%We remark that this result depends on the fact that
%$N$, the set of local names, is infinite.
%If both $N$ and $K$
%were finite, we could construct formulas that were c-valid but
%not o-valid.
%To see this, let $\phi^*$ be the formula
%$$\band_{\tk, \tk' \in K, \tn \in N} \neg (\tk \cert (\tn \bdto
%\tk')).$$
%Clearly if $(w,l,\tk) \sat \phi$ and $w = (\beta,c)$, then
%$\tn \bdto\tk'' \notin c(\tk')$, for all $\tk'$, $\tk''$, and $\tn$.
%It is easy to check that the local name assignment $l_0$ such that
%$l_0(\tk',\tn) = \emptyset$ for all $\tk' \in K$ and $\tn \in N$ is
%consistent with $w$.

%joe3: did lots of rewriting from here on in
%\subsection{Axioms and a Comparison with Abadi's Logic}
\subsection{A Complete Axiomatization}
\label{sec:llnc:axioms}

%joe3:
%As shown in the previous section, the open and closed semantics
%validate the same formulas. In this section we enumerate some of these
%formulas, and reconsider Abadi's rules and examples discussed above.
%The valid formulas discussed are listed in Figure~\ref{fig:ax:llnc}.
%joe11: rewrote; I believe that this is the place to talk about the
%cardinality of K
%Theorem~\ref{same} shows that the same formulas are o-valid and
%c-valid. Thus, both the open semantics and the closed semantics can be
%characterized by the same axiomatization,
%which is described in Figure~\ref{fig:ax:llnc}.
%%joe10: added; we may want to add another sentence here
%Note that the axioms depend somewhat on whether the set $K$ of
%principals is finite or infinite
We start this section by presenting a sound and complete axiomatization
for $\llnc$ with respect to the open semantics.  We then prove that the
open and closed semantics are characterized by the same valid formulas,
so that the axiomatization is also sound and complete with respect to
the closed semantics.

%joe11: some rewriting here in the next few paragraphs, so as to start
%with the infinite case.  I thought the story went better this way.

The axiomatization depends in part on whether the set $K$ of keys is
finite or infinite.  Figure~\ref{fig:ax:llnc} describes the axiom system
$\AXinf$ for the case where $K$ is infinite.

\begin{figure*}
\[ \begin{array}{ll}
%joe3: added next line
\mbox{Propositional Logic:} &\mbox{All instances of
propositional tautologies}\\
\mbox{Reflexivity:} & \tp \bdto  \tp \\
\mbox{Transitivity:} & (\tp \bdto \tq) \limp ((\tq \bdto \tr) \limp (\tp \bdto \tr))\\
\mbox{Left Monotonicity:} & (\tp \bdto \tq) \limp ((\tp\s \tr) \bdto (\tq\s \tr))\\
\mbox{Associativity:} & ((\tp\s \tq)\s \tr) \bdto (\tp\s (\tq\s \tr)) \\
                     & (\tp\s (\tq\s \tr)) \bdto ((\tp\s \tq)\s \tr) \\
\mbox{Key~Globality:} &  (\tk\s \tg) \bdto \tg
\mbox{ if $\tk \in K$ and $\tg \in G \union K$}\\
%joe3
%                 & \mbox{n $\tg$ is a global name or a key and $\tk$ is
%                 a key} \\
%joe11: added for infinite case
\mbox{Globality:} &(\tp\s\tk \bdto \tk) \rimp (\tp\s \tg \bdto \tg)
\mbox{ if } \tk \in K, \  \tg \in G \union K\\
%joe6
%\mbox{ if $\tk \in K$ and $\tg \in G$}\\
\mbox{Converse~of~Globality:} &  \tg \bdto (\tp\s \tg)
%joe3
%                 & \mbox{when $\tg$ is a global name or a key} \\
\mbox{ if $\tg \in K \union G$}\\
\mbox{Key Linking:} & (\tk \cert(\tn \bdto \tr)) \limp ((\tk\s \tn)
\bdto (\tk\s \tr))\\
              & \mbox{if $\tn$ is a local name} \\
%joe11: added for $\AXinf$
\mbox{Nonemptiness:} & \mbox{(a) } 
%ron11: add some spaces for clarity 
\quad \tp \bdto \tk_1 \rimp \tp\s \tk
\bdto \tk\\
& \mbox{(b) } \quad \neg (\tp \bdto \tq) \rimp \tq\s \tk \bdto \tk\\
& \mbox{(c) } \quad \tp\s \tq \bdto \tk_1 \rimp \tp\s \tk \bdto \tk\\
& \mbox{(d) } \quad (\tp\s \tk \bdto \tk \land \tk' \bdto \tp) \rimp (\tp \bdto
\tk')\\
%joe3: unnecessary; it follows from the others, given finiteness
%\mbox{Nonemptiness:}& ((\tp\s \tq \bdto \tk_1)) \land (\tk_2 \bdto \tp)) \limp (\tp \bdto \tk_2)\\
%ron8: you seem to have omitted cutting the following line as well
%                     & (\neg (\tq \bdto \tp) \land (\tk_2 \bdto \tp)) \limp (\tp \bdto \tk_2)\\
\mbox{Key Distinctness:} & \neg (\tk_1 \bdto \tk_2)
\mbox{ if $\tk_1$ and $\tk_2$ are distinct keys}\\
%\mbox{Witnesses:} &\neg(\tp \bdto \tq) \rimp \lor_{\tk \in K} (\neg(\tp
%\bdto \tk) \land (\tq \bdto \tk))\\
%                &(\tp\s \tq) \bdto \tk_1 \rimp \lor_{\tk \in K} ((\tp
%\bdto \tk) \land (\tk\s \tq \bdto \tk_1))\\
\mbox{Modus Ponens:} &\mbox{From $\phi$ and $\phi  \rimp \psi$ infer
$\psi$}.
\end{array}
\]
%joe11
%\caption{Axioms of \llnc}\label{fig:ax:llnc}
\caption{The axiom system $\AXinf$}\label{fig:ax:llnc}
\end{figure*}

It is interesting to compare the axioms in $\AXinf$ to Abadi's axioms.
Although we interpret $\bdto$ as superset and he interprets it as subset,
Reflexivity, Transitivity, Left-Monotonicity, and Associativity, hold in
both cases, for essentially the same reasons.  The switch from subset to
superset means that the Converse of Globality holds in our case.
Globality does not hold in general because the denotation of
$\tp\s \tg$ may be empty if the denotation of $\tp$ is empty
(as we observed, this is also why the Converse of
Globality does not hold in general for Abadi).  In fact, for our logic,
$\tp\s \tg \bdto \tg$ holds whenever the interpretation of $\tp$ is
nonempty.
We use $\tp\s \tk \bdto \tk$ as a canonical way of denoting that the
interpretation of $\tp$ is nonempty.  This explains the form of the
Globality axiom.  Since the interpretation of a key is always nonempty,
we also get Key Globality.
%joe11
%Key Globality is a special case of this, and it turns out to
%be all we need for completeness
%\footnote{We have essentially only two types of ways of saying
%that the denotation of $\tp$ is nonempty.  One is if
%$\tp\s \tq \bdto \tk$ holds, for some name $\tq$ and some key $\tk$,
%and
%the other is if $\neg (\tq \bdto \tp)$ holds, for some name $\tq$.  In
%both these cases, the fact that $\tp\s \tg \bdto \tg$ follows from Key
%Globality, using Witnesses.}

Key Linking is our analogue of Abadi's Linking axiom.  Of course, we use
$\certn$ whereas Abadi uses $\saysn$; in addition, only keys can certify
formulas for us.  While this axiom shows that there are some
similarities between $\certn$ and $\saysn$, there are some significant
differences.  We have no analogue of Abadi's Speaking-for axiom and,
unlike $\saysn$,
$\certn$
%ron3: added next line !
does not
satisfy the standard axiom and rule of modal logic:
$(\tk \cert (\phi \rimp \psi)) \land
(\tk \cert \phi)$ does not imply $\tk \cert \psi$ and $\tk \cert \phi$
is not valid even if $\phi$ is valid.
Interestingly, Abadi does not use these properties of Speaking-for in
proving that his name resolution rules $\NR$, used to capture REF2, are
sound.  As a result, (with very minor changes) we can show that the name
resolution
rules are also sound for \llnc, and hence we can prove analogues of
Propositions~\ref{pro:abadi1} and~\ref{pro:abadi2}.  However, we can
actually prove a much stronger result: whereas Abadi's logic is able
to draw conclusions about bindings that do not follow from REF2,
\llnc\ captures REF2 exactly (see Theorem~\ref{REF2thm}).

%joe11: rewrote
$\AXinf$ has two axioms that do not appear in Abadi's axiomatization:
Key Distinctness and
%ron11: Nonemptimess.....  umm, is that a joke? 
Nonemptiness.
Key Distinctness just captures the fact that we interpret
keys as themselves.
%ron11: The four parts of Nonemptiness capture various properties of nonempty
%interpretations.  
The first three parts of  Nonemptiness capture various ways that an 
expression can be seen to be nonempty. 
For example, part (a) says that if $\tp$ is bound to 
%joe14
(i.e., is a superset of) 
a key, then its interpretation must be nonempty and part (b)
says that if 
%joe14
%$\tp$ is bound to $\tq$, then $\tq$ must be nonempty.
$\tp$ is not a superset of $\tq$, then $\tq$ must be nonempty.
%ron11: added 
Part (d) of Nonemptiness says that if $\tp$ is nonempty and $\tk'$ is bound to 
$\tp$, then $\tp$ is bound to $\tk'$, i.e., $\tp$ and $\tk'$ have 
exactly the same interpretation. 

%ron11: 
\newcommand{\tl}{{\tt l}}

If $K$ is finite we need to add two further axioms to $\AXinf$.  Let
$\AXfin$ consist of all the axioms and rule in $\AXinf$ together with:
$$\begin{array}{ll}
\mbox{Witnesses:} &\neg(\tp \bdto \tq) \rimp \lor_{\tk \in K} (\neg(\tp
\bdto \tk) \land (\tq \bdto \tk))\\
                &(\tp\s \tq) \bdto \tk_1 \rimp \lor_{\tk \in K} ((\tp
\bdto \tk) \land (\tk\s \tq \bdto \tk_1))\\
%joe11: new axiom
%ron11: too many k's 
%\mbox{Current Principal:} &\lor_{\tk \in K} (\tn_\tk \bdto \tk_\tk \liff
%\tk\s \tn_\tk \bdto \tk_\tk)
\mbox{Current Principal:} &\lor_{\tk \in K} (\tn_\tk \bdto \tl_\tk \liff
\tk\s \tn_\tk \bdto \tl_\tk) \\
&\mbox{where $\tn_\tk \in N$ and $\tl_\tk\in K$ for each $\tk\in K$.}
\end{array}
$$

%joe11
The two axioms that make up Witnesses
%joe9: added more motivation
essentially capture our interpretation of $\bdto$ as containment.
They tell us that facts about
containment of principal expressions can be reduced to facts about
keys.  For example, the first one says that if $\tp$ does not
contain $\tq$, then there is a key bound to $\tq$ that is not
bound to $\tp$.
%joe9: softened
%These axioms are valid because of our technical
%assumption that the set of keys is finite.
%joe11: not quite true
%Although the exact statement of these axioms depends on our assumption
%that the set of keys is finite, somewhat analogous axioms would be
%required even if the set of keys were infinite.
%joe8: rewrote; I don't believe what we wrote before is true in any case
%(If we allowed infinitely
%many keys, we would drop this axiom, but we would have a more convoluted
%version of Key Globality, and other axioms would be needed.)
%joe9: cut this for now
%If we allowed infinitely many keys, we would need a somewhat weaker
%version of this axiom, Witnesses$'$:
%$$\begin{array}{l}
%\neg(\tp \bdto \tq\s \tk) \rimp (\neg(\tp
%\bdto \tk) \land (\tq\s \tk \bdto \tk))\\
%  (\tp\s \tk)\s \tq \bdto \tk_1 \rimp ((\tp\s \tk
%\bdto \tk) \land (\tk\s \tq \bdto \tk_1)).
%\end{array}
%$$
%It is easy to see that Witnesses$'$ follows from Witnesses (together
%with Key Globality, Key Distinctness, and Transitivity).
%joe5: I agree with Martin that this is irrelevant
%Abadi
%would have a similar property in his logic if he had identified keys
%with worlds in his semantics (as we suggested in our intuitive remarks)
%and also assumed that $K$ was finite.
Current Principal captures the fact that some key in $K$ must be the
current principal; if $\tk$ is the current principal, then for all
local names $\tn$ and keys $\tk'$, $\tn \bdto \tk' \liff \tk\s \tn
\bdto \tk'$ holds.  (This is actually
true not just for local names, but for all principal 
%ron11 expresions; it
expressions; it
suffices to state the axiom just for local names.)

While the properties captured by these two axioms continue to hold
even if $K$ is infinite, they can no longer be expressed in the logic,
since we cannot take a disjunction over all the elements in $K$.
Interestingly, we can drop Nonemptiness and Globality as axioms
in $\AXfin$.  These properties already follow from the other
properties in the presence of Witnesses.

As the following result shows,
these axiom systems completely characterize validity in the logic with
respect to the open semantics.

%joe3: added
%joe10: added label, so you can refer to it in the proof.  Actually, I'd
%prefer to give the system a name, like AX.  We'll actually have two
%systems, one for the finite case and one for the infinite case.
\thm\label{complete}
%joe11
%The axiom system in Figure~\ref{fig:ax:llnc} is a
$\AXinf$ (resp., $\AXfin$) is a sound
and complete axiomatization of \llnc\ with respect to
%joe10
%both the open and closed semantics.
the open semantics if $K$ is infinite (resp., $K$ is finite).
\ethm

%joe10: moved this here; I think the finite model result should be in
%the appendix
In the course of proving Theorem~\ref{complete}, we also prove a
``finite model'' result, which we cull out here.  Let $|\phi|$,
the {\em length\/} of $\phi$, be
the total number of symbols appearing in $\phi$.
%ron11: 
This result holds both when $K$ is finite and when $K$ is infinite.

%ron11: I think the bound can be improved: substituted for |\phi|^4 everywhere
\newcommand\bound{2\cdot|\phi|^2}

%joe11: new
\pro\label{finitemodel}  Let $K_\phi$ be the keys that appear in $\phi$
and let $C_\phi(\tk)$ consist of all bindings $\tn \bdto \tp$ such that
$\tk \cert \tn \bdto \tp$ is a subformula of $\phi$.
If $\phi$ is satisfiable with respect
to the open semantics, then for all sets $K'$ of keys such that $K_\phi
\subseteq K'$ and 
%ron11: $|K'| \le \min(|K|,\bound)$, .. the other way round surely! 
$|K'| \ge \min(|K|,\bound)$, 
there is a world $w = (\beta,c)$, local name
assignment $l$, and principal $\tk \in K'$ such that $w,l,\tk
\omodels\phi$ and 
%ron11: you need to say a bit more for model checking to be ptime 
%joe13: oops!  You're right, of course
%(a) $l(\tk',\tn) \subseteq K'$ for all $\tk' \in K$ and $\tn
%\in N$, (b) $l(\tk',\tn) = \emptyset$ if $\tk' \notin K'$, and (c)
%$c(\tk) \subseteq C_\phi(\tk)$ for all keys $\tk$.
(a) $l(\tk',\tn) \subseteq K'$ for all $\tk' \in K$ and $\tn
\in N$,
%joe13: removed some ``ands'' and made separate items for readability
(b) $l(\tk',\tn) = \emptyset$ if $\tk' \notin K'$, 
(c) $\beta(\tg) \subseteq K'$ for all $\tg\in G$,
(d) $\beta(\tg)= \emptyset$ if $\tg$ does not occur in $\phi$, and 
(e) $c(\tk) \subseteq C_\phi(\tk)$ for all keys $\tk$. 
\epro

\cor\label{decisionproc} The problem of deciding if a formula $\phi \in
\llnc$ is satisfiable with respect to the open semantics is NP-complete
(whether $K$ is finite or infinite).  \ecor

%ron11: do we really want the proof here?  a bit unbalanced. 

\prf The lower bound is immediate from the fact that we can
trivially embed satisfiability for propositional logic into
satisfiability for \llnc.  For the upper bound,
given $\phi$, choose $K'$ such that 
%ron11: $|K'| \le \min(|K|,\bound)$
$|K'| = \min(|K|,\bound)$
and $K' \supseteq K_\phi$.  Then guess $w,l,\tk$ as in
Proposition~\ref{finitemodel} and check whether $w,l,\tk \omodels \phi$.
Proposition~\ref{finitemodel} says that the guess is only polynomial in
$|\phi|$; it is clear that checking whether $w,l,\tk \omodels \phi$ can
also be done in time 
%ron11: polynoial 
polynomial 
in $\phi$.  
%ron11: I don't get this, for small K and large phi it does depend 
%joe13: rewrote as you suggested
%Note that the polynomial in
%this result is independent of $|K|$ if $K$ is finite. 
Note that for $|\phi| \leq |K|$ (which is likely to include all cases
of practical interest, given that $K$ will typically be a very large
%joe14
%set), the polynomial does not depend on $|K$. 
set), the polynomial does not depend on $|K|$. 
\eprf

%joe11
As we suggested earlier, the closed semantics and the open semantics are
characterized by exactly the same axioms.

\thm\label{same}
%joe3: we're now assuming this (although the assumptions about $K$ were
%unnecessary anyway)
%Suppose that both the set of local names $N$ and the set of keys $K$
%are countable and that the former is infinite.  Then
%Validity with
%respect to the open and closed semantics are equivalent,
The same formulas are c-valid and o-valid;
i.e., for all
formulas $\phi$, we have $\omodels \phi$ iff $\cmodels \phi$.
\ethm

%ron11: 
We remark that this result is sensitive
%joe14
to
the language under consideration. 
It may no longer hold if we move to a more expressive language. 

\cor\label{complete1} 
%ron11: The axiom system in Figure~\ref{fig:ax:llnc} 
$\AXinf$ (resp., $\AXfin$) 
is a sound and complete axiomatization of \llnc\ with respect to
the closed semantics
%ron11: 
when $K$ is infinite (rep., finite). 
\ecor

\cor\label{decisionproc1} The problem of deciding if a formula $\phi \in
\llnc$ is satisfiable with respect to the closed semantics is
NP-complete (whether $K$ is finite or infinite).  \ecor

%Like Abadi's logic, \llnc\ satisfies the standard axioms and rules of
%propositional logic. To make further comparisons we need to translate
%$\says$ to $\cert$ and $\spk$ to $\bdto$.  Since in \llnc\ only a key
%may certify a fomula, we also need to restrict occurrences of the principal
%expression $\tp$ in the form $\tp \says \phi$ to be a key $\tk$.
%When we do so, we find that whereas Abadi's logic
%treats the operator $\says$ as a modal operator, \llnc's semantics for
%$\cert$ is quite different. The operator $\cert$ does not satisfy the
%standard axiom and rule for modal logic.

%joe11: rewrote glue
%Considering the two contentious axioms discussed by Abadi, we find that
Let us now return to the contentious axioms discussed by Abadi.
Converse of Globality is valid in $\llnc$, as we observed earlier.
The generalization
of Linking considered by Abadi, restricted to be syntactically well
formed, amounts to
\[ (\tk \cert (\tp_1 \bdto \tp_2)) \limp
( \tk\s \tp_1 \bdto \tk\s \tp_2). \]
In general, this is not valid, since our semantics ignores certificates
stating $\tp_1 \spk \tp_2$ when $\tp_1$ is not a local name.
%[[ Argue either here or above that this is right! ]]
%joe3: did it below
Thus, we avoid the ``unreasonable'' conclusions
%joe3:
%Abadi was able to draw from these two contentious axioms.
that can be drawn from these axioms.  In particular,
it does not follow in our logic that $(\tk
\cert ({\tt DNS}!! \bdto \tk)) \limp {\tt DNS}!! \bdto \tk$.
However, the reason it does not follow in \llnc\ is quite different
from the reason it does not follow in Abadi's logic:
since ${\tt DNS}!!$ is a global name, a certificate such as
$\tk \cert ({\tt DNS}!! \bdto \tk)$
has no impact on the interpretation of global names.
%joe3: you're treating DNS as a key below.  I just cut this
%More
%precisely, let $w$ be a world in which the only certificate is ${\tt
%DNS}!! \bdto \tk$, issued by $\tk \neq {\tt DNS}!!$. Then for all $l$
%we have $\intension{{\tt DNS}!!}{w,l,\cp} = \{{\tt DNS}!!\}$, which is
%not a superset of $\intension{\tk}{w,l_w,\cp} = \{\tk\}$.
%Intuitively,
%this is right: in general, $\tk$ has no authority to bind global
%names, so its certificates concerning global names should not be taken
%at face value.
This captures the intuition that $\tk$ should not be trusted when making
assertions about bindings not under its control.  If we were willing to
trust $\tk$ on everything, then concluding that $\tk$ is bound to ${\tt
DNS!!}$ after $\tk$ certifies that it is would  not seem so unreasonable.

%ron3: Note that the same model shows that % there is no model now
%joe4
%We also do not obtain a variant of this
%formula,
The following formula is also not valid in \llnc:
\[
(\neg (\tk \cert \false) \land (\tk \cert ({\tt DNS}!! \bdto \tk)))
\limp {\tt DNS}!! \bdto
\tk.
\]
%joe4
%corresponding to the formula that we noted earlier is valid in Abadi's
%logic.
(This formula corresponds to the one that we noted earlier is valid in
Abadi's logic.)
Failure
to issue a certificate stating $\false$ has no more impact on global
names than does any other behavior of $\tk$. Nor would a
precondition asserting that the interpretation of $\tk$ is non-empty
validate the formula, since this is true in every world.
We can in fact
%establish the following result that gives a much
%Proposition~\ref{assurance} proved by Abadi.
prove the following generalization of Abadi's
Proposition~\ref{assurance}, which provides a
%joe3
%much
stronger statement of the safety of our logic than Abadi's result.

\pro\label{independent}
Let $\Gamma$ be any
%ron10:
c-satisfiable
boolean combination of formulas of the form $\tk
\cert \phi$, and let $\Delta$ be any boolean combination of formulas of
the form $\tp \bdto \tq$ where neither $\tp$ nor $\tq$ contains a
local name. Then $\cmodels \Gamma\limp \Delta$ iff
$\cmodels \Delta$.
\epro

%Informally, an assertion $\Delta$ about global names and keys is a
%consequence of the issuing of a set of certificates iff it is
%valid. As stated, this result concerns the language we have used, but
%it is
%in fact a direct consequence of a much more general semantic fact: for
%every key $\tk$, global name assignment $\beta$ and set of
%certificates $c$, there is a point $\la \beta,c\ra, \tk$ in the space
%used to define validity. Thus,
Informally, Proposition~\ref{independent} says that
facts about global names are completely
independent of facts about certificates; issuing certificates can
have no impact on the global name assignment.
%joe3
As we observed earlier, the analogous result does {\em not\/} hold for
Abadi's logic.

%joe3: rewrote and shrunk this section
\section{Name Resolution in \llnc}
\label{sec:nras}

In this section, we show that \llnc\ captures REF2 exactly.  Indeed,
we show
%ron3: much more. % not clear what the ``much more'' is referring to
that it does so for several distinct semantic interpretations.
Define the
order $\geq$ on worlds by $( \beta',c') \geq (\beta, c)$ if
\be
\item $\beta'(\tg) \supseteq \beta(\tg)$ for all global names $\tg$, and

\item $c'(\tk) \supseteq c(\tk)$ for all keys $\tk$.  \ee That is, $w'
\geq w$ when $w'$ contains more certificates than $w$ and the
bindings to global names in $w$ are a subset of those in
$w'$.
%ron10:
If $E$ is a set of formulas and $\phi$ is a formula, we write $E \omodels \phi$
if for all worlds $w$, local name assigments $l$
consistent with $w$ and all keys $\tk$, if
$w,l,\tk\omodels \psi$ for all $\psi$ in $E$ then $w,l,\tk \omodels \phi$.
Similarly,  $E \omodels \phi$
if for all worlds $w$  and all keys $\tk$, if
$w,\tk\cmodels \psi$ for all $\psi$ in $E$ then $w,\tk \cmodels \phi$.

\thm\label{REF2thm}
Suppose $\tk_1,\tk_2$ are principals, $w = (\beta,c)$ is a world, and
$\tp$ is a principal expression.   Let
%ron10: the formula $E_w$ consist of the conjunction of the
$E_w$ be the set of all
formulas $\tg \bdto \tk$ for all
global names $\tg$
%ron3: either the following or make $G$ finite above
%ron10: occurring in $\tp$ -- cut, adds pain to pf for little gain, see ron10*
and keys $\tk \in \beta(\tg)$ and the formulas
$\tk \cert \phi$ for all keys $\tk$ and formulas $\phi \in c(\tk)$.
The following are equivalent:
\be
\item
%ron3: $\tk_1\in \reft(w,\tk_2,\tp)$, % conform to the above
$\tk_1\in \reft(\tk_2,\beta,c,\tp)$,
%ron3: added the model checking version
\item $w,\tk_2 \cmodels \tp\bdto \tk_1$,
\item $w',\tk_2 \cmodels \tp\bdto \tk_1$ for all
worlds $w'\geq w$,
%ron10: \item $\models_c E_{w} \limp (\tk_2\s \tp \bdto \tk_1)$,
\item $E_w \models_c \tk_2\s \tp \bdto \tk_1$,
%joe3: added; this will be relevant to the later discussion
%ron10: \item $\models_o E_{w} \limp (\tk_2\s \tp \bdto \tk_1)$.
\item $E_w \models_o \tk_2\s \tp \bdto \tk_1$.
\ee
\ethm

%joe10: greatly expanded this material (pulling in stuff that we had cut
%for the CSFW version).  Started by adding next sentence.
This theorem gives a number of perspectives on name resolution in \llnc.
The equivalence between (1) and (2) in
this theorem tells us that $\reft$ is sound and complete
with respect to key binding, according to the semantics of \llnc.  That
is, $\reft(\tk,\beta,c,\tp)$
yields $\tk'$ iff $\tp \bdto \tk'$ is forced to be true by the bindings
of global names in $\beta$ and the certificates in $c$.
%joe10: added
Thus, viewed as a specification of the meaning
of SDSI names, the closed semantics and $\reft$ are equivalent.

%joe10: new paragraph
Informally, we have viewed $\reft$ as a
procedure that is run by an omniscient agent
with complete information about the interpretation of global names
and the certificates that have been issued. It is also
possible to understand $\reft$ as performing
a computation based on the limited information available to
a particular principal. Suppose that the world $w$
expresses the limited information this principal has about the
binding of global names and the certificates that have been issued.
Suppose that $w'$ describes the
actual bindings of global names and the certificates that have been
issued.  Assuming that all of the principal's information is correct,
then
$w \le w'$.
%ron10: added next two sentences
Thus, the set of $w'\geq w$ is the set of all worlds $w'$ that are
consistent with the information available to the principal.
(We could formalize this using the Kripke semantics for the
logic of knowledge in a distributed system 
%ron12: \cite{HM90}.)
\cite{HM1}.)
% Moreover,
%ron3: added next line
The equivalence between (2) and (3) essentially shows that it doesn't
matter whether we view the principal as having total or partial
information.
%ron10: cut - potentially misleading
%    ; more information will not change the truth of a formula of
%    the form $\tp \bdto \tk_1$.
%(and also between (4) and (5)) shows that
%in any world where
%ron5: cut: there are
%all the certificates in
%$c$ have been issued (and perhaps others)
%$\tp \bdto \tk'$ will be true in $\tk$'s name space.
%ron3:  The equivalence between 1 and 2 (and also between 3 and 4) emphasizes

%joe10: new paragraph
The implication from (1) to (4) in Theorem~\ref{REF2thm} is
analogous to Abadi's soundness result, Proposition~\ref{pro:abadi2}.
Of course, the converse implication gives us completeness, which, as
Abadi himself observed, does not hold for Abadi's logic (since
it validates  conclusions that do not follow from $\reft$).
Interestingly, although, as we have seen, there are significant
differences between $\llnc$ and Abadi's logic,
an examination of Abadi's soundness proof reveals that it does
not use the Speaking-for rule, the unrestricted form of Globality, or
the standard axiom and rule for the modal operator $\saysn$, which are
the main points of difference with our logic.  This
observation
says that the proof of the implication from (1) to (4) is
essentially the same for $\llnc$ and for Abadi's logic.

%joe10: cut
%These equivalences emphasize
%that we get the same conclusions
%whether we view the information given by $(\beta,c)$ as representing
%all the bindings or view it as partial information (perhaps one
%principal's view) of all the bindings that have in fact been created.

It is instructive to understand why the formulas
considered in Examples~\ref{counterexample1} and~\ref{counterexample2},
which give conclusions
%joe2
%beyond those derivable by REF2, do not hold in
in Abadi's logic beyond those derivable by REF2, are not valid in
\llnc.  It is easy to see why
the formula $\tk\s ({\tt Lampson}\s \tk') \speaksfor \tk'$ from
Example~\ref{counterexample1} (which,
by Associativity and Transitivity, is equivalent to $(\tk\s {\tt
Lampson})\s \tk' \speaksfor \tk'$) is not valid in \llnc.  This is
%joe11
%simply because Globality does not hold in general.
simply because the antecedent of (our version of) Globality does not
always hold.
Now consider the
formula in Example~\ref{counterexample2}.  The proof that this is valid
in Abadi's logic uses the
Speaking-for axiom, which does not hold for us (if we replace $\saysn$
by $\certn$).  To see that it is not valid in \llnc,
%Concerning Abadi's axioms for linked local name spaces in
%Figure~\ref{fig:aax},
%it is straightforward to check that the axioms Reflexivity,
%Transitivity, and Left-Monotonicity are also valid for \llnc. Our
%logic does not satisfy Abadi's Globality axiom, when we interpret this
%as being about global names.
%In particular, our logic does not validate the following formula,
%which Abadi noted as a point of variance between his logic and
%SDSI name resolution:
%${\tt (Lampson\s KR) \bdto KR}$ when
%{\tt Lampson} is a local name and {\tt KR} is a key.
consider a world
$w = (\beta,c)$ containing only the certificates forced by the
formulas
(i.e., $c(\tk) = \{{\tt Lampson}\speaksfor \tk_1,
{\tt Lampson}\speaksfor \tk_2\}$,
%ron5: added brackets to next two since they are sets
$c(\tk_1) =
\{
{\tt Ron}\speaksfor {\tt Rivest}
\}$,
$c(\tk_2) =
\{
{\tt Rivest} \speaksfor \tk_3
\}$).
Then it is easy to see
that $w,\tk \not\sat \tk\s ({\tt Lampson\s Ron}) \speaksfor \tk_3$,
since $\intension{\tk\s({\tt Lampson\s Ron})}{w,l_w,\tk} = \emptyset$
whereas $\intension{\tk_3}{w,l_w,\tk} = \{\tk_3\}$.

\section{Logic Programming Implementations of Name Resolution Queries}
\label{sec:lp}

The reader familiar with the theory of logic programming
% ron3: will have % if there were more detail in this version, perhaps
may have
noted a close resemblance of the results and constructions of the
preceding sections to the (now standard) fixpoint semantics for logic
programs developed originally by van Emden and Kowalski
\cite{vanEmdenKowalski86}.  Indeed,
%ron3: cut - we do enough of it in this version already, I think
%as we show
%%joe3
%in the full paper,
it is possible
to translate our semantics into the framework of logic programming.
%joe3: shrunk
%This has the benefit of allowing the wealth of research on the
%implementation and optimization of logic programs to be applied to
%developing implementations of SDSI query processing.
%
%A number of different translations of our semantics into logic
%programs are possible. One approach would be to introduce a function
%symbol $s$ to represent the composition operator ``$\s$'' and to
%translate a principal expression to a ground term. For example, the
%translation of the principal expression $(\tk\s \tn)\s \tm$ would be
%the term
%$s(s(\tk,\tn),\tm)$. We present another
In fact, we provide a translation that does not
require the use of function symbols
%The advantage of this translation
and thus produces a {\em Datalog} program, a restricted
type of logic program that has significant computational advantages over
unrestricted logic programs.
%For example, query processing using
%Datalog programs is decidable, whereas, in general, logic programs
%using function symbols are
%%joe3: what's Turing complete?
%%Turing complete.
%not.
Our translation allows us to take advantage of the significant body of
research on the optimization of Datalog programs
\cite{UllmanKB1,UllmanKB2}.
%for an exposition of this area.

\newcommand\name{{\tt name}}
%joe3:
The idea is to translate queries to formulas in a first-order language
over a vocabulary $V$ which consists of a constant symbol for
each element in $K \union G
\union N$ and a ternary predicate symbol $\name$.
%We begin by relating our semantics to first-order logic.  We work with
%a first-order language in which the elements of $K\union G \union N$
%are constants and there is a ternary predicate symbol .
%joe3: unnecessary
%(We will not make use of function symbols, so omit these from
%our exposition.)
Intuitively, $\name(x,y,z)$
%joe3
%will represent
says that, in
the local name space of key $x$, the
%ron10: name
basic principal expression (i.e., key, global name or local name)
$y$ is bound to key $z$.

Using $\name$, for
each principal expression $\tp$ and pair of variables $x,y$, we
define a first-order formula $\trans{x,y}{\tp}$ that, intuitively,
corresponds to the assertion ``$y\in \intension{\tp}{x}$,''
by induction on the structure of $\tp$:
\be
\item $\trans{x,y}{\tp} = \name(x,\tp,y)$ when $\tp \in K\union G\union
N$.
%joe4:
%\item $\trans{x,y}{\tp}$ is $\exists z(\trans{x,z}{\tq} \land
%\trans{z,y}{\tr})$ when $\tp$ is the principal expression $\tq\s \tr$.
%Here $z$ is required to be a fresh variable.
\item $\trans{x,y}{\tq\s \tr} = \exists z(\trans{x,z}{\tq} \land
\trans{z,y}{\tr})$, where $z \ne x, y$.
\ee
%

%ron10: moved this para here and revised
%Recall that a {\em Herbrand structure\/} $M = (U,I)$ over the
%vocabulary
%$V$ has as its domain $U = K \union G \union N$ and
%the interpretation $I$ is such that $I(c) = c$ for each constant $c \in
%K \union G \union N$.  Herbrand structures over $V$ differ in their
%interpretation of the predicate $\name$.  We can put a partial order
%over the Herbrand structures over $V$ by defining $M_1 = (U,I_1) \preceq
%M_2 = (U,I_2)$ iff $I_1(\name) \subseteq I_2(\name)$.  That is, $M_1
%\preceq M_2$ if the interpretation of the $\name$ relation in $M_1$ has
%at least as many triples as its interpretation in $M_2$.
%%joe4
%It turns out that minimal Herbrand models of $\Sigma_w$ correspond to
Recall that a {\em Herbrand structure\/} over the vocabulary $V$ is a
first-order structure
%joe11: cut; technically we should say it, but I think it would only
%confuse novices
%for this vocabulary,
that has as its domain the
set of constant symbols $K \union G \union N$
%joe11
%of the vocubulary and
in $V$ and
interprets each constant sybol as itself. Such a structure may be
represented as a set of tuples of the form $\name(x,y,z)$, where
$x,y,z\in K \union G \union N$. The subset relation on such sets
partially orders the Herbrand structures.

%ron10: new para and proposition
We say that a Herbrand structure $M$ over $V$ {\em represents\/}
a world $w=\la \beta,c\ra$ and local name assignment $l$
if, for all $x,y,z\in K\cup G\cup N$, we have  $\name(x,y,z)\in M$ iff
either
\be
\item $x,y, z\in K$ and $z=y$, or
\item $x\in K$, $y\in G$ and  $z\in \beta(y)$, or
\item $x\in K$, $y\in N$ and $z\in l(x,y)$.
\ee
Intuitively, $M$ represents $w$ and $l$ if it encodes all the
interpretations of basic principal expressions given by $w$ and
$l$.  The following result,
%joe11
whose straightforward proof is left to the reader,
shows that in this case
$M$ also captures the interpretation of all other
principal expressions, and expresses the correctness of our
translation of principal expressions.

\pro \label{prop:represents}
If $M$ represents $w$ and $l$ then, for all principal expressions
$\tp$ and $x,y\in K\cup G\cup N$, we have $M \models \trans{x,y}{\tp}$
iff
$x,y\in K$ and $w,l,x\models \tp\bdto y$.
\epro

%ron10:
We now show how a logic program can be used to capture the
relationsip between $w$ and $l_w$.
For each world $w = (\beta,c)$, we
%ron10: can then
define a theory (set of sentences) $\Sigma_w$ that characterizes $w$;
$\Sigma_w$ consists of the following sentences:
\be
\item
%ron3: typo: a sentence $\name(\tk_1,\tk_2,\tk_1)$, for  each
a sentence $\name(\tk_1,\tk_2,\tk_2)$, for  each
pair of keys $\tk_1,\tk_2\in K$, and
\item the sentence $\name(\tk_1,\tg,\tk_2)$, for  each
pair of keys $\tk_1,\tk_2\in K$ and global name $\tg\in G$ such that
%
%$\tk_2 \in \intension{\tg}{w,\tk_1}$, and
$\tk_2 \in \beta(\tg)$,
\item the sentence $\forall y (
%%joe3: typo
%%\trans{\tk,y}{\tp} \limp \name(\tk,\tn,y))$, for each
%\trans{\tk,q}{\tp} \limp \name(\tk,\tn,y))$, for each
%ron3: but that's also wrong ! - fixed
\trans{\tk,y}{\tq} \limp \name(\tk,\tn,y))$, for each
key $\tk$ and binding $\tn \bdto \tq$ in $c(\tk)$.
\ee

%joe10: reinstated; I hope I've got it right.
%ron10:
After some equivalence-preserving syntactic transformations (moving
the existentials in the body of these sentences to the front),
the theory $\Sigma_w$ is a {\em definite Horn theory},
i.e., it consists of formulas of the form
$\forall {\bf x} (B \limp H)$, where $B$ is a (possibly empty)
conjunction of
%joe11:
%atoms
{\em atoms\/} (that is, formulas of the form $\name(x,y,z)$ or $y=z$)
and $H$ is an atom. Well-known results from the theory
of logic programming show that such a theory $\Sigma$ has a
Herbrand model $M_\Sigma$ minimal with respect to
%ron10; $\preceq$.
the containment ordering on Herbrand structures.
Moreover,
this minimal Herbrand model
%ron10: corresponds to
captures
the minimal name assignments for $w$.

%ron10: revised statement and added Corollary
%       the new form better justifies the other queries we present
%\thm\label{Herbrand} There is a minimal Herbrand model $M_w$ of
%$\Sigma_w$ and $M_w \sat
%\trans{\tk_1,\tk_2}{\tp}$ iff  $w,l_w,\tk_1 \sat \tp \bdto \tk_2$.
%\ethm
%joe11: do we need to add a proof of this, or is it standard?
\thm\label{Herbrand} The minimal Herbrand model $M_w$ of
$\Sigma_w$ represents $w$ and $l_w$.
\ethm

Using Proposition~\ref{prop:represents}, we immediately obtain
the following corollary.

\cor\label{Herbrand-cor}
For all $x,y\in K\cup G\cup N$ and
principal expressions $\tp$, we have $M_w \models \trans{x,y}{\tp}$
iff $x,y\in K$ and $w,x \cmodels \tp \bdto y$.
\ecor

%joe10: reinstating more stuff
Because $\Sigma_w$ is a definite Horn theory, it corresponds to a logic
program.
%joe3
%As we explain in the full paper,
Moreover, for
{\em existential queries}, i.e., queries $\phi$ that are
sentences formed from atomic formulas using only conjunction,
disjunction and existential quantification (but not negation), we have
that $\Sigma$ entails $\phi$ iff $M_\Sigma \models \phi$.
%As an immediate consequence of Theorem~\ref{Herbrand},
This enables us to exploit logic programming technology to
obtain efficient implementations of several
types of queries, corresponding to different choices of bound and free
variables in the predicate ``$\name$''. We may even form complex queries
not corresponding in any direct way to the capacities of the
procedure REF2. Examples of this include the following:
\be
\item the query $\name(\tk_1,\tn,\tk_2)$ returns ``yes'' if $\tk_2$
is bound to the local name $\tn$ according to $\tk_1$;
\item the query $\name(X,\tn,\tk)$ returns the set of keys $X$ such that
$\tk$ is in $\tn$ according to $X$;
%ron3: cut to save soem space:
%joe10: reinserted; space no longer an issue
\item the query $\name(\tk_1,X,\tk_2)$ returns the set of global and
%ron10: local names containing $\tk_2$ according to $X$.
local names $X$ containing $\tk_2$ according to $\tk_1$.

%ron2: the following returns all keys, not all that interesting...
%\item the query $\exists z( \name(\tk_1,z,X) \land \name(\tk_2,z,X))$
%returns the set of keys $X$ for which $\tk_1$ and $\tk_2$ have a
%common local or global name.

\item  the query $\name(\tk_1,\tn,X) \land \name(\tk_2,\tn,X)$
returns the set of keys $X$ that $\tk_1$ and $\tk_2$ agree
to be associated with local name $\tn$.
\ee
%joe10: added
Many more possibilities clearly exist.  These observations show the
advantage of viewing name resolution in a logic programming framework.

\section{Self}\label{sec:self}

Abadi considers an extension of his logic
obtained by adding a special basic principal expression
$\self$,  intended to represent  SDSI's expression ({\tt ref:}).
(We remark that $\self$ is essentially the same as $I$ in the logic of
naming considered in \cite{GroveH2}.)
 Intuitively, $\self$ denotes the current principal.
The semantics given to $\self$ by Abadi extends the definition of the
set of principals associated with a principal expression by
taking $\intension{\self}{a} = \{a\}$ for each $a\in \W$.
This suffices to validate the following axiom.
\[\begin{array}{lll}
\mbox{Identity:}\ \ \
     &\self\s \tp \spk \tp & ~~~~\tp \spk \self\s \tp \\
      &\tp\s \self \spk \tp & ~~~~\tp \spk \tp\s \self
   \end{array}
\]
These axioms very reasonably capture the intuitions that
$\self$ refers to the current principal.

However, not all consequences of this semantics for $\self$ are so
reasonable.
%joe3: even for the full paper I would cut this
%Abadi discusses an example in which $\tk$ is a key and in
%the name space of the current principal, ${\tt Edward} \spk \tk$ holds,
%i.e., $\tk$ speaks for {\tt Edward}.  Since $\tk$ and $(\tk\s \self)$
%are bound to each other, it is reasonable to have $\tk \says (\tk \spk
%\self)$.  But it then follows that ${\tt Edward} \says (\tk
%\spk \self)$. According to Abadi, this is a ``suspicious statement'',
%but it should be ``harmless, since {\tt Edward} should not have the
%authority to convince others it speaks for $\tk$ when in fact
%{\tt Edward} is bound to $\tk$ but not vice versa.''
%
%We are not so sanguine, and feel that this example raises further
%doubts about Abadi's semantics.  Indeed, the problems go deeper than
%Abadi suggests. One peculiar aspect of this semantics is that although
%in a formula such as $\tk \says (\tk \spk \self)$, in the scope of the
%$\saysn$ operator we should intuitively have that ``$\self$'' and $\tk$
%refer to the same thing, in fact $\tk$ is semantically interpreted as
%$\beta(\tk)$, a set of worlds, whereas $\self$ is interpreted as a set
%containing a single world (varying over $\beta(\tk)$).  The semantics
%validates additional formulas that do not seem to be acceptable under
%a ``speaks-for'' interpretation of name binding.
For example,  the following is valid under Abadi's semantics:
\begin{equation}\label{badconc}
\begin{array}{l}
(\tk_P \says {\tt US} \spk \self) \land
   (\tk_P \says {\tt US}\spk \tk_{VP}) \\
\limp \tk_P \says (({\tt US} \says {\it false}) \lor
(\self \spk \tk_{VP}))
\end{array}
\end{equation}
Interpreting $\tk_P$ as the key of the president of the US and
$\tk_{VP}$ the key of the vice-president, this is clearly
unreasonable. It should not follow from the fact the the president
says that both he and the vice-president speak for the US that
according to the president, either the US speaks nonsense or the vice
president speaks for the president.

%[[could add that it makes some sense if names are required to refer to
%single individuals ]]

Abadi's suggested semantics for $\self$ works much better in the
context of the logic \llnc.  Suppose we extend this logic to include
$\self$, and like Abadi, define 
%ron11: $\intension{\self}{k} = \{k\}$ for keys $k\in K$.  
$\intension{\self}{\tk} = \{\tk\}$ for keys $\tk\in K$.  
This again validates the Identity axioms above.
%We note that we also obtain the following additional axiom under this
%semantics:
%joe11: rest of section rewritten
%In addition, we get the following axiom:
%%ron3: \[ (\self \bdto \tk ) \limp (\tk \bdto \self) \]
%% even better ..
%\[ (\self \bdto \tk ) \liff (\tk \bdto \self) \]
%where $\tk$ is a key.
To get completeness, we just need to add one axiom in addition to
Identity, which basically says that $\self$ acts like a key 
%ron11: 
(cf.~Nonemptiness (d)):
%joe11: this is unnecessary: we have self\s k \bdto k from the first
%axiom, so it follows from Left Monotonicity
%\item $\tp \bdto \self \rimp \tp\s \tk \bdto \tk$
$$\mbox{Self-is-key}\ \ \
\self \bdto \tp \land \tp\s \tk \bdto \tk \rimp \tp \bdto \self.$$

Let $\AXinfself$ (resp., $\AXfinself$) be the result of adding Identity
and Self-is-key to $\AXinf$ (resp., $\AXfin$).  Let \llncs\ be the
language that results when we add $\self$ to the syntax.
\thm\label{completeself}
$\AXinfself$ (resp., $\AXfinself$) is a sound
and complete axiomatization of \llncs\ with respect to
the open semantics if $K$ is infinite (resp., $K$ is finite).
\ethm

Propositions~\ref{finitemodel} and Theorem~\ref{same} hold with
essentially no change in proof for \llncs; it follows that $\AXfinself$
(resp., $\AXinfself$) is also complete with respect to the closed
semantics and the satisfiability problem is NP-complete.

Interestingly, the proof of completeness shows that once we add Identity
and Self-is-key to the axioms, we no longer need
Current Principal as an axiom in the finite case.
%joe11: may want to cut this
Here is a sketch of
the argument: From Identity we get that $\self\s \tk \bdto
\tk$ is provable for any key $\tk$.  Now applying Witnesses, we get
that $\lor_{\tk \in K} \self \bdto \tk$ is provable.  Together with
Self-is-key, this says that $\self$ is one of the keys in $K$.  Identity
(together with Transitivity) tells us that for that key $\tk$ that is
Self, $\tn \bdto \tk' \liff \tk\s \tn \bdto \tk'$ holds, giving us
Current Principal.

%joe3: unnecessary; this is already
%Our semantics would need amendment to
%take into account certificates using $\self$, but
%it is clear that
%ron3: moved some stuff to prevent the mailer munging ``from''!
Note that with our semantics for $\self$, the counterintuitive
conclusion (\ref{badconc}) does not follow. From 
$\tk_P \cert {\tt US} \bdto \self$ and $\tk_P \cert {\tt US}\bdto
\tk_{VP}$ it follows that $\intension{{\tt US}}{\tk_P} \supseteq
\{ \tk_P, \tk_{VP}\}$.  Thus, we have neither $\intension{{\tt
US}}{\tk_P} = \emptyset$ nor $\{\tk_{VP}\} \supseteq \intension{{\tt
US}}{\tk_P}$, which would be required to get a conclusion similar to
that drawn by Abadi's logic.

\section{Conclusions}\label{sec:concl}

%ron3: [[STILLL TO COME]]
%joe4: added
We have introduced a logic LLNC for reasoning about SDSI's local name
spaces and have argued that it has some significant advantages over
Abadi's logic.  Among other things, it provides a complete
characterization
of SDSI's REF2, has an elegant complete axiomatization, and its
connections with Logic
Programming lead to efficient implementations of many queries of
interest.

%joe4
%While, by contrast to Abadi's logic, LLNC provides a complete
%characterization of SDSI's procedure REF2,
We believe that some of the
dimensions in which Abadi's logic differs from SDSI warrant further
investigation.
%A number of intuitive explanations of examples such as
For example, under some
% ron4: qute reasonable
sensible
interpretations, the conclusions
reached by Abadi's logic in Example~\ref{counterexample2}
%joe4
%are possible.
are quite reasonable.
%joe4
%One would be that while
One such interpretation is that while
local names may be bound to more than one key, they are intended to
denote a single individual.
%ron5: added sentence and rewrote following one slightly
If $\tk$ knows that $\tk_1$ and $\tk_2$ are two keys used by the
one individual Lampson, and Lampson uses $\tk_1$ to certify
that his
%joe6: it seems funny say of that Lampson uses ... his name Ron
%name {\tt Ron} is bound to his name {\tt Rivest},
local name {\tt Ron} is bound to the name {\tt Rivest},
and also uses his key $\tk_2$ to certify
that his
%joe6
local
name {\tt Rivest} is bound to $\tk_3$, then it
is very reasonable to conclude that $\tk\s {\tt Lampson}\s
{\tt Ron}$ is bound to $\tk_3$.
%joe10:  Do we want to say more about this stuff?  Or perhaps cut it?
Another interpretation supporting this
conclusion would be that $\saysn$ aggregates
the certificates issued using a number of distinct keys (possibly
belonging to distinct individuals) much in the way that the notion of
%joe6: changed FHMV95 to FHMV, as in z.bib
{\em distributed knowledge} \cite{FHMV} from the literature on
reasoning about knowledge aggregates the knowledge of a collection of
agents.  We believe that our semantic framework, which, unlike Abadi's,
makes the set of certificates issued explicit, provides an appropriate
basis for the study of such issues.

%joe4: added
Our semantic framework also lends itself to a number of generalizations,
which we are currently exploring.  These include reasoning about the
beliefs of principals and reasoning about permission, authority, and
delegation.  We hope to report on this work shortly.
%joe10: should put more here at some point; also our revocation paper.

%joe10
\appendix

\section{Proofs}
In this appendix, we prove all the technical results stated in the main
text.  For ease of exposition, we repeat the statements of the results
here.

\bigskip

\othm{thm:min:lna}
Given a world $w$, there exists a unique local name assignment
$l_w$ minimal in the set of all local name assignments consistent
with $w$.
Moreover, if $\tp$ is a principal expression and $\tk_1$ and
$\tk_2$ are keys, then
$w,l_w,\tk_1 \omodels \tp \bdto \tk_2$ iff,
for all local name assignments $l$
consistent with $w$, we have $w,l,\tk_1 \omodels \tp \bdto \tk_2$.
\eothm

\medskip

\prf This result can be established
using standard results from the theory of fixed points.
Suppose $(X,\leq)$ is a complete partial order.
Denote the least upper bound of a set $Y \subseteq X$
by $\lub Y$.
A mapping $T:X \rightarrow X$ is said to be {\em monotonic} if
for all $x\leq y$ in $X$ we have $T(x) \leq T(y)$. Such a mapping $T$
is said to be
{\em continuous} if for all
infinite increasing sequences $x_0 \leq x_1 \leq \ldots$ in $X$ we
have $T(\lub \setcpr{x_i}{i\in \Nat} ) =
\lub  \setcpr{T(x_i)}{i\in \Nat}$. Note that continuity implies
monotonicity. To establish continuity of a monotonic mapping $T$, it
suffices to show that $T(\lub \setcpr{x_i}{i\in \Nat} ) \leq
\lub  \setcpr{T(x_i)}{i\in \Nat}$,
%joe3
since the opposite containment is immediate from monotonicity.

%ron10: added para and lemma
For a fixed expression $\tp$, world $w$ and key $\tk$, the
expression $\intension{\tp}{w,l,\tk}$ is easily seen to be
monotonic in $l$, i.e., if $l \leq l'$ then
$\intension{\tp}{w,l,\tk} \subseteq \intension{\tp}{w,l',\tk}$.
%joe11: it seems strange to observe monotonicity, then make continuity
%a  lemma and not provide a proof.
%Using this, it can be seen that
Moreover, it is also continuous in $l$.

\begin{lemma} \label{lem:int:lim}
Suppose $l_0 \leq l_1 \leq \ldots$ is an increasing sequence of local
name assignments and let $l_\omega = \sqcup_{m\in \Nat} l_m$. For all
principal expressions $\tp$, we have $\intension{\tp}{w,l_\omega,\tk}
= \bigcup_{m\in \Nat} \intension{\tp}{w,l_m,\tk}$.
\end{lemma}

%joe11: added
\prf By a straightforward induction on the structure of $\tp$. \eprf

Given the world $w= \la \beta,c\ra$, we define an operator $T_w$ on the
space of local name assignments $\lna$. For a local name assignment
$l$, we define $T_w(l)$ to be the local name assignment such that for
all $\tk\in K$ and $\tn\in N$, the set $T_w(l)(\tk,\tn)$ is the union
of the sets $\intension{\tp}{w,l,\tk}$ such that the formula $\tn\bdto
\tp$ is in $c(\tk)$.
%joe3
% We establish the following results concerning this operator.
%ron10: The following two lemmas are almost immediate from the definitions.
%joe11
%Using Lemma~\ref{lem:int:lim}, we may easily establish the following:
The following lemma is follows easily from Lemma~\ref{lem:int:lim}.

\lem \label{lem:tw:cont}
The mapping $T_w$ is a continuous operator on $(\lna,\leq)$.
\elem

%ron10:
The following lemma is almost immediate from the definitions.

\lem \label{lem:tw:pfp}
A local name assignment $l$ is consistent with a world $w$
iff $T_w(l) \leq l$.
\elem

Suppose $(X,\leq)$ is a complete partial order with minimal element
$\bot$.
%joe10
%A point $x$ in $X$ is said to be a
An element $x \in X$ is said to be a
{\em pre-fixpoint} of an operator $T$ on $X$ if $T(x) \leq x$;
%joe10
%A point $x$ is said to be a {\em fixpoint} of $T$ if $T(x) =x$.
$x$ is a {\em fixpoint} of $T$ if $T(x) =x$.
Given an operator $T$ on $X$, define a sequence of
%joe10
%points
elements
%joe3: \alpha is overloaded
$T\uparrow \gamma$, where $\gamma$ is an ordinal, as follows.
%joe10
%For the base case $\gamma = 0, put $T\uparrow \alpha = \bot$.
For the base case, let $T \uparrow 0 = \bot$.
For successor ordinals $\gamma+1$, define $T\uparrow \gamma+1 =
T(T\uparrow \gamma)$. For limit ordinals $\gamma$, define $T\uparrow
%joe3: I think this is backwards and missing a \lub
%\gamma = \setcpr{\delta <\gamma}{T\uparrow \delta}$. A well known
\gamma = \lub \setcpr{T\uparrow \delta}{\delta <\gamma}$. 
%joe13: rewritten as you suggested
%A well-known result of Knaster and Tarski
%\cite{Tar} states that if $T$ is monotone, then
%this sequence converges to the least pre-fixpoint of $T$. Moreover, a
%%joe10: need Kleene ref
%%joe11: Ron, do you know the appropriate ref here?
%version of this result due to Kleene \cite{} states that when $T$ is
%continuous the sequence has converged by $\gamma = \omega$
%%joe10
%and that $T \uparrow \omega$ is in fact a fixed point of $T$.
 A well-known result (see \cite{LNS82} for a discussion of its history) 
  states that if $T$ is continuous then then this sequences converges 
  to the least pre-fixpoint of $T$, that convergence has taken place by 
  $\gamma = \omega$, 
  and that $T \uparrow \omega$ is in fact a fixed point of $T$.
Thus, we
obtain as a corollary of Lemma~\ref{lem:tw:cont} and
Lemma~\ref{lem:tw:pfp} that there exists a minimal local name
assignment consistent with $w$, and that this local name assignment
equals $T_w\uparrow \omega$.
%joe11: added
The second half of Theorem~\ref{thm:min:lna} is immediate from the
earlier observation that $\intension{\tp}{w,l,\tk}$ is monotonic in $l$.
%This completes the proof of Theorem~\ref{thm:min:lna}.
\eprf

\bigskip

\othm{complete}
%joe11
%The axiom system in Figure~\ref{fig:ax:llnc} is a
%sound
%and complete axiomatization of \llnc, with respect to both the open and
%closed semantics.
$\AXfin$ (resp., $\AXinf$) is a sound
and complete axiomatization of \llnc\ with respect to
the open semantics if $K$ is infinite (resp., 
%ron11:$A$ 
$K$ 
is finite).
\eothm

\medskip

\prf
%joe11: added some glue here
We start with the completeness proof for $\AXinf$, so that we assume
that $K$ is infinite.  We then show how to deal with
$\AXfin$.  As usual, it suffices to show that if $\phi$ is
$\AXinf$-consistent, then $\phi$ is satisfable.   In fact, we put
a little extra work into our proof
that $\phi$ is satisfiable so that we can prove
Proposition~\ref{finitemodel} as well.

%Define an {\em atom\/} to be either a formula of the form $\tp \bdto
%\tq$ or one of the form $\tk \cert (\tn \bdto \tp)$.  The atoms are
%the primitive formulas in \llnc.  A {\em literal\/} is either an atom
%or the negation of an atom.
%
%Since $\AXinf$ includes all instances of
%propositional tautologies and modus ponens, as usual, for completeness,
%%Since we include all propositional logic axioms,
%%ron9: and know the open and closed semantics to be equivalent,
%it suffices to show that every $\AXinf$-consistent conjunction of
%literals is satisfiable with respect to the open semantics.
%%ron9: added treatment of certs
% The role of $\certn$ literals is trivial, so it suffices to focus on
% containments.
%joe11: rewrote this approach slightly
%We show that if a finite set $S$ of
%%containment literals
%literals
%is $\AXinf$-consistent then it is satisfiable.
%joe11: no longernecessary
%Simultaneously, we establish
%that satisfiability may be established by a specific algorithm.

%joe11: expanded this discussion
%Because AX contains the associativity and transitivity
%axioms, we may identify associative variants of principal expressions.

\newcommand{\Sub}{\mbox{Sub}}

Let $\Sub(\phi)$ consist of all subformulas of
$\phi$.
We say that a principal expression $\tp'$ is a {\em variant\/} of $\tp$
if $\tp \bdto \tp'$ and $\tp' \bdto \tp$ are both provable using only
Reflexivity, Associativity, and Transitivity.
The {\em left-associative\/}
variant of a principal expression $\tp$ is the one where we associate
all terms to the left.  Thus, $((\tn_1\s \tn_2)\s \tn_3)\s \tn_4$ is the
left-associative variant of $\tn_1\s ((\tn_2\s \tn_3)\s \tn_4)$.

\newcommand\conte{\rightarrow}
\newcommand\namee[1]{\stackrel{#1}{\rightarrow}}

Define $P$ to be the smallest set of principal expressions such that
\be
\item if $\tp \bdto \tq$
is in $\Sub(\phi)$ then $\tp$ and $\tq$ are in $P$,
\item if $\tk \cert (\tn \bdto \tp) \in \Sub(\phi)$
then $\tk\s \tn$ and $\tk\s \tp$ are in $P$,
%joe11: added
\item if $\tp \in P$ and $\tp'$ is the left-associative variant of
$\tp$, then $\tp' \in P$,
% \item if $\tp \s (\tq \s \tr) \in P$ then $(\tp \s \tq)\s \tr \in P$,
%joe11: simplified; I don't think it hurts
%\item if $(\tp \s \tq)\in P$ then $\tp \in P$,
%\item if $\tp \in K\cup G\cup N$ occurs in any principal expression in
%$P$, then $\tp\in P$, and
\item $P$ is closed under subexpressions, so that if $\tp\s \tq \in P$,
then so are $\tp$ and $\tq$,
\item if $\tk\in P$ is a key and $\tn \in P$ is a local name, then
$\tk\s \tn \in P$.
\ee

For Proposition~\ref{finitemodel}, it is necessary to get an upper bound
on the size of $P$ in terms of $|\phi|$.

\lem\label{length} $|P| < \bound$. \elem

\prf Let $|\tp|$ be the
total number of expressions in $G \union K \union N$ that
appear in $\tp$, counted with multiplicity.
An easy proof by induction on structure shows that a principal
expression $\tp$ has at most $|\tp|$ subexpressions, at least one of
which must be in $G \union K \union N$.  For every other 
%ron11: subexpressions
subexpression
$\tq$, there is a unique left-associative variant $\tq'$, which has at
most $|\tq'| = |\tq| \le |\tp|$ subexpressions, each of which is
associated to the left.  Thus, starting with a principal expression
$\tp$, the
least set closed under clauses 3 and 4 above contains at most
$|\tp|^2$
elements.  Now a straightforward induction on the structure of $\phi$
shows that the least set $P'$ closed under clauses 1-4 above has at most
$|\phi|^2$ expressions.  Finally, it is easy to see that closing off
under 5 
%ron11: given 
gives 
us $P$, since the set that results after closing off under
5 is still closed under 1--4.  
%ron11: Moreover, $|P| < |P'|^2 < \bound$.
Moreover, this final step adds at most $|\phi|^2$ expressions
$\tk\s \tn$, since both $\tk$ and $\tn$ must 
%joe13
%originate in $\phi$. 
be subexpressions of $\phi$.
\eprf

%joe11: I don't use this algorithm any more
%The algorithm for satisfiability is nondeterministic and operates in
%two steps.  In the first step, we add a nondeterministically selected
%set of literals $E$ to $S$ in order to obtain a set of literals that
%%ron9: determines concerning
%decides
%the emptiness or non-emptiness of every
%principal expression in $P$. In the second step, we close $S \cup E$
%under a particular set of inference rules.  The algorithm returns
%``no'' if, for all choices of $E$ in the first step, the second step
%generates some atom $\tp \bdto \tq$ such that $\neg(\tp \bdto \tq) \in
%S\cup E$.  Otherwise the algorithm returns ``yes''.

Let $\tk_0$ be some key not occurring in $P$.  We use
$\tk_0$ both to express emptiness of expressions in $P$ and as the
``current principal''.  Define $P_1$ to be the set of
principal expressions
%ron9: containing $P$, the key $\tk_0$, and all associative variants
%of expressions in
%$ \{ \tp\s \tk_0 ~:~ \tp \in P\}\cup \{ \tk_0 \s \tp ~:~ \tp \in P\}$.
% ... last part not needed now
$P\cup \{\tk_0\} \cup \{ \tp\s \tk_0 ~:~ \tp \in P\}$.
% The set $E$ used in the first step of the algorithm
%is required to be any of the sets of atoms obtained by including
%joe11
Let $E$ be consist of the formulas
%ron9: simplify this to just the emptiness atoms,
%      one item deleted, the other moved to the closure rules
%\be
%\item
%for each principal expression $\tp \in P$ either the literal
$\tp\s \tk_0 \bdto \tk_0$ for each $\tp \in P$.
Note that all principal expressions occurring in the formulas in $E$
are in $P_1$.
Let $S$ be an $\AXinf$-consistent set containing $\phi$ and, for every
formula $\psi \in \Sub(\phi) \union E$, either $\psi$ or $\neg\psi$.
%joe11
%and $S \union E$ is $\AXinf$-consistent.
Since $\phi$ is $\AXinf$-consistent, there
must be some $\AXinf$-consistent set $S$ of this form.
%
%ron9: moved to closure rules
%\item for each distinct pair of keys
%%ron9: $\tk, \tk' \in P_1$,
%% .. don't seem to need k_0, and probably don't want it there
%%    when we add Self! .. [[check]]
%$\tk, \tk' \in P$,
%the atom $\neg (\tk \bdto \tk')$
%
%ron9: ... can now do without the following
%\item the atoms $\tp \bdto \tk_0\s \tp$ and $\tk_0\s \tp \bdto \tp$
%for each $\tp \in P$. \\
%~~~ [[ darn, there is a small bug in the proof because of this. See proof of
%the theorem at the end. I claim that this can be fixed by working with the logic
%that includes $\self$ and using here instead
%$\tp \bdto \self\s \tp$ and $\self\s \tp \bdto \tp$
%for each $\tp \in P$. Its worth doing the self logic anyway.]]
%\ee
%joe11: cut; I think it's a distraction
%(Clearly, when $\tp \in P$ is a key, the atom $\neg(\tp\s \tk_0 \bdto
%\tk_0)$
%is inconsistent. Therefore, we may assume without loss of generality
%that we choose $\tp\s \tk_0 \bdto \tk_0$ for such $\tp$.)

%joe11
%\newcommand\clse{S_1}
\newcommand\clse{S^+}

%The closure operation applied in the second step is as follows.
Define $\clse = Cl(S, P_1)$ to be the
smallest set of formulas containing $S$
%joe11: moved to list of clauses
%all instances of Reflexivity
%%ron9: and Associativity,
%involving expressions in $P_1$ and
closed under
%joe11
Reflexivity,
Transitivity, Left Motonocity, Converse of Globality, Globality, and
Nonemptiness, in the sense that
\begin{itemize}
%joe11: added
\item[(ClR)] if $\tp \in P_1$, then $\tp \bdto \tp \in \clse$,
\item[(ClT)] if $\tp \bdto \tq$ and $\tq \bdto \tr$ are both in
$\clse$, then $\tp \bdto \tr \in \clse$,

\item[(ClLM)] if $\tp \bdto \tq \in \clse$, $\tp\s \tr \in P_1$, and $\tq\s \tr
\in P_1$, then
$\tp\s \tr \bdto \tq\s \tr \in \clse$,

\item[(ClCG)] if $\tp\s \tg \in P_1$ for $\tg \in K\cup G$
then
%ron9: $\tk \bdto \tp\s \tg \clse$,
 $\tg \bdto \tp\s \tg \in \clse$,

\item[(ClG)] if $\tp\s \tk \bdto \tk \in \clse$ for some key $\tk$ and $\tp\s
\tg \in P_1$,
%ron9: added
where $\tg \in K\cup G$,
then $\tp\s \tg \bdto \tg \in \clse$,

%ron9: new case for certs
\item[(ClKL)] if $\tk \cert (\tn \bdto \tp) \in \clse$
then $(\tk\s \tn \bdto \tk\s \tp) \in \clse$,

\item[(ClK)] if $\tp\bdto \tk' \in \clse$ and $\tp\s \tk \in P_1$, then
$\tp\s \tk \bdto \tk \in \clse$,

\item[(ClN)] if $\neg(\tp\bdto \tq) \in \clse$ and $\tq\s \tk \in P_1$, then
$\tq\s \tk \bdto \tk \in \clse$,

\item[(ClC)] if $\tp\s\tq\bdto \tk_1 \in \clse$ and $\tp\s \tk \in P_1$, then
$\tp\s \tk \bdto \tk \in \clse$,

\item[(ClNE)] if $\tp\s \tk \bdto \tk$ and $\tk' \bdto \tp$ are both in
$\clse$, then $\tp \bdto \tk' \in \clse$,

%ron9: moved this from E to here
\item[(ClKD)]
%joe11: rewrote so that it reads like the other clauses
%for each distinct pair of keys
%ron9: $\tk, \tk' \in P_1$,
% .. don't seem to need k_0, and probably don't want it there
%    when we add Self!
if $\tk$ and $\tk'$ are distinct keys in $P$,
%joe11: not an atom!
%the atom $\neg (\tk \bdto \tk')\in \clse$.
then $\neg (\tk \bdto \tk')\in \clse$,
%joe11: added
%ron11: revised, this would have allowed \tp and \tp' to be 
%       in different equivalence classes, leading to a bug in the model 
%       the revised form implies what you had, using ClT 
%joe13: actually, I think the two versions are equivalent, because of 
%(ClR).
%\item[(ClLV)] If $\tp'$ is the left-associative variant of $\tp$, then
%$\tp \bdto \tq \in \clse$ iff $\tp' \bdto \tq \in \clse$.
\item[(ClLV)] If $\tp'$ is the left-associative variant of $\tp\in P$, then
$\tp \bdto \tp' \in \clse$ and  $\tp' \bdto \tp \in \clse$.

\end{itemize}
%ron9: added sentence below
%joe11: cut (and moved computation to decision procedure proof)
%Recall that we view associative variants of principal expressions
%as identified, so the above rules should be treated as being
%applicable if each atom in their left hand side is equal to
%some atom obtained from one in $\clse$ by substituting an associative
%variant for each of its principal expressions.
%Note that all principal expressions occurring in $Cl(S\cup E, P_1)$ are
%in $P_1$. Hence the closure computation must terminate in at
%most $|P_1|^2$ steps.

%joe11: added
It is easy to see that $\clse$ is $\AXinf$-consistent, since $S$ is
and each of the closure rules emulates an axiom in $\AXinf$.
Our goal now is to show that there exists a triple $w, l, \tk$ such that
$w,l, \tk \sat \psi$ for all $\psi \in S$ (and thus, in particular,
$w,l,\tk \sat \phi$).

%ron9: new lemma
\begin{lemma} \label{k0inclse}
%joe11
%All literals in $\clse$ containing $\tk_0$ are of the form
%$\tp \bdto \tq$ where
If $\tk_0$ appears in the formula $\tp \bdto \tq \in \clse$, then
$\tk_0$ appears in both $\tp$ and $\tq$.
\end{lemma}

\prf
An easy induction on the construction of $\clse$,
%joe11
using the fact that all principal expressions occurring in $\clse$ are
in $P_1$ and $\tk_0$ appears only as the right most expression in a
principal expression in $P_1$.
\eprf

By Lemma~\ref{k0inclse}, if $\tp \bdto \tq \in \clse$ and
one of the expressions $\tp, \tq$ is in $P$
%joe11: added
(and thus does not mention $\tk_0$)
then so is the other.
%joe11
Define a binary relation $\approx$ on $P$ by defining
%If $\tp$ and $\tq$ are expressions in $P$,
%we write $\tp \approx \tq$ when
$p \approx q$ if both $\tp \bdto \tq$ and $\tq \bdto
\tp$ are in $\clse$.  It is immediate from transitivity and
reflexivity that $\approx$ is an equivalence relation on
%ron9: $P_1$. Given $\tp\in P_1$,
$P$. Given $\tp\in P$,
we write $[\tp]$ for the equivalence class of $\tp$
under $\approx$.

We classify the expressions in $P$ as follows.  Say that an
expression $\tp$ in $P$ is {\em empty}
%joe11
%in $S\cup E$
(with respect to $\clse$)
if $\neg(\tp\s \tk_0 \bdto \tk_0)$ is in $\clse$.
%joe11
Say that $\tp$ is {\em
key-equivalent} if it is not empty and $\tk \bdto \tp$ is in $\clse$
for some key $\tk$ (by
%ron9: added
%joe11: cut
%definition of $E$ and
(ClNE) this implies $\tp \approx \tk$).
%joe11: added
Intuitively, the interpretation of an empty expression will be the
empty set and the interpretation of a key-equivalent expression $\tp$
such that $\tk \bdto \tp \in \clse$ will be $\{\tk\}$.
If $\tp$
is neither empty nor key-equivalent, we say it is {\em open}.
Clearly, every expression in $P$ is either empty, key-equivalent, or
open. Moreover, by (ClLM) and (ClT), if $\tp \approx \tq$ then $\tp$ is
empty, key-equivalent or open iff $\tq$ is. In particular, we may
sensibly refer to open $\approx$-equivalence classes of expressions
%ron9: added
in $P$.

%Say that a set $T$ of literals is {\em contradiction-free} if it does
%not contain a contradictory pair of literals of the form
%%ron9: -- could also have cert contradiction
%% $\tp \bdto \tq$ and $\neg (\tp \bdto \tq)$.
%$L$ and $\neg L$.
%We now show that if $\clse$ is contradiction-free
%%joe11: no longer necessary (since \clse = S_E)
%%(for some choice of $E$)
%then $S$ is satisfiable
%%joe11
%%.  In order to
%%establish this we show how to construct a model for $S$ from $\clse$.
%by explicitly constructing a model for $S$ from $\clse$.

%joe11: cut; said it above
%Clearly, any expression in $P$ that is empty
%should have empty intension in the
%model, and any expression equivalent to a key $\tk$ should have
%%intension equal to the singleton $\{\tk\}$.
%should be interpreted as $\{ \}$ in the model
%and any expression equivalent to a key $\tk$ should be interpreted as
%$\{\tk\}$. That leaves the open expressions.
Let $O$ be the set of open equivalence classes of expressions in $P$.
Note that if $K_\phi \subseteq K$ consists of all the keys in $K$ that
appear in $\phi$, then there are 
%ron11 $< \bound - |K_\phi|$ 
fewer than  $\bound - |K_\phi|$ 
equivalence
classes of open expressions.   For each class
$c \in O$, let $\tk_c$ be a fresh key.
Intuitively, the key $\tk_c$ will act as a
canonical representative of the keys in the interpretation of an
expression 
%ron11: $p\in c$, 
$\tp\in c$, 
in the sense that the interpretations of
$\tp\s \tq$ and $\tk_c\s \tq$ will be the same for certain
expressions $\tq$.  Since $K$ is infinite, we are guaranteed that we can
always find keys $\tk_c$, but the argument works even if $K$ is
finite, as long as $|K| \ge \bound$.  (We also need to have a key in
%ron11: filled gap in sentence 
$K\setminus K_\phi$
to be $\tk_0$.)

%joe11
%\newcommand\clsem{S_2}
\newcommand\clsem{S^*}

Define $\clsem$ to be consist of $\clse$ together with,
for all $c\in O$,
\be
\item the formula $\tk_c \bdto \tk_c
%ron9: \in \clsem
$, and
\item the formulas $\tp \bdto \tk_c
%ron9: \in \clsem
$, where
for some $\tq \in c$ we have $\tp \bdto \tq \in \clse$.
\ee
%ron9: added
%joe11: reworded
It is easy to show that $\tk_0$ does not appear in any formula in
$\clsem - \clse$:
Clearly $\tk_0$ does not appear in the formulas $\tk_c \bdto \tk_c$
added by
clause 1.  If $\tp \bdto \tq$ is a formula added by clause 2, then there
is some equivalence class $c$ and expression $\tq \in c$ such that $\tp
\bdto \tq \in \clse$.  Since $c$ is an
equivalence class of expressions in $P$, none of which
contain $\tk_0$, the expression $\tq$ does not contain $\tk_0$.
It follows from Lemma~\ref{k0inclse} that $\tp$ does not
contain $\tk_0$.
%we consider only expressions $\tq \in P$, which do not contain $\tk_0$.
%Hence, by Lemma~\ref{k0inclse}, we introduce no new formulas
%containing $\tk_0$ in this step,
Since $\clsem - \clse$ contains no formulas involving $\tk_0$,
$\clsem$ also satisfies the
property stated for $\clse$ in Lemma~\ref{k0inclse}.

Define the local name assignment $l$ as follows.
Given a key $\tk$ and local name $\tn$,
%joe11
%we put
\be
%ron9: added
\item $l(\tk_0,\tn) = \{ \tk'\in K~|~ \tn \bdto \tk' \in \clsem \}$,
\item $l(\tk,\tn) = \{ \tk'\in K~|~ \tk\s \tn \bdto \tk' \in \clsem
\}$ if $\tk \in
%ron9: P_1$.
P$,

\item $l(\tk,\tn) = \{ \tk'\in K~|~ \tp\s \tn \bdto \tk' \in \clsem
~\mbox{and}~ \tp \in c \}$ if $\tk = \tk_{c}$ for some $c\in O$,
%ron9: added
\item $l(\tk,\tn) = \emptyset$ for all other $\tk$.
\ee

Define the world $w = \la \beta,c \ra$ by
taking $\beta(\tg) = \{ \tk \in K~|~ \tg \bdto \tk \in \clsem\}$
%ron9: add certs
and defining $c(\tk)$, for each key $\tk$, to be the
set of formulas $\tn \bdto \tp$ such that $(\tk\cert (\tn\bdto \tp)) \in S$.
%joe11
Note for future reference that there exists a finite subset $K_1$ of $K$
such that 
%ron11: $l(n,\tk) \subseteq K_1$, .. and below 
$l(\tn,\tk) \subseteq K_1$, 
% $l(n,\tk) = \emptyset$ 
$l(\tn,\tk) = \emptyset$ 
for $\tk \notin K_1$,
%joe13: moved and added new clause
$\beta(\tg) \subseteq K_1$, and
$\beta(\tg) = \emptyset$ if $\tg$ does not appear in $\phi$.  
Indeed, $K_1$ consists of
the keys that appear in $S$, $\tk_0$, and the keys $\tk_c$ for $c \in
O$.

%joe11: added; you seemed to be implicitly assuming this lemma
%implicitly.
Let $I(\tp) = \{ \tk \in K~|~ \tp \bdto \tk \in \clsem\}$.
\lem\label{empty} If $\tp \in P$, then
$\tp$ is empty iff $I(\tp) = \emptyset$. \elem

\prf  If $\tp$ is not empty, then it is either key-equivalent or open.
If it is key-equivalent, we have already observed that there must exist
some key $\tk'$ such that $\tp \bdto \tk' \in \clsem$, so $I(\tp) \ne
\emptyset$.  If it is open, suppose it is in equivalence class $c$.
Then $\tp \bdto \tk_c \in\clsem$, since $\tp \bdto \tp \in \clse$ by
(ClR).  Again, it follows that $I(\tp) \ne \emptyset$.

Conversely, suppose that $I(\tp) \ne \emptyset$.  Thus, $\tp \bdto \tk
\in \clsem$ for some key $\tk$.  If $\tp \bdto \tk \in \clse$, then
by (ClK), $\tp\s \tk_0 \bdto \tk_0 \in \clse$, so $\tp$ is not empty.
If $\tp \bdto \tk \notin \clse$, then $\tk= \tk_c$, and there is some 
%ron11 $q \in c$ ... couple more below 
$\tq \in c$
such that $\tp \bdto \tq \in \clse$.  Since $\tq$ is open, $\tq$ cannot
be empty, so $\tq\s \tk_0 \bdto \tk_0 \in \clse$.  Moreover, by
(ClLM), $\tp\s \tk_0 \bdto \tq\s \tk_0 \in \clse$.  Thus, by (ClT),
$\tp\s
\tk_0 \bdto \tk_0 \in \clse$, so $\tp$ is nonempty.
\eprf

\begin{lemma} \label{lem:intp}
For all expressions $\tp\in P$, we have
$\intension{\tp}{w,l,\tk_0} = I(\tp)$.
\end{lemma}

\prf
We proceed
%joe11: added
by induction on $|\tp|$ (as defined in Lemma~\ref{length}).
The claim is immediate from the definitions in case $\tp$ is a global
name
%ron9:
or a local name.
Suppose that
$\tp$ is a key $\tk_1$.  Then
$\intension{\tp}{w,l,\tk_0} = \{ \tk_1\}$.
%joe11
%Note that $\tk_1 \in \{ \tk \in K
%~|~ (\tp \bdto \tk) \in \clsem\}$ because we apply reflexivity to keys
%%ron9: both in the construction of $\clse$ and $\clsem$.
%in the construction of $\clse$.
Since $\tk_1 \bdto \tk_1 \in \clsem$ by construction, it follows that
$\tk_1 \in I(\tk_1)$.  It remains to show
that
%joe11
%if $(\tk_1 \bdto \tk) \in \clsem$ then $\tk = \tk_1$.
%ron11: $I(\tk_1) = \{\tk_1\}$. .. more to the point 
$I(\tk_1) \subseteq \{\tk_1\}$.
%ron9: For, suppose that $\tk \neq \tk_1$.
%ron9:
Suppose $(\tk_1 \bdto \tk) \in \clsem$.
By Lemma~\ref{k0inclse}, we cannot have $\tk = \tk_0$.
%joe11
%Since $S$ is contradiction-free and
%contains the key distinctness literals for all keys in
Since $\clse$ is 
%ron11: $\AXfin$-consistent 
$\AXinf$-consistent 
and closed under (ClKD),
%ron9: $P_1$,
if $\tk\in P$ we must have $\tk_1 = \tk$.
% at least one of the keys $\tk, \tk_1$ must be equal to a key $\tk_c$ for some
% $c\in O$.
%ron9: added
The remaining possibility  for
$\tk$, that it equals $\tk_c$ for some $c\in O$, cannot happen.
%Now we cannot have $\tk = \tk_c$.
For if so, only the
second clause of the definition of $\clsem$ could explain $(\tk_1
\bdto \tk) \in \clsem$. But then we have $(\tk_1 \bdto \tq)\in \clse$
for some $\tq \in c$. This contradicts the assumption that $c$ is an
equivalence class of open expressions.
%ron9: the following not necessary since $\tk_1 \in P$
%In the remaining possibility,
%i.e., $\tk_1 = \tk_c$, only the first clause of the definition of
%$\clsem$ can explain $(\tk_1 \bdto \tk) \in \clsem$, so we indeed have
%$\tk_1 = \tk$.

%ron9: definition revised ....
%Consider next the case that $\tp$ is a local name $\tn \in P$.   By definition of $E$, we
%have $\tn \bdto \tk_0\s \tn $ and $\tk_0\s \tn \bdto \tn$ in $\clse$.
%It follows from this, and the fact that $\clse$ is closed under
%transitivity, that $(\tk_0\s \tn \bdto \tk') \in \clsem$ iff $(\tn \bdto
%\tk') \in \clsem$. Thus,
%$\intension{\tp}{w,l,\tk_0} = l(\tk_0,\tn) = \{ \tk' ~|~
%(\tk_0\s \tn \bdto \tk') \in \clsem\} =
%\{ \tk' ~|~
%(\tn \bdto \tk') \in \clsem\} $, as required.

%joe11: largely rewrote argument
%Consider finally compound principal expressions of the form $\tp\s
%\tr$. Here we note that (by Lemma~\ref{??})
%Finally, suppose that $\tp$ has the form $\tq\s \tr$.
Finally, suppose that $|\tp| > 1$.
Let $\tp'$ be the
left-associative variant of $\tp$.  It is clear from the semantics that
%$\intension{(\tp \s \tq)\s \tr}{w,l,\tk_0} =
%\intension{\tp \s (\tq\s \tr)}{w,l,\tk_0}$
$\intension{\tp}{w,l,\tk_0} = \intension{\tp'}{w,l,\tk_0}$.
%ron11: Morover, (ClLV) guarantees that $I(\tp) = I(\tp')$.  
% I don't think this was right for your version 
Morover, (ClLV) and (ClT) guarantee that $I(\tp) = I(\tp')$.  
Thus, it suffices
to prove that $I(\tp') = \intension{\tp'}{w,l,\tk_0}$.
%%ron9:
%and we have identified associative variants.
%%Moreover, because $\clse$ is closed under associativity, we have that
%%$\{ \tk' ~|~ (((\tp \s \tq)\s \tr) \bdto \tk') \in \clsem\} = \{ \tk'
%%~|~ ((\tp \s (\tq\s \tr)) \bdto \tk') \in \clsem \}$.
%Thus, it
%suffices to consider expressions that are associated to the left,
%i.e. of the form
%%ron9: $(\ldots( (\tp_1 \s \tp_2) \s \tp_3) \s \ldots\tp_{n-1}) \s
%\tp_n$.
%$(\ldots( (\tp_1 \s \tp_2) \s \ldots\tp_{n-1}) \s \tp_n$
%where the $\tp_i$ are basic.
%Suppose by way of induction hypothesis that
Suppose that $\tp' = \tq\s \tr$.  The definition of
length guarantees that $|\tp'| = |\tp| > |\tq|$,
so the induction hypothesis applies to $\tq$.  Since
$\tp'$ is associated to the left, $\tr \in G \union K \union N$.

Suppose that $\tr = \tg \in G \union K$.
%We handle the cases where $\tr= \tg$ is a key or a global
%name together.
Note that $\intension{\tq\s \tg}{w,l,\tk_0} = \emptyset$
if $\intension{\tq}{w,l,\tk_0} = \emptyset$ and
$\intension{\tq\s \tg}{w,l,\tk_0} = \intension{\tg}{w,l,\tk_0}$
if $\intension{\tq}{w,l,\tk_0} \ne \emptyset$. We consider these two
cases separately.

Suppose first that $\intension{\tq}{w,l,\tk_0} = \emptyset$, so
$\intension{\tp'}{w,l,\tk_0} = \emptyset$.  By the induction hypothesis,
$I(\tq) = \emptyset$.  To show that $I(\tp') = \emptyset$,
%joe11: I couldn't follow your proof (see below), so I rewrote it
we show that $\tp'$ is empty.  Suppose not.  Then $(\tp')\s \tk_0
\bdto \tk_0 \in \clse$.  Since $\clse$ contains either
$\tq\s \tk_0 \bdto \tk_0$ or $\neg(\tq\s \tk_0 \bdto \tk_0)$ and
$\clse$ is $\AXinf$-consistent, by Nonemptiness(c), Associativity,
and Transitivity, we
must have $\tq\s \tk_0 \bdto \tk_0 \in \clse$.  Thus, $\tq$ is not
empty.  By Lemma~\ref{empty}, $I(\tq) \ne \emptyset$, a contradiction.
Hence, $\tp'$ is empty.  It now follows from
Lemma~\ref{empty} that $I(\tp') = \emptyset$, as desired.

%joe11: I didn't see why the openness of \tp' (your \tp) immediately
%implied that \tp\s \tk_0 \bdto \tk_0 \in \clse, as your proof suggests
%so $\tp\s \tk_0 \bdto \tk_0 \in \clse$.  Thus in either case, $\tp$ is
%not empty, hence either key-equivalent or open.  If $\tp$ is
%key-equivalent, then $\tp\bdto \tk' \in \clse$ for some key
%$\tk'$. But then, by the induction hypothesis, we get $\tk' \in
%\intension{\tp}{w,l,\tk_0}$, contradicting the assumption that this
%set is empty. If $\tp$ is open, then we have $\tp \bdto \tk_{[\tp]}
%\in \clsem$, leading to a similar contradiction.  Thus, $\{ \tk\in K
%~|~ \tp\s \tg \bdto \tk\in
%\clsem\}= \emptyset = \intension{\tp\s\tg}{w,l,\tk_0}$ when
%$\intension{\tp}{w,l,\tk_0} = \emptyset$.

Consider next the case where $\intension{\tq}{w,l,\tk_0} \ne\emptyset$,
so $\intension{\tp'}{w,l,\tk_0} =
\intension{\tq\s \tg}{w,l,\tk_0} =
\intension{\tg}{w,l,\tk_0}$. To show that
$\intension{\tp'}{w,l,\tk_0} = I(\tp')$, we show that $I(\tp') = I(\tg)$.
The result then follows from the induction hypothesis.
%First, we show
%$\{ \tk' \in K~|~ \tp\s \tg \bdto \tk'\in \clsem \}=
%\{ \tk' \in K~|~ \tg \bdto \tk'\in \clsem \}$.

By the induction hypothesis, $I(\tq) \ne \emptyset$, so by
Lemma~\ref{empty}, $\tq$ is not empty.  It follows from
%joe11: this argument requires ClG, not ClCG, which is why I think you
%need the lemma
%by ClCG, $\tp\s \tg \bdto \tg \in \clse$. It follows from this that
(ClG) that $\tq\s \tg \bdto \tg \in \clse$.  Suppose that $\tk \in
I(\tg)$.  If $\tk \in P_1$, then
$\tg \bdto \tk \in \clse$, so by (ClT), $\tq\s \tg \bdto \tk \in \clse$
and $\tk \in 
%ron11: I((\tp')$.  
I(\tp')$.  
If $\tk = \tk_c$ for some $c \in O$,
then $\tg \bdto \tq'\in \clse$ for
some $\tq' \in c$.  Thus, $\tp' \bdto \tq'\in \clse$ by (ClT) and we
obtain that $\tp' \bdto \tk\in \clsem$ by construction of
$\clsem$.  Thus, $I(\tg) \subseteq I(\tp')$.

For the opposite containment, note that
%joe11
%by the inductive hypothesis
%and the assumption that $\intension{\tp}{w,l,\tk_0}$ is nonempty,
%we have $\tp \bdto \tk \in \clsem$ for some key $\tk$.
%It follows that $\tp$ is either key-equivalent or open, hence
%$\tp\s \tk_0 \bdto \tk_0 \in \clse$. Thus, by CLG, we have
by (ClCG) we have
$\tg \bdto \tq\s \tg \in \clse$. Arguing as above, we obtain using
(ClT) that $I(\tg) \supseteq I(\tp')$.
%$\{ \tk' \in K~|~ \tp\s \tg \bdto \tk'\in \clsem\} \subseteq
%\{ \tk' \in K~|~ \tg \bdto \tk' \in \clsem\}$.
This completes the proof that $I(\tp') = I(\tg)$.

%joe11: this is just the inductive hypothesis!  Unless I'm missing
%something, you've just repeated the first part of the proof.
%To show $\{ \tk' \in K~|~ \tg \bdto \tk'\in \clsem\}=
%\intension{\tg}{w,l,\tk_0}$, we consider the cases that $\tg \in G$ and
%$\tg\in K$ separately.  If $\tg \in G$, then the claim is immediate
%from the construction of $w$. Suppose therefore that $\tg = \tk\in K$:
%here we need to show that $\{ \tk' \in K~|~ \tk \bdto \tk'\in
%\clsem\}= \{\tk\}$. It is immediate from the construction
%that $\tk \in \{ \tk' \in K~|~ \tk \bdto \tk'\in
%\clsem\}$, so it sufffices to show that if
%$\tk \bdto \tk'\in\clsem$ for a key $\tk'$ then $\tk' = \tk$.  Suppose
%$\tk \bdto \tk'\in\clsem$. If $\tk' = \tk_0$, then $\tk \bdto
%\tk'\in\clse$, so also $\tk = \tk_0$ by Lemma~\ref{k0inclse}. If $\tk'
%\in P$, then also $\tk \bdto \tk'\in\clse$ and we must have $\tk =
%\tk'$ by consistency of $\clse$ and the fact that this set contains
%the key distinctness atoms for keys in $P$.  Finally, the case that
%$\tk' = \tk_c$ for an open class $c\in O$ cannot happen. For then we
%would have $\tk \bdto \tq \in \clse$ for some $\tq\in c$.  But this
%means that $\tq$ is empty or key-equivalent, contradicting the
%assumption that $c$ is a class of open expressions.  Thus, in all
%possible cases we have $\tk = \tk'$.  This completes the proof for the
%case $\tp\s \tg$ where $\tg \in K\cup G$.

It remains to deal with the case that $\tp'$ has of the form
$\tq\s \tn$, where $\tn$ is a local name.
%joe11: simplified argument
There are three possibilities: $\tq$ is empty, key-equivalent or
open.
If $\tq$ is empty, then by Lemma~\ref{empty} and the induction
hypothesis, $I(\tq) = \emptyset$ and $\intension{\tq}{w,l,\tk_0} =
\emptyset$.  It follows that $\intension{\tp'}{w,l,\tk_0} = \emptyset$.
Moreover, using Nonemptiness(c), Associativity, and Transitivity as
above, it follows that $\tp'$ is empty and hence by Lemma~\ref{empty},
$I(\tp') = \emptyset$, as desired.

If $\tq$ is key-equivalent, say $\tq \approx \tk_1$, then $\tq \bdto
\tk_1 \in \clse$ and $\tk_1 \bdto \tq \in \clse$.  Using
Key Distinctness and the consistency of $\clse$, it easily follows that
$I(\tq) = \{\tk_1\}$.  By the induction hypothesis,
$\intension{\tq}{w,l,\tk_0} = \{k_1\}$.   Thus,
$\intension{\tp'}{w,l,\tk_0} = l(\tk_1,\tn)$.  By construction,
$l(\tk_1,\tn) = I(\tk_1\s \tn) = I(\tp')$, as desired.

Finally, suppose that $\tq$ is open.   If $\tk \in I(\tp')$, then it
is immediate from the construction that
that $\tq \bdto \tk_{[\tq]} \in \clsem$ and $\tk \in
l(\tk_{[\tq]},\tn)$.  By the induction hypothesis, $\tk_{[\tq]} \in
\intension{\tq}{w,l,\tk_0}$, so $\tk \in
\intension{\tp'}{w,l,\tk_0} = \union_{\tk' \in
\intension{\tq}{w,l,\tk_0}} l(\tk',\tn)$.
Thus, $I(\tp') \subseteq \intension{\tp'}{w,l,\tk_0}$ if $\tp'$ is
open.

For the opposite containment, suppose that $\tk \in
\intension{\tp'}{w,l,\tk_0}$.  This means that there is some key $\tk'$
such that
%By the characterization of $\intension{\tq\s \tn}{w,l,\tk_0}$ noted
%above, we have
$\tk' \in \intension{\tq}{w,l,\tk_0}$ and $\tk \in l(\tk',\tn)$.  By
the induction hypothesis, $\tk' \in I(\tq)$, so
$\tq \bdto \tk' \in \clsem$.
If $\tk'\in P_1$, then $\tq \bdto \tk'\in \clse$ and
$(\tk')\s \tn \bdto \tk\in \clse$.
%ron9: added
(Since $\tq \in P$ and $\tq \bdto \tk'\in \clsem$,
we cannot have $\tk'=\tk_0$, by Lemma~\ref{k0inclse}.)
By (ClLM), $\tq\s
\tn \bdto (\tk')\s \tn \in \clse$, so by (ClT) we get $\tq\s \tn \bdto
\tk \in
\clse$.  Hence, $\tk \in I(\tp')$.
\newcommand\ts{{\tt t}}
If $\tk' = \tk_c$, where $c$ is an open
equivalence class, then from $\tq \bdto \tk'\in \clsem$ it follows that
$\tq \bdto \tq' \in \clse$ for some $\tq' \in c$.  From $\tk\in
l(\tk_c, \tn)$ it follows that $(\tr') \s \tn \bdto \tk \in \clsem$ for
some $\tr'
\in c$. 
%ron11: clarified reasoning 
%Since $\tr' \approx \tq'$, we must have $(\tr')\s \tn \in
%P_1$ and
%%Since both $\tr$ and $\tq$ belong to $c$, we have
%$\tq' \bdto
%\tr' \in \clse$. 
By construction of $\clse$ we must have $(\tr')\s \tn \in
P_1$, and since $\tr' \approx \tq'$, we have $\tq' \bdto \tr' \in \clse$. 
By (ClT) we obtain $\tq \bdto \tr' \in \clse$, and
hence
by (ClLM) that $\tq\s \tn \bdto (\tr')\s \tn \in \clse$.  Now notice
that
it follows from $\tq\s \tn \bdto (\tr')\s \tn \in \clse$ and
$(\tr')\s \tn
\bdto \tk \in \clsem$ that $\tq\s \tn \bdto \tk \in \clsem$. If
$\tk\in P_1$, this is immediate from (ClT). In case $\tk = \tk_d$
for some open class $d$, we have $(\tr') \s \tn \bdto \ts \in \clse$
for some $\ts \in d$. But then $\tq\s \tn \bdto \ts\in \clse$
%ron9: and again we get by (ClT) and definition of $\clsem$ that
by (ClT); by definition of $\clsem$ we get that
$\tq \s \tn \bdto \tk \in \clsem$. This completes the proof.
\eprf

\begin{lemma} \label{lem:mod}
%joe11
%For all literals $L\in S$, we have
For all formulas $\psi \in \Sub(\phi) \union E$, we have
$\psi \in S$ iff $w,l,\tk_0
%ron9: \models
\omodels
\psi$.
\end{lemma}

\prf
%ron9:
%joe11: we proceed by induction on the structure of $\psi \in \clsem$.
We first show that by induction on the structure of $\psi \in
\Sub(\phi) \union E$ that
$\psi \in S$ iff $w,l,\tk_0 \models \psi$, and then
show that the assignment $l$ is consistent with $w$.

%ron9: assuming we have restricted the syntax as suggested by the reviewer...
It is immediate from the construction of $w$ that
%joe11:
%$w,l,\tk_0 \models \psi$ for all $\psi \in S$ of the form
%$\tk\cert(\tn\bdto \tp)$ or $\neg(\tk\cert(\tn\bdto \tp))$.
$w,l,\tk_0 \models \psi$ iff $\psi \in S$
for $\psi$ of the form $\tk\cert(\tn\bdto \tp)$.

If $\psi$ has the form $\tp \bdto \tq$, note that $w,l,\tk_0 \models \tp
\bdto \tq$ iff $\intension{\tp}{w,l,\tk_0} \supseteq
\intension{\tq}{w,l,\tk_0}$ iff (by Lemma~\ref{lem:intp})
iff $I(\tp) \supseteq I(\tq)$.  Thus, it suffices to show that $I(\tp)
\supseteq I(\tq)$ iff $\tp \bdto \tq \in \clse$, for $\tp, \tq \in P$.

The ``if'' direction is immediate from
(ClT):  If $\tk \in I(\tq)$ then $\tq \bdto \tk \in \clsem$,
so by (ClT) and the construction of $\clsem$, $\tp \bdto \tk \in \clsem$
and thus $\tk \in I(\tp)$.

For the ``only if'' direction, suppose by way of contradiction that
$I(\tp) \supseteq I(\tq)$ but $\tp \bdto \tq \notin \clse$.  Then, by
construction, $\neg(\tp \bdto \tq) \in \clse$.  We consider three
cases, depending on whether $\tq$ is empty, key-equivalent, or open.

Note first that $\tq$ cannot be empty:
$\neg (\tp \bdto \tq) \in \clse$, so by (ClN) we have
$\tq\s \tk_0 \bdto \tk_0 \in \clse$.

Suppose that $\tq$ is key-equivalent, with
%ron9: $\tk \bdto \tq$.
$\tk \bdto \tq\in \clse$.
If $\tp \bdto \tk \in \clse$ then, by (ClT),
$\tp \bdto \tq \in \clse$, but this is not possible
because $\clse$ is $\AXinf$-consistent.  Thus
$\tp \bdto \tk \notin \clse$.
%ron9: added next line
Since $\tk\in P$, $\tp \bdto \tk \notin \clsem$,
and thus $\tk \in  I(\tp) - I(\tq)$, giving us the desired
contradiction.

Finally, suppose $\tq$ is open.
By construction, $\tq \bdto \tk_{[\tq]} \in
\clsem$.
Moreover, we cannot have $\tp \bdto \tk_{[\tq]} \in \clsem$, for
then there would exist $\tr \approx \tq$ such that
$\tp \bdto \tr \in \clse$. Using (ClT), it would follow that
$\tp \bdto \tq \in \clse$, which is impossible
since $\clse$ is $\AXinf$-consistent.  Thus, $\tk_{[\tq]} \in I(\tp) -
I(\tq)$, giving the required contradiction, and completing the proof in
the case that $\psi$ is of the form $\tp \bdto \tq$.

If $\psi$ is of the form $\neg \psi'$ or $\psi_1 \land \psi_2$, the
result is immediate from the induction hypothesis (in the latter case,
we need the fact that if $\psi_1 \land \psi_2 \in \Sub(\phi) \union E$,
then in fact $\psi \land \psi_2 \in \Sub(\phi)$, so $\psi_1, \psi_2
\in \Sub(\phi)$ and the induction hypothesis applies).
This completes the induction proof.

%ron9: added for certs
%This completes the proof that for all literals $L\in S$,
%we have $w,l,\tk_0 \models L$.
To show that the assignment $l$ is consistent with $w$,
%joe11: added
suppose that 
%ron11: $\tp \bdto \tp \in c(\tk)$.  
$\tn \bdto \tp \in c(\tk)$.  
Then, by construction,
$\tk\cert (\tn \bdto \tp) \in S$. By
(ClKL), we have
$\tk\s \tn \bdto \tk\s \tp \in \clse$. By what we have just
shown $w,l,\tk_0 \models \tk\s \tn \bdto \tk\s \tp$.
It follows that $w,l,\tk \models \tn \bdto \tp$.
Thus,  $l$ is consistent with $w$.
\eprf

Thus, we have shown that $\phi$ is satisfiable, completing
the proof of Theorem~\ref{complete} in the case that $K$ is infinite.
The same argument works without change if $K$ is finite but $|K| \ge
\bound$.
%ron11: (Note that 
(A consequence of this is that 
we do not need to use the axioms Witnesses
and Current Principal to derive a valid formula $\phi$ in $\AXfin$ if
$\bound \le |K|$.)
Moreover, the proof shows that Proposition~\ref{finitemodel} holds if
$|K| \ge \bound$.

Now suppose that $K \le \bound$.  We show
that if $\phi$ is $\AXfin$-consistent, then $\phi$ is satisfiable.  The
proof is in the spirit of that in the case of $\AXinf$, but simpler.

%joe11: other theorems no longer necessary; moved to after \end{document}
Now let $P$ be the least set of principal expressions containing all
principal expressions that appear in $\phi$ and closed under
subexpressions.  Let $F$ consist of all formulas of the form
$\tp \bdto \tk'$ and $\tk\s \tp \bdto \tk'$, where $\tp \in P$ and $\tk,
\tk' \in K$.
Let $S$ be an $\AXfin$-consistent set containing $\phi$ and, for every
formula $\psi \in \Sub(\phi) \union F$, either $\psi$ or $\neg\psi$.
Since $\phi$ is $\AXfin$-consistent, there
must be some $\AXfin$-consistent set $S$ of this form.

There must be some key $\tk_0 \in K$ such that for every local name in
$P$ and key $\tk \in K$, we have $\tn \bdto \tk \in S$ iff $\tk_0\s \tn
\bdto \tk \in S$.  For otherwise, for each key $\tk$, there is some
local name $\tn_\tk$ and key $\tk_\tk$ such that either both $\tn_\tk
\bdto
\tk_\tk$ and $\neg(\tk\s\tn_\tk \bdto \tk_\tk)$ are in $S$ or
both $\neg (\tn_\tk \bdto
\tk_\tk)$ and $\tk\s\tn_\tk \bdto \tk_\tk$ are in $S$.  This means that
$S$ is inconsistent with the axiom Current Principal.
Define the local assignment $l$ so that $l(\tk,\tn) = \{\tk': \tk\s \tn
%joe12: typo!  please correct ... %ron11: done 
%\bdto \tk \in S\}$.  Similar to the case for $\AXinf$, define
\bdto \tk' \in S\}$.  Similar to the case for $\AXinf$, define
the world $w = \la \beta,c \ra$ by
taking $\beta(\tg) = \{ \tk \in K~|~ \tg \bdto \tk \in
%ron11:  \clsem\}$  ... no S^* in this construction, right? 
\clse\}$
and defining $c(\tk)$, for each key $\tk$, to be the
set of formulas $\tn \bdto \tp$ such that $\tk\cert (\tn\bdto \tp)
\in S$.

Now we have the following analogue to Lemma~\ref{lem:mod}.
\begin{lemma} \label{lem:mod1}
For all formulas $\psi \in \Sub(\phi) \union F$, we have
$\psi \in S$ iff $w,l,\tk_0 \omodels \psi$.
\end{lemma}

\prf Again we first show that by induction on the structure of $\psi \in
\Sub(\phi) \union E$ that
$\psi \in S$ iff $w,l,\tk_0 \models \psi$, and then
show that the assignment $l$ is consistent with $w$.

It is immediate from the construction of $w$ that
$w,l,\tk_0 \models \psi$ iff $\psi \in S$
for $\psi$ of the form $\tk\cert(\tn\bdto \tp)$.

We next show that the result holds if $\psi$ is of the form $\tp \bdto
\tk'$, for $\tp \in P$, by induction on the structure of $\tp$.  We
strengthen the induction hypothesis to also show that
$w,l,\tk_0 \models \tk\s \tp \bdto \tk'$ iff
$\tk\s\tp \bdto \tk \in S$.  If $\tp$ is a key $\tk_1$, then $w,l,\tk_0
\models \tk_1 \bdto \tk'$ iff $\tk' = \tk_1$ and by Reflexivity and
Key Distinctness, $\tk_1 \bdto \tk' \in S$ iff $\tk_1 = \tk'$.
Similarly, $w,l,\tk_0 \models \tk\s \tk_1 \bdto \tk'$ iff
$w,l,\tk_0 \models \tk_1 \bdto \tk'$ iff $\tk_1 \bdto \tk' \in S$ iff
$\tk\s \tk_1 \bdto \tk' \in S$, by Transitivity, Key Globality, and
Converse of Globality (using the fact that $S$ is $\AXfin$-consistent).

If $\tp$ is a global identifier $\tg$, $w,l,\tk_0
\models \tg \bdto \tk'$ iff 
%ron11: $\tg \bdto \tk'$ 
$\tg \bdto \tk' \in S$ 
by the definition of
$\beta$.  The argument for $\tk\s \tg \bdto \tk'$ is identical to the
case that $\tp = \tk$.

If $\tp$ is the local name $\tn$, then $w,l,\tk_0
\models \tn \bdto \tk'$ iff $\tk' \in l(\tk_0,\tn)$ iff $\tk_0\s \tn
\bdto \tk' \in S$ iff $\tn \bdto \tk' \in S$, by choice of $\tk_0$.
Similarly, $w,l,\tk_0 \models \tk\s \tn \bdto \tk'$ iff
$\tk' \in l(\tn,\tk)$ iff $\tk\s \tn \bdto \tk' \in S$.

Finally, if $\tp$ is of the form $\tq\s \tr$, then $w,l,\tk_0
\models \tq\s \tr \bdto \tk'$ iff there exists a key $\tk''$ such that
$w,l,\tk_0 \models \tq \bdto \tk''$ and $w,l,\tk_0 \models (\tk'')\s \tr
\bdto \tk'$ iff (by the induction hypothesis) there exists a key
$\tk''$ such that $\tq \bdto \tk''
\in S$ and $(\tk'')\s \tr \bdto \tk' \in S$ iff $\tq\s \tr \bdto \tk'
\in S$.  The ``only if'' direction of the last equivalence follows using
Left Monotonocity and Transitivity; the ``if'' direction follows from
Witnesses.  The argument for $\tk\s (\tq\s \tr) \bdto \tk'$ is
identical, using Associativity: $w,l,\tk_0
\models \tk\s(\tq\s \tr) \bdto \tk'$ iff there exists a key $\tk''$ such
that
$w,l,\tk_0 \models \tk\s \tq \bdto \tk''$ and $w,l,\tk_0 \models
(\tk'')\s \tr \bdto \tk'$ iff there exists a key
$\tk''$ such that $\tk\s \tq \bdto \tk''
\in S$ and $(\tk'')\s \tr \bdto \tk' \in S$ iff $\tk\s(\tq\s \tr) \bdto
\tk' \in S$.

We now continue with our induction in the case that $\tp \bdto \tq$.
Note that $w,l,\tk_0 \models \tp \bdto \tq$ iff $w,l,\tk_0 \models \tq
\bdto \tk'$ implies $w,l,\tk_0 \models \tp \bdto \tk'$ for all $\tk' \in
K$ iff (by the induction hypothesis) $\tq \bdto \tk' \in S$ implies $\tp
\bdto \tk' \in S$ iff $\tp \bdto \tq \in S$.  The ``only if'' direction
of the last equivalence follows immediately from Transitivity; the
``if'' direction follows from Witnesses.

We complete the induction proof by observing that if $\psi$ is of the
form $\neg \psi$ or $\psi_1 \land \psi_2$, the result follows
immediately from the induction hypothesis.

To show that $l$ is consistent with $w$, suppose that $\tn \bdto \tp \in
c(\tk)$.  By construction, this means that
$\tk\cert (\tn \bdto \tp) \in S$.  By Key Linking, we must also have
$\tk\s \tn \bdto \tk\s \tp \in S$. By what we have just
shown, $w,l,\tk_0 \models \tk\s \tn \bdto \tk\s \tp$.
It follows that $w,l,\tk \models \tn \bdto \tp$.
Thus,  $l$ is consistent with $w$.
\eprf

This completes the proof of Theorem~\ref{complete} in the case that $K$
is finite.  Note that since we can assume without loss of generality
that $|K| \le \bound$ here (otherwise the argument for the case that
$K$ is infinite applies) the proof also shows that
Proposition~\ref{finitemodel} holds. \eprf

\bigskip

\othm{same}
The same formulas are c-valid and o-valid;
i.e., for all
formulas $\phi$, we have $\omodels \phi$ iff $\cmodels \phi$.
\eothm

\medskip

\prf
We show that $\neg \phi$ is o-satisfiable iff $\neg \phi$ is
c-satisfiable, which is equivalent to the claim.  The direction from
c-satisfiability to o-satisfiability is straightforward:  Since for
every world $w$ the local name assignment $l_w$ is $w$-consistent, it
follows from $w, \tk\cmodels \neg \phi$ that $w, l_w,\tk\omodels
%joe10: proof rewritten
\neg \phi$. Thus, it remains to show that if $\neg \phi$ is
o-satisfiable, then it is c-satisfiable.

So suppose that $\neg \phi$ is o-satisfiable.  By
Proposition~\ref{finitemodel}, there is a world $w = (\beta,c)$, local
name assignment $l$, and principal $\tk$ such that $w,l,\tk \sat_o \neg
\phi$ and a finite subset $K'$ of $K$ such that $l(\tk',\tn) \subseteq
K'$ for all $\tk' \in K$ and $\tn \in N$,
%ron10:
and $\beta(\tg) \subseteq K'$ for all global names $\tg$.
By standard propositional reasoning,
$\neg \phi$ is equivalent to a disjunctive normal
form expression in which the atoms are of the form $\tp \bdto \tq$ and $\tk_1
\cert \psi$, where $\tp$ and $\tq$ are principal expressions, $\tk_1$ is a
key,
and $\psi$ is a formula. If $w,l,\tk\omodels \neg \phi$ then one of the
%joe3: alpha is overloaded
%disjuncts $\alpha$ is satisfied, i.e., $w,l,\tk\omodels \alpha$.
disjuncts $\sigma$ is satisfied, i.e., $w,l,\tk\omodels \sigma$.
Suppose that $\sigma$ is the conjunction of the formulas in the set $A
\union B$, where
\be
\item $A$ is a set of formulas of the form $\tp \bdto \tq$
or $\neg (\tp \bdto \tq)$,
\item $B$ is a set of formulas  of the form $\tk_1 \cert \psi$
or $\neg (\tk_1 \cert \psi)$.
\ee

\newcommand{\tpn}[1]{\tp_{#1,\phi}}

Let $K_\phi$ be the set of
%ron10: principals
keys
that appear in the formula $\phi$
together with $K'$ and $\tk$.  Let $N_\phi$ be the set of local names
that appear in $\phi$.  Define the world $w'= \la \beta',c'\ra$ as follows.
Take the interpretation of global names $\beta'$ to be equal to
$\beta$, the
interpretation of global names in $w$.  Define $c'$ by taking the set
of certificates $c'(\tk')$ to be the empty if $\tk' \notin K_\phi$ and
to consist of $c(\tk')$ together with all certificates of the form $\tn \bdto
%ron10: \tk''_\phi ... here and elsewhere
\tpn{\tk''}$ if $\tk' \in K_\phi$, $\tn \in N_\phi$,
and $\tk'' \in l(\tn,\tk')$, where $\tpn{\tk''}$ is a
principal expression of the form $(\tk'')\s(\tk'')\s\ldots (\tk'')$
that does not appear in $\phi$.  (Clearly we can make the expression
sufficiently long so as to ensure it does not appear in $\phi$.)
Clearly $\union_{\tk' \in K} c(\tk')$ is finite.
%joe10: Here's the old version.  Note that it resulted in an infinite
%set of certificates for each principal if l(n,k_2) is infinite.
%Thus, we would have needed your proof even if we had assumed infinitely
%many names!
%Define the world $w'= \la \beta',c'\ra$ as follows.  Take
%the interpretation of global names $\beta'$ to be equal to $\beta$, the
%interpretation of global names in $w$.  Define $c'$ by taking the set
%of certificates $c'(\tk_1)$ issued by principal $\tk_1$ to be equal to
%$c(\tk_1)$, together with, for each name $\tn$ not in $M$
%and principal $\tk_2$
%such that $\tk_2 \in l(\tn,\tk_1)$, the two certificates $\tn \bdto
%\tm_{\tn}$ and $\tm_{\tn} \bdto \tk_2$. Here $\tm_{\tn}$ is
%required to be a local name in $M$, i.e. one that occurs neither in
%$\phi$ nor in any certificate in $w$. Moreover, a different local
%name $\tm_{\tn}$ is required for each local name $\tn$ not in $M$.

We show that $w',\tk \cmodels \sigma$. It follows from this that $w',\tk
\cmodels \neg \phi$. Note first that from the fact that $c(\tk')
\subseteq c'(\tk')$ for all $\tk'$, it follows that $w',\tk \cmodels
\tk'\cert \psi$ for all formulas $\tk'\cert \psi$ in $B$. Moreover,
if $\neg (\tk'\cert \psi)$ is in $B$ then, since the expressions
$\tpn{\tk''}$ on the right-hand side of the certificates in
$c'(\tk') - c(\tk)$ do not appear in $\phi$ it follows that
$w',\tk \cmodels \neg (\tk'\cert \psi)$.
Thus $w',\tk \cmodels B$.

It remains to show that the formulas in $A$ are satisfied.
To show this, we show that
\begin{equation}\label{fact1}
\mbox{$l_{w'}(\tn,\tk') = l(\tn,\tk')$ for all
$\tn \in N_\phi$ and $\tk' \in K_\phi$.}
\end{equation}
%For this, we show that for each principal $\tk' \in K_\phi$ and each
%characterization of $l_{w'}$:
%%
%\be
%%
%\item for all the local names $\tm_\tn\in M$ and all keys $\tk_1$,
%we have $l_{w'}(\tm_\tn,\tk_1) = l(\tn,\tk_1)$
%\item $l_{w'}(\tn,k_1) = l(\tn,k_1)$ for all local
%names $\tn$ not in $M$ and all keys $\tk$.
%%
%\ee
%
It easily follows from (\ref{fact1}),
%ron10:
the fact that all keys in $\phi$ are in $K'$,
and the fact that global names
have the same interpretation in $w$ and $w'$
%ron10:
%joe11: this is hard to parse;  what interpretation?  (I realize you
%mean the ones that we're about to define.)  I think cutting it will
%actually be less confusing than saying it.
%this interpretation being a subset of $K'$,
%and the definition of $\intension{p}{}$
that $\intension{p}{w',l_{w'},\tk'} = \intension{p}{w,l,\tk'}$ for
all principal expressions $p$ occurring in $A$ and all keys $\tk' \in
K_\phi$. This in turn is easily seen to imply that $w',\tk \cmodels A$.

%joe3: I think the remainder of this proof can be simplified.  I'll deal
%with it later though.
%follows straightforwardly from the fact that the set of
%certificates in $w'$ binding the names $\tm_\tn$ is precisely the
%%set of certificates $\tm_\tn \bdto \tk'$ for $\tk'\in
%l(\tn,\tk_1)$.
It remains to prove (\ref{fact1}).
%So suppose $\tn \in N_\phi$ and $\tk' \in K_\phi$.
It is almost immediate from the
definition of $l'$ that
%$M$. Note that it follows from claim (1) and the
%fact that $\tk_1$ issues a certificate $\tn \bdto \tm_\tn$ that
$l_{w'}(\tn,\tk') \supseteq  l(\tn,\tk')$
for all $\tn \in N_\phi$ and $\tk' \in K_\phi$.
For the opposite containment, we prove
by induction on $j$ that $(T_{w'}\uparrow j)(\tn,\tk')  \subseteq
l(\tn,\tk')$
for all $j\in \Nat$, $\tn \in N_\phi$, and $\tk' \in K_\phi$.
The base case $j=0$ is trivial. For the induction
step, suppose that $j = j'+1$ and $\tk'' \in (T_{w'}\uparrow
j)(\tn,\tk')$.  Thus,
%ron10: $\tk'' \in T_{w'}(T_{w'} \uparrow j'))(\tn,\tk')$, ... paren
        $\tk'' \in (T_{w'}(T_{w'} \uparrow j'))(\tn,\tk')$,
which means that $\tk'' \in \intension{\tp}{w',T_{w'}\uparrow j',\tk'}$
for some
%ron10: principle
principal
expression $\tp$ such that $\tn \bdto \tp \in
c'(\tk')$.  There are two possibilities:  (1) $\tn \bdto \tp \in
c(\tk')$ or (2) $\tn \bdto \tp \in c'(\tk') - c(\tk')$.  In case (2),
$\tp$ must be of the form
%ron10: $\tpn{\tk''}$, -- resolved confusion here about k''
$\tpn{\tk_1}$
so
%ron10:$\intension{\tp}{w',T_{w'}\uparrow j',\tk'} = \tk''$ and $\tk_1 =\tk''$.
$\intension{\tp}{w',T_{w'}\uparrow j',\tk'} = \{\tk_1\}$ and $\tk_1 =\tk''$.
But in this case, by construction, $\tk'' \in l(\tn,\tk')$.
In case (1), using the induction hypothesis and the fact that global
names
%ron10:
and keys in $\tp$
have the same interpretation in $w$ and $w'$
%ron10:
%joe11: left it in, but put it in parens
(this interpretation being a subset of $K'$),
we get that
$\intension{\tp}{w',T_{w'}\uparrow j',\tk'} \subseteq
\intension{\tp}{w,l,\tk'}$.  Thus, $\tk'' \in
\intension{\tp}{w,l,\tk'}$.
Because $l$ is $w$-consistent and $\tn \bdto \tp \in c(\tk')$,
we again obtain that  $\tk''\in l(\tn,\tk')$, as required.
%
%\be
%
%\item[] {\bf Case 1:} There exists a certificate $\tn\bdto p$ in
%$c(\tk_1)$ such that $\tk' \in \intension{p}{w',T_{w'}\uparrow j',
%\tk_1}$.  We have by the inductive hypothesis that $T_{w'}\uparrow j'
%\leq_{N\setminus M} l$. Thus, using Lemma~\ref{lem:int:mon}
%and the fact that all local names in $p$ are in $N\setminus M$
%it follows that $\tk' \in \intension{p}{w',l,\tk_1}$.
%Moreover, since global names have the same interpretation in $w$
%and $w'$ it follows that $\tk' \in \intension{p}{w,l,\tk_1}$.
%Because $l$ is $w$-consistent we obtain that  $\tk'\in l(\tn,\tk_1)$,
%as required.
%\item[] {\bf Case 2:} There exists a certificate $\tn \bdto
%\tm_{n,\tk_1}$ in $c'(\tk_1)$ and $\tk'\in \intension{\tm_\tn}
%{w',T_{w'}\uparrow j',\tk_1}$. Thus, $\tk'\in (T_{w'}\uparrow
%j')(\tm_\tn,\tk_1)$.  By the observation above,
%$ (T_{w'}\uparrow j')(\tm_\tn,\tk_1) \subseteq l(\tn,\tk_1)$,
%so again $\tk'\in l(\tn,\tk_1)$.

%\ee
%
Since $l_{w'}(\tn,\tk')$ is the union of the $( T_{w'}\uparrow
j)(\tn,\tk')$, it follows that $l_{w'}(\tn,\tk') =
l(\tn,\tk')$. This completes the proof of (\ref{fact1}).
\eprf

\bigskip

\opro{independent}
Let $\Gamma$ be any
%ron10:
c-satisfiable
boolean combination of formulas of the form $\tk
\cert \phi$, and let $\Delta$ be any boolean combination of formulas of
the form $\tp \bdto \tq$ where neither $\tp$ nor $\tq$ contains a
local name. Then $\cmodels \Gamma\limp \Delta$ iff
$\cmodels \Delta$.
\eopro

\medskip

\prf Clearly $\cmodels \Delta$ implies $\cmodels \Gamma \limp \Delta$.
For the converse, suppose by way of contradiction that $\cmodels
\Gamma \limp \Delta$ and there is a world $w= \la\beta,c\ra$ and a
principal $\tk$ such that $w,\tk \cmodels \neg \Delta$.
%ron10: corrected proof for this bit
%Our goal now is to construct a world
%$w'$ such that $w',\tk \cmodels \Gamma \land \neg \Delta$.  This is
%easy.  Suppose that $w = (\beta,c)$.
%Let $\Gamma_{\tk'}$ consist of all the formulas $\psi$ such that $\tk'
%\cert \psi \in \Gamma$ and define $w' = (\beta,c')$, where
%$c(\tk')  = \Gamma_{\tk'}$.  By construction, $w',\tk \cmodels \Gamma$.
%Moreover, if $\tp$ is a principal expression that appears in $\Delta$,
%then $\intension{\tp}{w,l_w,\tk} = \intension{\tp}{w,l_{w'},\tk}$, since
%$\tp$ does not contain a local name, by assumption.  Thus, for every
%formula of the form $\tp \bdto \tq$ that appears in $\Delta$, we have
%have that $w,\tk \cmodels \tp \bdto \tq$ iff $w',\tk \cmodels \tp \bdto \tq$.
%It follows that $w',\tk \cmodels \neg \Delta$, giving us our desired
%contradiction.
Since $\Gamma$ is assumed to be c-satisfiable, there exists a world
$w'=\la \beta',c'\ra$ and a principal $\tk'$ such that
$w',\tk'\cmodels \Gamma$. Let $w''$ be the world $\la \beta,c'\ra$.
Then a straightforward induction shows that for all principal
expressions $\tp$ not containing a local name, we have
$\intension{\tp}{w'',l_{w''},\tk} = \intension{\tp}{w,l_{w},\tk}$.
Moreover, for all keys $\tk_1$ and formulas $\phi$, we have $w'',\tk
\cmodels \tk_1 \cert \phi$ iff $w',\tk'\cmodels \tk_1\cert \phi$. It
follows that $w'',\tk \cmodels \Gamma \land \neg \Delta$, giving us
our desired contradiction.
%ron10: by using \tk as the key in the above, it works for self as well
\eprf

\bigskip

\othm{REF2thm}
Suppose $\tk_1,\tk_2$ are principals, $w = (\beta,c)$ is a world, and
$\tp$ is a principal expression.
%ron10: Let the formula $E_w$ consist of the conjunction of the
Let $E_w$ be the set of all the
formulas $\tg \bdto \tk$ for all
global names $\tg$
%ron10: occurring in $\tp$ -- OK, but it adds pain to the proof for little gain
%       see ron10*
and keys $\tk \in \beta(\tg)$ and the formulas
$\tk \cert \phi$ for all keys $\tk$ and formulas $\phi \in c(\tk)$.
The following are equivalent:
\be
\item $\tk_1\in \reft(\tk_2,\beta,c,\tp)$,
\item $w,\tk_2 \cmodels \tp\bdto \tk_1$,
\item $w',\tk_2 \cmodels \tp\bdto \tk_1$ for all
worlds $w'\geq w$,
%ron10: \item $\models_c E_{w} \limp (\tk_2\s \tp \bdto \tk_1)$,
%       \item $\models_o E_{w} \limp (\tk_2\s \tp \bdto \tk_1)$.
\item $E_w \models_c \tk_2\s \tp \bdto \tk_1$,
\item $E_w\models_o \tk_2\s \tp \bdto \tk_1$.
\ee
\eothm

\medskip

\prf
%The equivalence between (1) and (2) follows by a straightforward
%induction on the structure of $\tp$.  That is, we prove by induction
%on the structure of $\tp$ that for all principals $\tk_1, \tk_2$,
%we have $\tk_1\in \reft(\tk_2,\beta,c,\tp)$ iff
%$w,\tk_2 \cmodels \tp\bdto \tk_1$.  We leave details to the reader.
%ron10: no, a simple induction on p breaks for p=n, which requires
%       a call to REF2 for an arbitrary size q s.t. k cert n\bdot q
%       put back in the discussion from the Jan version & revised
The presentation of $\reft$ in Figure~\ref{fig:reft}
is still slightly informal, combining
recursion and nondeterminism. To make it fully precise,
define a  \emph{computation tree\/} of $\reft$ to be a finite tree
labelled by expressions of the form ``$\tk_1\in \reft(\tk_2,\beta,c,\tp)$'',
such that if $N$ is a
node so labelled, then
%joe11
%either
one of the following four conditions holds:
\be
\item $\tp$ is a key $\tk$, we have $\tk=\tk_1=\tk_2$, and $N$ is a leaf of the tree,
%joe11: removed: or
\item $\tp$ is a global name $\tg$ and $\tk_1 \in \beta(\tg)$,
\item $\tp$ is a local name $\tn$ and
$c(\tk_2)$ contains a formula $\tn\bdto \tq$
and $N$ has exactly one child, labelled
``$\tk_1\in \reft(\tk_2,\beta,c,\tq)$'',
\item $\tp$ is of the form $\tq\s \tr$ and $N$ has exactly two children,
labelled
``$\tk\in \reft(\tk_2,\beta,c,\tq)$'' and
``$\tk_1\in \reft(\tk,\beta,c,\tr)$'', for some key $\tk$.
\ee
We take $\tk_1\in \reft(\tk_2,\beta,c,\tp)$ to mean that
there exists a computation tree of $\reft$ with root labelled
``$k_1\in \reft(\tk_2,\beta,c,\tp)$''.

Given a world $w= \la\beta,c\ra$ and $m\in \Nat$,
%joe11: write ->
let
$l_m = T_w\uparrow m$.
The following result establishes a correspondence between the stages of
the computation of $l_{w}$ and the computation trees of \reft.
The proof is by a straightforward induction on $m$, with a
%joe11
%subargument by induction on the structure of $\tp$.
subinduction on the structure of $\tp$.

\begin{lemma} \label{lem:int:tree}
For all $m\in \Nat$, keys $\tk_1,\tk_2$,
worlds $w=\la \beta,c\ra$,
and principal expressions
$\tp$, we have $\tk_1 \in \intension{\tp}{w,l_m,\tk_2}$ iff
there exists a computation tree of $\reft$ of height at most $m$
whose root is labelled {\rm ``}$\tk_1\in \reft(\tk_2,\beta,c,\tp)${\rm
''}.
\end{lemma}

Using
%ron10:
the fact that $l_w = \lub \setcpr{l_m}{m\in \Nat}$,
Lemma~\ref{lem:int:lim}, and Lemma~\ref{lem:int:tree},
we obtain the equivalence between (1) and (2).

%ron10: The implication from (2) to (3) also follows by
The proof of the implication from (2) to (3) is by
a straightforward induction
on the structure of $\tp$; that is, for fixed $w' \ge w$, we show by
induction on the structure of $\tp$ that if $w,\tk_2 \cmodels \tp\bdto
\tk_1$ then $w',\tk_2 \cmodels \tp\bdto \tk_1$.  The opposite
implication from (3) to (2) is trivial, since $w \ge w$.
For the implication from (3) to (4), suppose that (3) holds and (4) does
not.
%ron10: Then $w',\tk_2 \cmodels E_{w} \land \neg(\tk_2\s \tp \bdto \tk_1)$.
Then for some world $w'$ and key $\tk$ we have $w',\tk \cmodels E_{w}$
and $w', \tk\cmodels \neg(\tk_2\s \tp \bdto \tk_1)$.
The latter implies $w', \tk_2\cmodels \neg(\tp \bdto \tk_1)$.
%ron10*: following would be not quite right were g restricted to p
%        fix tk_2 -> tk an independent issue
%Since $w',\tk_2 \cmodels E_w$, it follows that $w' \ge w$.
Since $w',\tk \cmodels E_w$, it follows that $w' \ge w$.
Thus, by (3), $w',\tk_2 \cmodels \tp\bdto \tk_1$, contradicting our
assumption.  The implication from (4) to (3) is immediate, since
$w',\tk_2 \cmodels E_w$ for all $w' \ge w$.  Finally, the equivalence
between (4) and (5) is just a special case of Theorem~\ref{same}.
\eprf

%ron10: added proofs for logic programming representation
\opro{prop:represents}
If $M$ represents $w$ and $l$ then for all principal expressions
$\tp$ and $x,y\in K\cup G\cup N$ we have $M \models \trans{x,y}{\tp}$ iff
$x,y\in K$ and $w,l,x\models \tp\bdto y$.
\eopro

\prf
By a straightforward induction on the structure of $\tp$.
The base cases, where $\tp\in K\cup G\cup N$, are immediate
from the definition of ``represents'' and the semantics of the logic.
The inductive case, where $\tp = \tq\s \tr$, is
%joe11
%direct
immediate
from the semantics and the definition of the translation.
\eprf

\othm{Herbrand} The minimal Herbrand model $M_w$ of
$\Sigma_w$ represents $w$ and $l_w$.
\eothm

\prf (Sketch)
%ron10: handwaving a bit, I don't want to go through the
% whole of logic programming theory here!
The proof proceeds by showing a direct correspondence between the
construction of the minimal Herbrand model of $\Sigma_w$ and
the fixpoint construction of $l_w$.

%joe11: Ron, can you send me this .bib item?
The theory of logic programming \cite{Lloyd} associates
with the Horn theory $\Sigma_w$ an operator $\Phi_w$ on the
space of Herbrand models on the vocabulary $V$, defined by
$\name(x,y,z) \in \Phi_w(M)$ if there exists a substitution instance
of a formula in $\Sigma_w$ of the form $B\rimp \name(x,y,z)$
such that $M \models B$. The least Herbrand model
$M_w$ of $\Sigma_w$ is then equal to $\Phi_w\uparrow \omega =
\bigcup_{m\in \Nat} \Phi_w \uparrow m$,
where $\Phi_w\uparrow 0 = \emptyset$ and
$\Phi_w\uparrow m+1 = \Phi_w(\Phi_w\uparrow m)$ for $m \geq 0$.

Let $T_w$ be the operator on local name assignments defined in the
proof of Theorem~\ref{thm:min:lna}. Using
Proposition~\ref{prop:represents} to handle the rules in $\Sigma_w$
corresponding to certificates, we may then show by a straightforward
induction on $m$ that for all $m\geq 1$, the Herbrand
model $\Phi\uparrow m$ represents the world $w$ and the local name
assignment $T_w \uparrow m$. It follows that
$M_w = \Phi\uparrow \omega$ represents $l_w = T_w \uparrow \omega$.
\eprf

\bigskip

\othm{completeself}
$\AXinfself$ (resp., $\AXfinself$) is a sound
and complete axiomatization of \llncs\ with respect to
the open semantics if $K$ is infinite (resp., $K$ is finite).
\eothm

\medskip
\prf The argument is very similar to that in the proof of
Theorem~\ref{complete}.  First suppose that $K$ is infinite.

We add the following clauses to the definition of $P$:
\be
\item[6.] $\self\in P$,
\item[7.] if $\tn\in P$ is a local name then $\self\s \tn \in P$.
\ee
We also add the following clauses to the definition of $\clse$,
corresponding to the new axioms for $\self$.
\be
\item[(ClSP)] if $\self\s \tp \in P$ then
$\self\s \tp \bdto \tp \in \clse$ and $\tp \bdto \self\s \tp\in \clse$,
\item[(ClPS)] if $\tp\s \self\in P$ then
$\tp\s \self\bdto \tp \in \clse$ and  $\tp \bdto \tp \s \self \in \clse$,
%joe11: unncessary
%\item[(ClSN)]
% if  $\tp \bdto \self \in \clse$ and $\tp\s \tk \in P_1$
%then $\tp\s \tk \bdto \tk \in \clse$
\item[(ClSE)] if $\self \bdto \tp\in \clse$ and
$\tp\s \tk \bdto \tk \in \clse$ then $\tp \bdto \self\in \clse$.
\ee

Lemma~\ref{k0inclse} still applies. The definitions following this
lemma, up to and including that of $\clsem$ are unchanged.
\newcommand{\tks}{\tk_*}
However, the construction of the model changes slightly. We no longer
use $\tk_0$ to represent the ``current principal'', instead, we use
the key $\tks$ that the construction associates with $\self$. This
could be either a key in $P_1$ or one of the keys $\tk_c$ for
$c\in O$, depending on whether $\self$ is key-equivalent or open.
Note that we cannot have $\self$ empty
%joe11
(thanks to the Identity axiom).
%For, we have
%$\self \bdto \self \in \clse$, so by (ClSN) we have $\self\s \tk_0 \bdto
%\tk_0 \in \clse$.
If $\self$ is key-equivalent, then by (ClKD) it is
equivalent to at most one key $\tk\in P$. In this case,
we define $\tks = \tk$.  If
$\self$ is open we define $\tks$ to be $\tk_c$,
where $c=[\self]$.

We now define $w$ and $l$ exactly as before, except that we now
set $l(\tk_0, \tn) = \emptyset$, since we no longer use $\tk_0$ as the
``current principal.''
The following lemma is the analogue of Lemma~\ref{lem:intp}.

\begin{lemma} \label{lem:intps}
For all expressions $\tp\in P$, we have
$\intension{\tp}{w,l,\tks} = I(\tp)$.
\end{lemma}

\prf
The proof is very similar to that of Lemma~\ref{lem:intp}; we
just describe the modifications required. The base cases for $\tp$ a
global name or a key are identical.

When $\tp = \tn$ is a local name, we proceed as follows.  There are
two possibilities, depending on whether $\tks\in P$ or not.  Suppose
first that $\tks\in P$. Then we have $\tks \approx \self$ and, by
(ClLM) and (ClSP), $\tks\s \tn \approx \self\s\tn \approx \tn$.  It
%joe12: \tk' -> \tk
then follows by (ClT) and construction of $l$ that $\tn \bdto \tk\in
\clsem$ iff $\tks\s \tn \bdto \tk\in \clsem$ iff $\tk \in l(\tks,
\tn)$, as required.

%joe12: rewrote slightly
If $\tks= \tk_c$ for $c$ an open class,
we proceed as follows.
%joe12: typo + overloaded p
%$\tk'\in \intension{\tp}{w,l,\tks}$ iff $\tp\s
%\tn \bdto \tk'$ for some $\tp \approx \self$.
If $\tk \in I(\tn)$, then we consider two cases, depending on whether
$\tk \in P_1$.  If $\tk \in P_1$, then
% and prove that $\tk'\in
%\intension{\tp}{w,l,\tks}$. There are two possibilities, depending on
%whether $\tk'\in P_1$. If $\tk'\in P_1$, then
$\tn \bdto \tk\in
\clse$ and it follows that $\self\s \tn \bdto \tk$ by (ClSP) and
(ClT). Since $\self \approx \self$ it is immediate that $\tk\in
\intension{\tp}{w,l,\tks}$.  Alternatively, if $\tk = \tk_d$, for
$d\in O$, then we have $\tn \bdto \tq \in \clse$ for some $\tq \in d$.
By (ClSP) and (ClT) it follows that $\self\s \tn \bdto \tq \in \clse$,
hence $\self\s \tn \bdto \tk \in \clsem$. As before, this implies that
$\tk\in \intension{\tn}{w,l,\tks}$.

For the opposite inclusion, suppose that $\tk\in
\intension{\tn}{w,l,\tks}$.  Since we are assuming that $\self$ is open,
%joe12 p overloaded again; also missing "\in \clsem"
%we have $\tp\s\tn \bdto \tk'$ for some $\tp \approx \self$.
there must be some $\tq \approx \self$ such that $\tq\s\tn \bdto \tk
\in \clsem$.
By (ClLM), we have $\self\s \tn \bdto \tq\s \tn \in \clse$.
It follows using (ClT) that 
%ron11: $\self\s \tn \bdto \tk' \in \clsem$, hence $\tn \bdto \tk' \in \clsem$.
% no primes here 
$\self\s \tn \bdto \tk \in \clsem$, hence $\tn \bdto \tk \in \clsem$.
This completes the argument
for the base case of $\tn $ a local name.

There is now an additional base case for $\tp = \self$.
Here, note that $\intension{\self}{w,l,\tks} = \{\tks\}$.
We therefore need to show that $\self \bdto \tk \in \clsem$ iff
$\tk = \tks$. When $\tks \in P_1$, we have $\self \approx \tks$,
so $\self \bdto \tk \in \clsem$ iff $\tks \bdto \tk$, and
the claim follows by (ClKD) and (ClT) as in the base case for keys.
The alternative is that $\tks = \tk_c$ for $c= [\self] \in O$.
Since have $\self \bdto \tk_c \in \clsem$ by construction of
$\clsem$,  it remains to prove that if $\self \bdto \tk \in \clsem$
then $\tk = \tk_c$. Now we cannot have $\self \bdto \tk \in \clsem$
for $\tk \in P_1$, for then by the argument above that
$\self$ is nonempty and (ClSE), we have $\tk \bdto \self \in \clse$,
contradicting the assumption that $c$ is open.
Thus, we must have $\tk = \tk_d$ for some $d\in O$.
In this case, there exists $\tq \in d$ such that $\self \bdto \tq \in \clse$.
Since $d$ is open, we have $\tq\s \tk_0 \bdto \tk_0 \in \clse$, hence
$\tq \bdto \self \in \clse$ by (ClSE). Thus, $\self \approx \tq$, and
it follows that $d = c$, hence $\tk = \tks$ as required.
This completes the argument for the base case where $\tp = \self$.

The inductive case is exactly as before, except that we need to
consider the new case $\tp\s \self$.
Here, we note that $\intension{\tp\s \self}{w,l,\tks} =
\intension{\tp}{w,l,\tks}$. Thus, by the induction hypothesis,
we are required to prove that $\tp\bdto \tk \in \clsem$ iff $\tp\s
\self \bdto \tk \in \clsem$.  This follows using (ClPS) and (ClT).
\eprf

The remainder of the proof in the case that $K$ is infinite proceeds as
before, using $\tks$ in place of $\tk_0$.

%joe12: added my few words
%If $K$ is finite, \ldots [[I'LL ADD A FEW WORDS HERE]].
If $K$ is finite, the proof is even closer to that for the logic without
$\self$.  As sketched in the main text, because $S$ is consistent,
it follows from Identity, Witnesses, and Self-is-key that there must be
some key $\tks \in K$ such that $\self \bdto \tks \in S$.  For this
key $\tks$, we must have $\tks\s \tn \bdto \tk \in S$ iff $\tn \bdto \tk
\in S$.  Thus, $\tks$ plays the role of $\tk_0$ in the earlier argument.
(Note that we now no longer need Current Principal to ensure the
existence of $\tk_0$.)  The rest of the argument is unchanged.)
\eprf
%joe10

%joe14: moved this here per journal style
\section*{Acknowledgments}
%ron12: acknowledge UTS 
Work on this paper was done while the second author was with the School
of Computing 
Sciences, University of Technology, Sydney. 
This work was supported in part by NSF under
grant IRI-96-25901 and by a UTS internal research grant.
%joe10: added
A preliminary version of this 
%ron1:  
paper 
appeared in the {\em Proceedings of the
12th IEEE Computer Security Foundations Workshop}, 1999, pp.~111--122.

\bibliographystyle{alpha}
%joe6: IEEE style
%\bibliographystyle{latex8}
%joe6: now put everything in joe.bib, to reduce number of bib files
\bibliography{joe,z}

\begin{thebibliography}{HvdMS99}

\bibitem[Aba98]{Abadi98}
M.~Abadi.
\newblock On {SDSI}'s linked local name spaces.
\newblock {\em Journal of Computer Security}, 6(1-2):3--21, 1998.

\bibitem[ABLP93]{ABLP93}
M.~Abadi, M.~Burrows, B.~Lampson, and G.~D. Plotkin.
\newblock A calculus for access control in distributed systems.
\newblock {\em ACM Transactions on Programming Languages and Systems},
  15(4):706--734, 1993.

\bibitem[BFL96]{BFL96}
M.~Blaze, J.~Feigenbaum, and J.~Lacy.
\newblock Decentralized trust management.
\newblock In {\em Proceedings 1996 IEEE Symposium on Security and Privacy},
  pages 164--173, 1996.

\bibitem[Bir67]{Birkhoff67}
G.~Birkhoff.
\newblock {\em Lattice Theory}.
\newblock American Mathematical Society, Providence, R.I., 3rd edition edition,
  1967.

\bibitem[EK76]{vanEmdenKowalski86}
{M.H. van} Emden and R.~A. Kowalski.
\newblock The semantics of predicate logic as a programming language.
\newblock {\em Journal of the ACM}, 23(4):733--742, 1976.

\bibitem[FHMV95]{FHMV}
R.~Fagin, J.~Y. Halpern, Y.~Moses, and M.~Y. Vardi.
\newblock {\em Reasoning about Knowledge}.
\newblock MIT Press, Cambridge, Mass., 1995.

\bibitem[GH93]{GroveH2}
A.~J. Grove and J.~Y. Halpern.
\newblock Naming and identity in propositional logics, {P}art {I}: the
  propositional case.
\newblock {\em Journal of Logic and Computation}, 3(4):345--378, 1993.

\bibitem[Gro98]{SPKI}
SPKI~Working Group.
\newblock Simple public key infrastructure, internet draft.
\newblock at {\tt http://www.ietf.org/html.charters/spki-charter.html}, 1998.

\bibitem[HM90]{HM1}
J.~Y. Halpern and Y.~Moses.
\newblock Knowledge and common knowledge in a distributed environment.
\newblock {\em Journal of the ACM}, 37(3):549--587, 1990.
\newblock A preliminary version appeared in {\em Proc.~3rd ACM Symposium on
  Principles of Distributed Computing}, 1984.

\bibitem[HvdM99]{HM00}
J.~Y. Halpern and R.~van~der Meyden.
\newblock Adding revocation and timestamps to a logic for sdsi's linked local
  name spaces.
\newblock unpublished manuscript, 1999.

\bibitem[HvdMS99]{HMS}
J.Y. Halpern, R.~van~der Meyden, and F.~Schneider.
\newblock Logical foundations for trust management.
\newblock manuscript, 1999.

\bibitem[LABW92]{LABW92}
B.~Lampson, M.~Abadi, M.~Burrows, and E.~Wobber.
\newblock Authentication in distributed systems: Theory and practice.
\newblock {\em ACM Transactions on Computer Systems}, 10(4):265--310, 1992.

\bibitem[Llo87]{Lloyd}
J.~W. Lloyd.
\newblock {\em Foundations of Logic Programming}.
\newblock Springer-Verlag, Berlin, 2nd edition edition, 1987.

\bibitem[LNS82]{LNS82}
J.-L. Lassez, V.~L. Nguyen, and E.~A. Sonenberg.
\newblock Fixed point theorems and semantics: a folk tale.
\newblock {\em Information Processing Letters}, 14(3):112--116, 1982.

\bibitem[RL96]{RL96}
R.L. Rivest and B.~Lampson.
\newblock {SDSI} --- a simple distributed security infrastructure.
\newblock at {\tt http://theory.lcs.mit.edu/$\sim$cis/sdsi.html}, 1996.

\bibitem[Ull88]{UllmanKB1}
J.~D. Ullman.
\newblock {\em Principles of Database and Knowledge Base Systems, Volume I}.
\newblock Computer Science Press, 1988.

\bibitem[Ull89]{UllmanKB2}
J.~D. Ullman.
\newblock {\em Principles of Database and Knowledge Base Systems, Volume II:
  The New Technologies}.
\newblock Computer Science Press, 1989.

\end{thebibliography}

\end{document}